%% file: main.tex
\documentclass[manuscript=article]{achemso}

\usepackage{amsmath,amssymb,bm}
\usepackage{booktabs}
\usepackage{tabularx}
\usepackage{longtable}
\usepackage{float}
\usepackage{placeins}
\usepackage{graphicx}
\usepackage[version=4]{mhchem}
\usepackage{xcolor}
\usepackage{tikz}
\usetikzlibrary{arrows.meta,positioning,shapes.geometric,fit}
\usepackage[hidelinks]{hyperref}

\newcolumntype{Y}{>{\raggedright\arraybackslash}X}

\title{Continuous perception and adaptive metrics for scalable structure
refinement}

\author{Vincenzo Barone}
\affiliation{Consorzio Interuniversitario Nazionale per la Scienza e Tecnologia
dei Materiali (INSTM), via G. Giusti 9, 50121 Firenze, Italy}
\email{vincebarone52@gmail.com}


\begin{document}

\begin{tocentry}
\centering
\includegraphics[width=\linewidth]{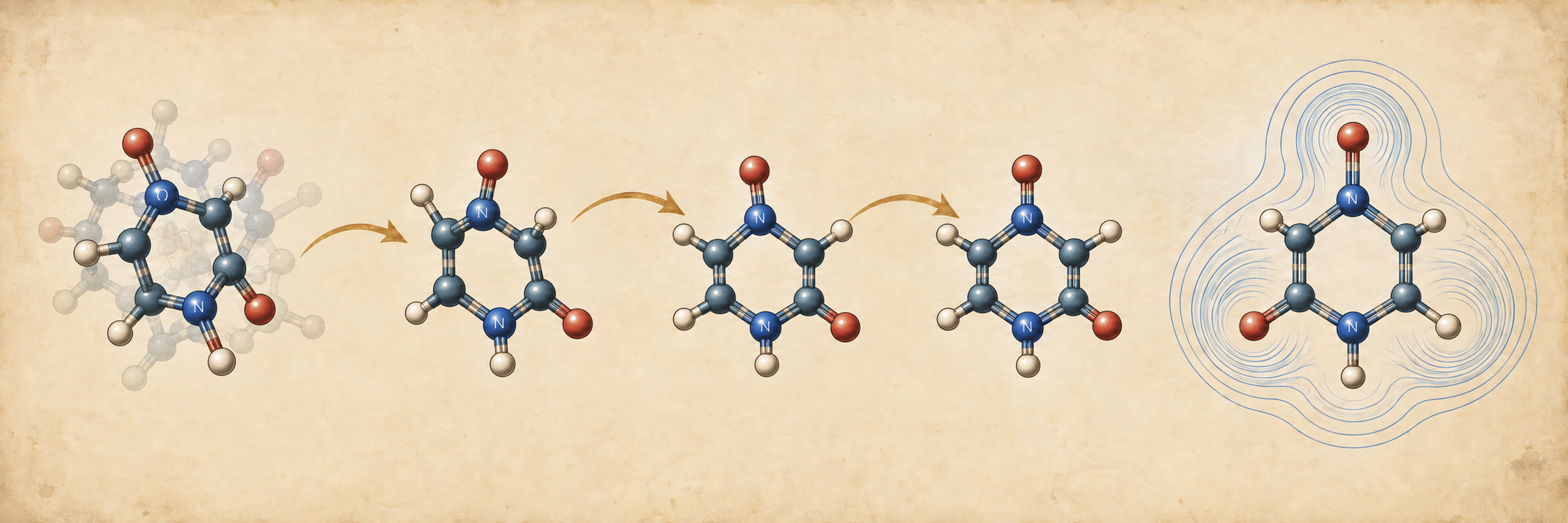}
\end{tocentry}

\begin{abstract}
We present a geometry-refinement workflow that adapts to the cost of an
energy--gradient evaluation and the reliability of molecular topology.  An
initial Cartesian calculation and continuous bonding diagnostics determine
whether to begin in projected Cartesian coordinates or in chemically adapted
internal coordinates.  Chemical perception generates local redundant
measurements; sparse rank and conditioning tests combine them into
nonredundant, symmetry-adapted coordinates; and a common optimizer uses the
selected representation with resident or external calculators.

For inexpensive gradients, the projected Cartesian route limits coordinate
overhead; a single monitored internal-coordinate step can then provide
chemical interpretability.  When gradients are costly and topology is
certified, symmetry-adapted internal coordinates reduce expensive
energy--gradient calls.  A quadraticized map improves localized anharmonic
cases, while mixed Cartesian/internal charts cover uncertain or complex
topologies.  Local sparse operators and resident force-field and tight-binding calculators
extend both routes to large systems. The same Cartesian calculator contract
supports multilevel energy and gradient composition. The measured coordinate
pipeline and recurrent resident tight-binding update show near-linear scaling
over the tested size ranges; this statement applies to those qualified
components, not to arbitrary electronic-structure calculations or cold starts.
For certified, directly separable internal charts, the optimizer can use
chemically designed Morse or Gaussian maps to quadraticize the local step
problem. The quadraticized branch uses a short geometry-local history, while
delocalized charts remain unchanged. This reduces iterations in difficult
minima without changing the electronic potential or coordinate chart.
\end{abstract}

Computational modelling is now routine across chemistry.  Researchers need
powerful tools that let them focus on the chemical question while making the
numerical choices intelligible.  Geometry refinement tests this principle:
the best coordinates depend on both the calculation and the starting
structure.

The first choice is economic.  With an expensive quantum-mechanical gradient,
fewer evaluations can repay the construction of chemically adapted
coordinates.  With a force field or fast semiempirical model, coordinate
transformation and nonlinear realization can instead dominate the calculation
\cite{Schlegel2011Review,Pulay1980DIIS,Broyden1970,Fletcher1970,
Goldfarb1970,Shanno1970,Nocedal1980,DennisMore1977,Goedecker1999,
BowlerMiyazakiGillan2002,RubenssonRudberg2011,MauriGalliCar1993}.  The second
choice is chemical: uncertain bonds or coordination can make an internal
coordinate chart ill conditioned or incorrect.  We call the symmetry-projected
Cartesian space, with overall translations and rotations removed, SYCART;
SONIC denotes symmetry-oriented nonredundant internal coordinates.  FF denotes
a force field and xTB an extended tight-binding model.
Table~\ref{tab:decision} relates these choices.

\begin{table}[t]
\caption{Coordinate and calculator choice from the two independent diagnostics
used by the adaptive workflow.}
\label{tab:decision}
\centering
\footnotesize
\begin{tabularx}{\linewidth}{@{}lYY@{}}
\toprule
 & Topology reliable & Topology uncertain \\
\midrule
Low-cost gradient & Resident FF/SYCART for rapid preparation, or direct
SYCART when no preparation is needed & Resident xTB/SYCART until chemical and
numerical topology tests pass \\
\midrule
High-cost gradient & SONIC, to minimize expensive evaluations; use the
quadraticized map for localized anharmonic charts & Lower-level Cartesian
preparation followed by perception and high-level SONIC; use a mixed
SYCART/SONIC chart for uncertain or reactive regions \\
\bottomrule
\end{tabularx}
\end{table}

One initial Cartesian energy and gradient diagnose distance from a stationary
region.  Continuous bond orders, graph stability, and a prospective coordinate
certificate establish whether the topology supports internal coordinates.
Otherwise a resident semiempirical calculator prepares the structure in
SYCART.  The test is repeated after accepted geometry changes and can classify
different regions of one system independently.

Three tools implement this decision.  ORACLE records chemical state and local
redundant measurements \cite{Barone2026ORACLE}; SMITH selects nonredundant
coordinates with the required symmetry and degree of delocalization
\cite{Barone2026SONIC}; LINK optimizes in the selected chart using a common
Cartesian calculator interface \cite{Barone2026LINK}.

Quantum backends supply energies, Cartesian derivatives, and, when needed,
orbitals or densities.  SWITCH converts between SMILES and Cartesian structures;
APOC extracts CM5 charges, Mayer bond orders, and related observables from the
electronic density \cite{Barone2026ORACLE}.  Chemical perception therefore
remains independent of the quantum backend.

Earlier LINK calculations refined structures whose topology and coordinate
chart remained stable.  Distorted or exploratory structures require the chart
to change with the geometry.  Our adaptive metric selects the least
delocalized, well-conditioned representation in each certified region and
assigns uncertain regions to SYCART.

The present implementation also introduces a local quadraticization for direct
SONIC optimization. Morse maps are used for nonperiodic stretches and other
coordinates with a finite local range, Gaussian maps for bounded angular
coordinates, and periodic maps only for coordinates explicitly marked
periodic by ORACLE. The analytic pullback metric preserves the physical
Cartesian step and the trust-region acceptance test. Because a delocalized
chart does not provide separable local variables, it is deliberately excluded
from quadraticization. The corresponding short GDIIS memory prevents history
points from different local quadratic models from being mixed.
The notation used below is summarized in Table~\ref{tab:glossary}.

\begin{table}[t]
\caption{Notation and acronym glossary used throughout the manuscript.}
\label{tab:glossary}
\centering
\small
\begin{tabularx}{\linewidth}{@{}lX@{}}
\toprule
Term & Meaning \\
\midrule
QM; TB; FF & Quantum-mechanical; tight-binding; and force-field calculations. \\
\midrule
ORACLE; SMITH; LINK & The chemical-perception, sparse-coordinate, and optimization tools, respectively; their persistent roles are defined in the Methods. \\
\midrule
SONIC; SYCART & Symmetry-oriented nonredundant internal coordinates; and symmetry-projected Cartesian coordinates, obtained after removal of overall translation and rotation. \\
\midrule
GFN-FF; GFN2-xTB; TBLite & Resident force field\cite{Spicher2020}; resident extended tight-binding model; and its in-process implementation. \\
\midrule
ANCopt & Native xTB optimizer used as the external control\cite{xtb671ModelHessian}. \\
\midrule
SH; GIC; CSR & Primitive-diagonal starting-Hessian model; generalized internal coordinate; and compressed-sparse-row representation used for Wilson operators. \\
\midrule
QM1; QM2; QM/MM; ONIOM & One-level; two-level; quantum/classical embedded; and subtractive multilevel calculations. \\
\midrule
RFO; GDIIS; L-BFGS; SVD & Rational-function optimization; geometry direct inversion in the iterative subspace; limited-memory Broyden--Fletcher--Goldfarb--Shanno (L-BFGS); and singular-value decomposition. \\
\midrule
CNA; SOOP; EEQ & Continuous coordination number; selective out-of-plane; and electronegativity equilibration. \\
\midrule
D3; D4; FMM3D & Third- and fourth-generation dispersion corrections; and fast multipole evaluation. \\
\midrule
AO; SCC; DFT & Atomic orbital; self-consistent charge; and density-functional theory. \\
\midrule
RMSD; RSS & Root-mean-square deviation; and resident set size. \\
\midrule
$\mathbf{x}$; $\mathbf{q}$; $\mathbf{B}$; $\mathbf{P}$; $\mathbf{G}$ & Cartesian coordinates; active coordinate vector; Wilson Jacobian $\mathbf{B}=\partial\mathbf{q}/\partial\mathbf{x}$; rigid-motion projector; and row-normalized Wilson metric defined below. Gradients are column vectors, and coordinate selection and conditioning use no atomic masses. \\
\bottomrule
\end{tabularx}
\end{table}

\begin{figure*}[t]
\centering
\resizebox{\textwidth}{!}{\input{figures/pipeline.tex}}
\caption{Adaptive three-tool architecture from perception to local refinement.}
\label{fig:pipeline}
\end{figure*}
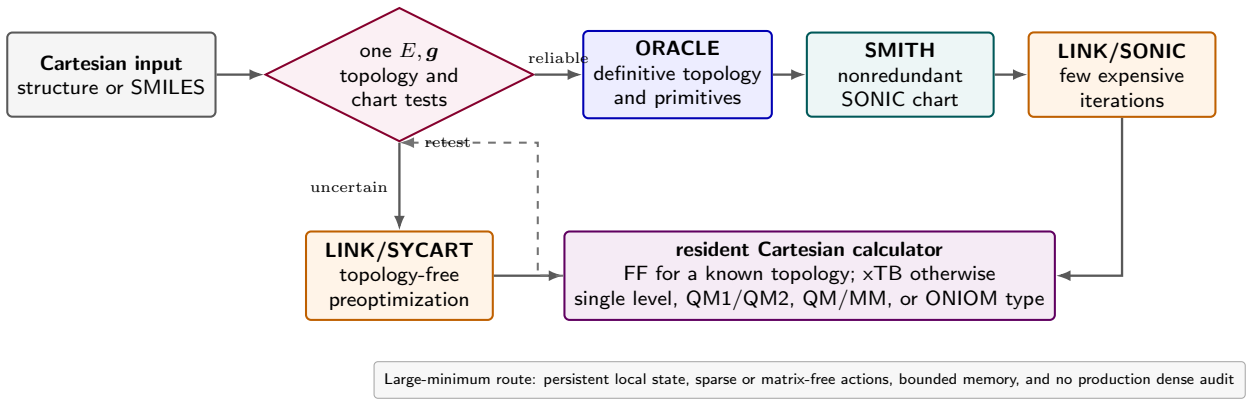

Earlier work addressed reliable topology with costly gradients, where the
construction of SONIC coordinates was inexpensive by comparison.  Here the
extension is structural: ORACLE selectively updates an extensible local
primitive atlas; SMITH builds and certifies sparse charts with only the
delocalization needed for conditioning; and LINK changes representation with
bounded-memory, matrix-free actions.  Resident force-field and tight-binding
calculators use local sparse kernels and supply both low-cost preparation and
the low-level component of multilevel calculations
\cite{WarshelLevitt1976,FieldBashKaru1989,Vreven2006ONIOM,
Grimme2010D3,Grimme2011D2,PrachtGrantGrimme2020}.  These changes are compared
with the preceding work in Table~\ref{tab:boundary};
Figure~\ref{fig:pipeline} shows the complete flow.

\begin{table}[t]
\caption{Established three-tool capabilities and the present extensions.}
\label{tab:boundary}
\centering
\footnotesize
\begin{tabularx}{\linewidth}{@{}lYY@{}}
\toprule
Layer & Foundation & Present extension \\
\midrule
ORACLE & Persistent chemical state; overlapping-fragment route to large
molecules & Selective local reperception, neighbor/graph kernels, a broader
extensible primitive atlas, sparse primitive rows, and a non-dense
geometry-identity test; xTB seed construction and shared ANC treatment \\
\midrule
SMITH & Exact-rank typed SONIC; symmetry, rings, multicenter and fragment-pose
families; sparse Wilson rows & Incremental nonredundant-chart construction,
exact sparse support certificate, bounded locality partition, and an explicit
Cartesian substrate block in composite charts; SH congruence from the ORACLE
primitive diagonal \\
\midrule
LINK & Backend-neutral full-metric optimization, nonlinear realization,
state transport, and calculator portability & Adaptive SYCART/SONIC selection,
profile-guided recurrent SONIC construction, SH initialization, bounded-memory
matrix-free steps, partial optimization, and composite-gradient consumption \\
\midrule
\bottomrule
\end{tabularx}
\end{table}

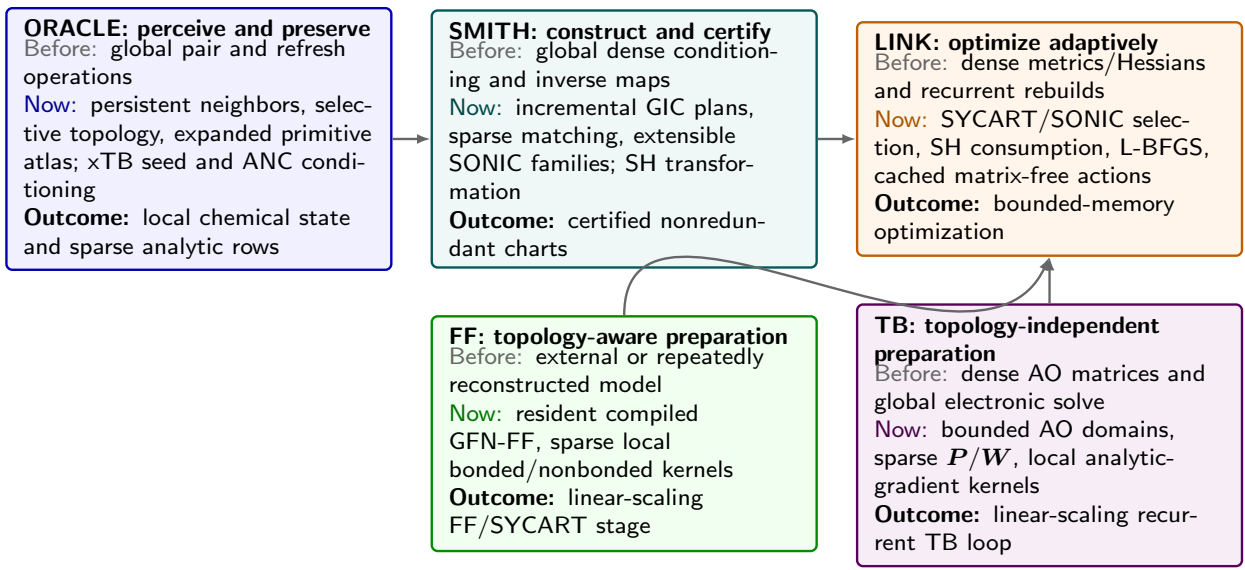
\begin{figure*}[t]
\centering
\resizebox{\textwidth}{!}{\input{figures/layer_modifications.tex}}
\caption{Structural changes across the three tools and resident calculators.}
\label{fig:layers}
\end{figure*}

Linear scaling requires every recurrent layer to avoid global dense Hessians,
Wilson metrics, and realization maps.  Figure~\ref{fig:layers} identifies
these changes; the Methods give the scientific contracts and refer to the
Supporting Information (SI) for implementation details.

The same requirement applies to the energy--gradient calculator. A resident
force field is needed first to remove process, file-I/O, serialization, and
data-transfer overhead from inexpensive FF/SYCART steps; at that cost scale,
an external executable can be more expensive than the force-field evaluation
itself. GFN-FF is therefore compiled and called in process, with its topology
and local interaction records retained between steps. For GFN2-xTB the issue is
also algorithmic: the conventional xTB implementation used by ANCopt contains
global dense electronic-structure operations and is not linear scaling. We
therefore reimplemented the recurrent TBLite path with bounded atom-centred
orbital domains, sparse local contractions, and bounded neighbor lists. This
removes global overlap inverses, eigenvector matrices, and dense atom-pair
intermediates from the large-system route. The resulting resident calculators
are what make an end-to-end near-linear \emph{tested workflow} possible; the
external xTB/ANCopt implementation remains essential as a numerical reference
with the same physical model but a different, nonlinear asymptotic behavior.

\section{Methods}

The ORACLE--SMITH--LINK path is revisited after accepted geometry changes.
This section defines its chemical and numerical contracts; validation and
implementation details appear in the SI.

\subsection{Entry-point qualification and route selection}

At entry, one Cartesian energy and gradient measure proximity to a stationary
region.  ORACLE also evaluates continuous bonding and local graph stability;
SMITH tests the rank and conditioning of a prospective SONIC chart.  These
chemical and numerical diagnostics govern the initial route.

When both tests pass, ORACLE freezes a versioned topology and its redundant
primitives; SMITH builds and certifies a nonredundant chart for LINK.  Otherwise
LINK starts in matrix-free SYCART or, when local connectivity is usable, in
simple redundant valence primitives.  An uncertain topology calls for
TBLite/GFN2-xTB rather than a topology-dependent force field.  At bounded
intervals, significant local displacements trigger selective reperception and
a prospective SONIC step.  Handoff requires stable topology, adequate rank
and conditioning, and an acceptable predicted step; the boundary energy and
gradient are reused.

For a validated SMILES topology, a poor three-dimensional embedding can first
be prepared with a resident force field in SYCART.  The policy is also local:
certified regions contribute SONICs while uncertain or reactive regions remain
in SYCART or simple primitives.  State transitions and failure codes are
specified in the SI.

\subsection{ORACLE: local chemical-state construction}

ORACLE remains the sole owner of chemical perception and of redundant
primitive definitions.  At each accepted topology checkpoint it freezes atom
identity, continuous and discrete bonding, symmetry, fragments, local
references, and the evidence supporting them \cite{Barone2026ORACLE}.
An ORACLE primitive is a scalar measurement with frozen support, units,
orientation convention, domain of validity, and analytic Cartesian derivative.
Its registry is not limited to the conventional valence set: any differentiable
function of one or more registered primitive supports can be added, and both
the family and function registries can be extended without changing the
ORACLE--SMITH contract, provided that a strict support invariant is respected.
Every intrafragment primitive contains either one central atom and only its
first neighbors, or one central bond and only atoms directly bonded to its two
ends. The only special primitives outside these two templates describe an
interaction between two fragment centers (either center may reduce to one
atom), or the rigid rotation of a fragment. ORACLE never delocalizes internal
deformations over a fragment. Rings, faces, cages, haptic sites, and similar
objects are ORACLE structural elements used to organize the later construction;
they are not themselves delocalized intrafragment primitives.
Table~\ref{tab:oracle-primitives} gives the current
production inventory and distinguishes numerical operators from their chemical
roles.

The coordinate-atlas input marks families or individual candidates as
required, optional, or forbidden, globally or within named atom domains.
Consequently the complete inventory need not be activated for every molecule.
For a problematic domain the conservative input profile retains only ordinary
distances, valence and linear bends, out-of-plane/improper coordinates, and
the complete set of proper dihedrals, including both endocyclic and exocyclic
ones.  Supports used only for ring, fullerene, haptic, oriented-contact, or
fragment-pose coordinates are then suppressed.  This is a deliberate
restriction of the same extensible atlas, not a separate perception engine.

\scriptsize
\setlength{\tabcolsep}{3pt}
\renewcommand{\arraystretch}{1.12}
\begin{longtable}{@{}>{\raggedright\arraybackslash}p{3.0cm}
>{\raggedright\arraybackslash}p{4.1cm}
>{\raggedright\arraybackslash}p{7.55cm}@{}}
\caption{ORACLE elementary primitives and their chemical roles.}
\label{tab:oracle-primitives}\\
\toprule Measurement & Operator ($K_{\rm SH}$) & Roles and qualifications \\
\midrule\endfirsthead
\toprule Measurement & Operator ($K_{\rm SH}$) & Roles and qualifications \\
\midrule\endhead
Atom--atom distance & \texttt{R}: $1.000$ covalent; $0.100$ contact; $0.080$ pseudobond & Stretches, contacts, pseudobonds, and frame anchors. \\
\midrule
Valence angle & \texttt{A}: $0.250$; pseudobond $0.080$ & Valence, multicenter, and contact bends. \\
\midrule
Linear bend & \texttt{L}: $0.100$ & Two transverse components with a frozen gauge. \\
\midrule
Proper or improper dihedral & \texttt{D}: $n_jn_k[0.0023+0.07S_{jk}(b_{jk})]$; \texttt{IMPD}: $0.120$ & Torsions, impropers, and bridge planes. \\
\midrule
Out-of-plane angle/height & \texttt{U}/\texttt{H}: $0.120$ & Oriented angular and height measures. \\
\midrule
Selective out-of-plane flap & \texttt{SOOP}: $0.050$ & Signed local out-of-plane displacement. \\
\midrule
Elementary angular flap & local dihedral support: $0.050$ & Signed angle between local planes. \\
\midrule
Signed face flap & \texttt{FACE\_FLAP}: $0.050$ & Oriented height relative to a local face. \\
\midrule
Center--center/atom distance & \texttt{FC\_DIST}, \texttt{FCA\_DIST},
\texttt{CENTER\_ATOM\_DIST}: $0.100$ & Fragment and haptic-center distances. \\
\midrule
Center--face distance & \texttt{CENTER\_FACE\_}\newline\texttt{DIST}: $0.100$ & Fragment/face-center pair distance. \\
\midrule
Face--direction angle & \texttt{FACE\_DIRECTION\_}\newline\texttt{ANGLE}: $0.080$ & Face normal versus center direction. \\
\midrule
Adjacent-face hinge & \texttt{FACE\_HINGE}: $0.050$ & Signed angle at a shared edge. \\
\midrule
Oriented contact angle & \texttt{CONTACT\_CENTER\_}\newline\texttt{ANGLE}: $0.080$ & Donor/acceptor and halogen-contact orientation. \\
\midrule
Relative translation & \texttt{FTRANS}: $0.080$ & Three components in a local fragment frame. \\
\midrule
Relative rotation & \texttt{FROT}: $0.025$ & Three quaternion exponential-map components. \\
\midrule
Linear-body pose & \texttt{FLIN\_TRANS}: $0.080$; \texttt{FAXIS}: $0.025$ & Axial translation and axis orientation. \\
\bottomrule
\end{longtable}
\renewcommand{\arraystretch}{1.0}
\setlength{\tabcolsep}{6pt}
\normalsize

The atomic synthon describes the environment of atom $i$ by charge $q_i$,
geometric compactness $C_i$, the incident bond-order product $D_i$, and
angular strain $S_i$:
\begin{equation}
 \bm s_i=\left(q_i,C_i,D_i,S_i\right).
 \label{eq:atomic-synthon}
\end{equation}
Continuity through bond formation and dissociation begins with the continuous
coordination number.  With $\widetilde{\mathcal N}(i)$ the weighted neighbor set,
\begin{equation}
 N_{\mathrm{CN},i}=\sum_{j\ne i}\frac12\left[1+\operatorname{erf}\!\left(\alpha_{\rm CN}
 (R^0_{ij}-R_{ij})\right)\right],
 \label{eq:synthon-cna}
\end{equation}
where $R^0_{ij}$ uses smoothly interpolated coordination-dependent covalent
radii.  The discrete graph is assigned after this evaluation.  Defining
$R_i^{\rm cov}$ as the interpolated covalent radius and
$\bar R_i=|\widetilde{\mathcal N}(i)|^{-1}\sum_jR_{ij}$,
\begin{equation}
 C_i=\frac{1}{|\widetilde{\mathcal N}(i)|}\sum_{j\in\widetilde{\mathcal N}(i)}
 \frac{R_i^{\rm cov}+R_j^{\rm cov}}{R_{ij}+R_i^{\rm cov}+R_j^{\rm cov}},
 \qquad
 \Delta_i^R=\frac{1}{|\widetilde{\mathcal N}(i)|\bar R_i}
 \sum_{j\in\widetilde{\mathcal N}(i)}|R_{ij}-\bar R_i| .
 \label{eq:synthon-compactness}
\end{equation}
The first quantity measures radial compactness and the second its dispersion.
Charge and bond orders come from one consistent electronic level.  Angular
strain is
\begin{equation}
 S_i=\left[\frac{1}{|\mathcal P_i|}
 \sum_{(j,k)\in\mathcal P_i}\left(\cos\theta_{ijk}
 -\cos\theta_i^{\mathrm{ref}}(N_{\mathrm{ED},i})\right)^2\right]^{1/2}
 \label{eq:synthon-strain}
\end{equation}
where $\mathcal P_i$ is the set of distinct pairs of accepted neighbors and
$N_{\mathrm{ED},i}$ comprises coordination, lone-pair, and open-shell domains.
Smooth Hermite interpolation between reference domain
geometries makes Eq.~\ref{eq:synthon-strain} applicable to every coordination,
including hypervalent and highly coordinated centers rather than only the
usual two-, three-, and four-coordinate organic cases.

Radial dispersion remains a geometric diagnostic.  Delocalization is measured
by the joint incident bond-order product
\begin{equation}
 D_i=\prod_{j\in\mathcal N(i)}\mathrm{BO}_{ij}.
 \label{eq:synthon-delocalization}
\end{equation}
Equation~\ref{eq:synthon-delocalization} distinguishes, for example, an
acetylene-like carbon ($3\times1$), the central carbon of a cumulene
($2\times2$), and an aromatic carbon ($1.5\times1.5\times1$).  Bond orders remain
continuous observables within a certified chemical state, and a topology
change creates a new versioned state rather than silently changing the meaning
of a stored synthon.

The effective atomic number combines an electrostatic shift,
$\Delta Z_i^{\rm el}=-q_i$, with a shape contribution from standardized
$(C_i,D_i,S_i)$.  A common principal-component metric was calibrated on the
328 accurate LCB25 structures\cite{LazzariCrisciBarone2025LCB25} with CM5
charges and Mayer bond orders, pooling carbon, nitrogen, and oxygen centers.
\begin{equation}
 \begin{aligned}
 \mathbf{x}^{\rm CDS}_i={}&
 \left((C_i,D_i,S_i)-\boldsymbol{\mu}_{\rm CDS}\right)
 \oslash\boldsymbol{\sigma}_{\rm CDS},
 \\
 \xi_i^{\rm shape}={}&\mathbf{v}_{\rm CDS}^{\mathsf T}\mathbf{x}^{\rm CDS}_i,
 \\
 Z_i^{\rm eff}={}&Z_i+\Delta Z_i^{\rm el}+\Delta Z_i^{\rm shape}
 =Z_i-q_i+\Delta Z_i^{\rm shape}.
 \end{aligned}
 \label{eq:synthon-zeff}
\end{equation}
Here $\boldsymbol\mu_{\rm CDS}=(0.483,1.719,0.180)$ and
$\boldsymbol\sigma_{\rm CDS}=(0.016,0.675,0.159)$ are the pooled
LCB25 mean and standard-deviation vectors, and
$\oslash$ denotes elementwise division.  The common shape direction is
$\mathbf v_{\rm CDS}=(0.613,0.607,-0.506)$.  It explains 73.5\% of
the shape-descriptor variance; the first two shape components explain 93.1\%.
One direction and normalization serve all three elements; their score
distributions reflect their chemical environments.  In the bounded
implementation,
$\Delta Z_i^{\rm shape}$ is the smooth, element-preserving map of
$\xi_i^{\rm shape}$, and is reported as the exact residual
$Z_i^{\rm eff}-Z_i+q_i$ so that the two contributions add to the plotted
effective number.  Charge is stored independently, so a compensating shape
contribution remains recoverable as $Z_i^{\rm eff}-Z_i+q_i$.  For hydrogen and
elements outside the calibrated C/N/O domain, $Z_i^{\rm eff}=Z_i-q_i$ while
all synthon descriptors remain available.  Extending the shape calibration
requires comparably annotated reference data.  The transition-metal tests
therefore assess topology and coordinates, not extrapolated shape scores.
Table~\ref{tab:synthon-carbon-examples} shows the resulting distributions for
the three elements; the broad ranges demonstrate that the common metric
retains chemically meaningful local variation.

\begin{table}[H]
\caption{LCB25 calibration distributions of the common electrostatic-plus-shape effective atomic number.}
\label{tab:synthon-carbon-examples}
\centering
\scriptsize
\begin{tabular}{@{}lrrrrrr@{}}
\toprule
Element & centers & 5th percentile & median & 95th percentile & mean & min--max \\
\midrule
Carbon & 1507 & 5.56 & 6.01 & 6.33 & 5.91 & 5.50--6.46 \\
\midrule
Nitrogen & 184 & 6.67 & 6.90 & 7.15 & 6.91 & 6.50--7.38 \\
\midrule
Oxygen & 180 & 7.59 & 7.97 & 8.29 & 7.97 & 7.51--8.34 \\
\midrule
\multicolumn{7}{@{}l@{}}{\scriptsize Values use the common C/N/O PCA model; hydrogen atoms are excluded.} \\
\bottomrule
\end{tabular}
\end{table}

\begin{figure}[H]
\centering
\includegraphics[width=\linewidth]{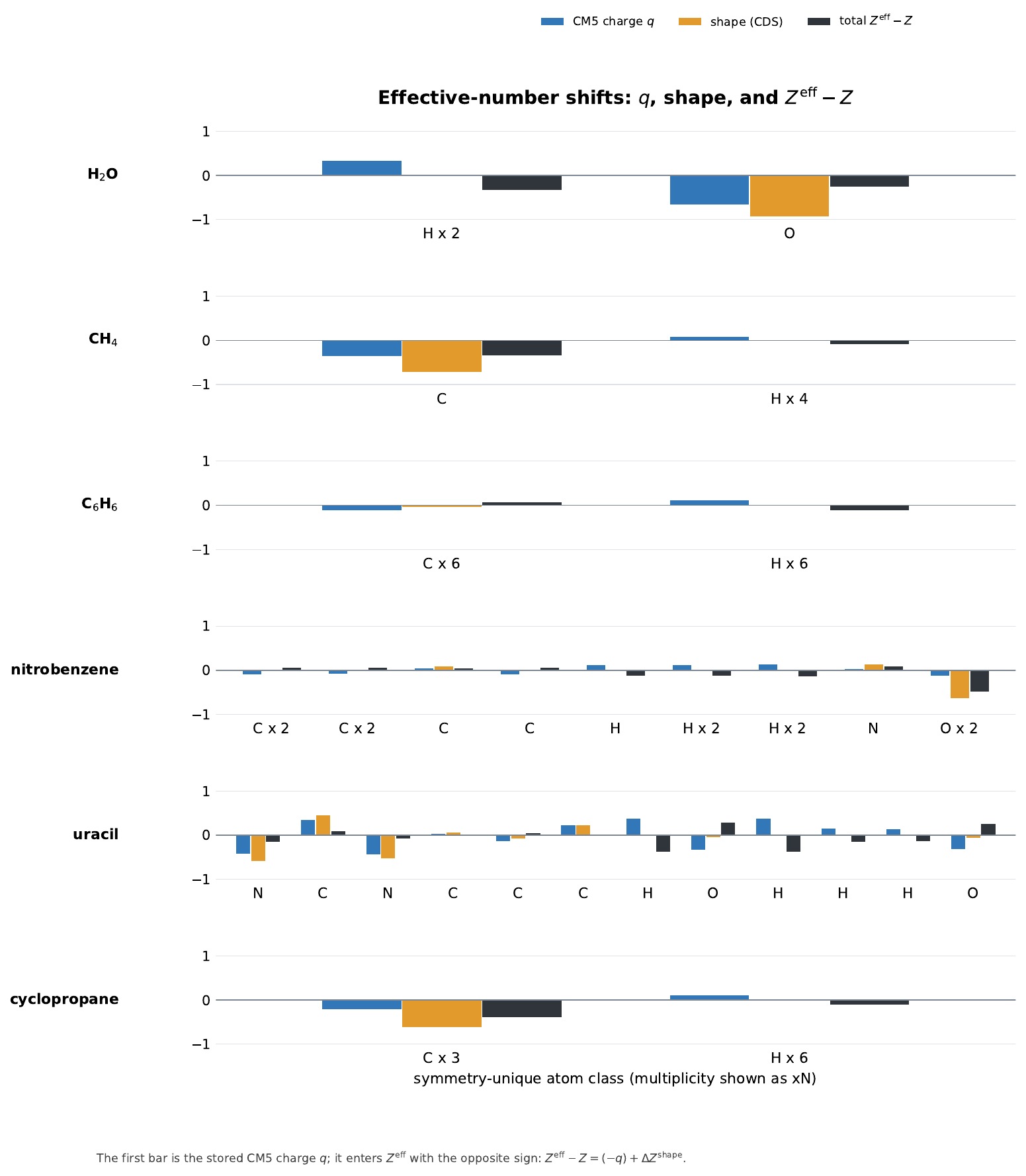}
\caption{Symmetry-unique LCB25 atom classes: charge, shape, and effective-number shifts.}
\label{fig:zeff-examples}
\end{figure}

Figure~\ref{fig:zeff-examples} displays one value for each symmetry-equivalent
atom class in representative LCB25 molecules; the multiplier gives the class
population.  The separate bars show the stored charge, the common shape
metric, and their resulting effective-number shift for C, N, and O.

In the workflow, $Z_i^{\rm eff}$ and the synthon vector are used for
reproducible atom typing, similarity matching, and lookup of local force-field
and SH-curvature records. They guide adaptive coordinate and parameter
selection; they are descriptors, not additional energy terms.

The synthon and $Z_i^{\rm eff}$ jointly define the continuous atom-similarity
metric from which reproducible atom classes are obtained. ORACLE attaches
atomic numbers, charges, effective atomic numbers, and continuous bond orders
to primitive supports for chemical typing and subsequent force-field work.
These descriptors remain available to the retained local-curvature atlas.
The current initialization uses one SH model owned by ORACLE. ORACLE assigns
one local curvature to every selected redundant primitive and stores the
result as a primitive-space diagonal. The standard distance, angle, linear-
bend, torsional, and pseudobond entries follow the xTB model, while additional
local primitive families use the same extensible registry. No curvature is
assigned directly to a ring, fullerene, or other derived SONIC coordinate.
SMITH transforms this single primitive model to whichever nonredundant chart
is selected, and LINK only consumes the resulting action (see below).

Neighbor lists, bounded graph domains, compiled primitive batches, and state
fingerprints make this representation local; continuous bond orders are reused
within a certified state and integer connectivity is the explicit force-field
fallback.  Storage and recurrent work are therefore linear for bounded local
coordination.  Certificates and restart rules are given in the SI.

\subsection{SMITH: certified sparse SONIC charts}

SMITH converts frozen ORACLE measurements into derived nonredundant
coordinates.  Local flaps become ring puckering, butterfly, condensed-ring,
or fullerene motions; interfragment measurements become relative distances
and poses.  Table~\ref{tab:smith-families} lists the current families.  New
functions use the same rank and symmetry registries.  In rings, SMITH combines
local selective out-of-plane (SOOP) flaps before rank selection.

% Start the complete family table on a fresh page so that its longtable header
% and all rows remain visually grouped rather than splitting at a page bottom.
\clearpage
\scriptsize
\setlength{\tabcolsep}{3pt}
\renewcommand{\arraystretch}{1.12}
\begin{longtable}{@{}>{\raggedright\arraybackslash}p{4.65cm}
>{\raggedright\arraybackslash}p{10.0cm}@{}}
\caption{SMITH representations and SONIC families derived from ORACLE
primitives.}
\label{tab:smith-families}\\
\toprule Family or representation & Primitive sources and interpretation \\
\midrule\endfirsthead
\toprule Family or representation & Primitive sources and interpretation \\
\midrule\endhead
\texttt{STRETCH}, \texttt{LOCAL\_XH\_STRETCH} & Distance and local X--H symmetry combinations. \\
\midrule
\texttt{BEND}, \texttt{LINEAR\_BEND}, \texttt{COUPLED\_ANGULAR} & Angular combinations and transverse pairs. \\
\midrule
\texttt{TORSION} & Proper torsions combined on a common central bond. \\
\midrule
\texttt{OUT\_OF\_PLANE}, \texttt{IMPROPER\_DIHEDRAL} & Local out-of-plane and improper combinations. \\
\midrule
\texttt{SOOP}, \texttt{SOOP\_RING\_PUCKER} & Frozen-frame out-of-plane and ring-puckering combinations. \\
\midrule
\texttt{CYCLIC\_BEND}, \texttt{SPIRO\_BEND} & Ring and shared-center bends. \\
\midrule
\texttt{RING\_PUCKER\_COMPONENT} & Triangular-flap and height-based ring modes. \\
\midrule
\texttt{CYCLIC\_TORSION}, \texttt{CONDENSED\_RING\_TORSION},
\texttt{BUTTERFLY} & Ring, fused-ring, and interplane modes. \\
\midrule
\texttt{PSEUDO\_CYCLE\_BEND}, \texttt{PSEUDO\_CYCLE\_TORSION} & Contact-cycle bends and torsions. \\
\midrule
\texttt{FULLERENE\_CAGE}, \texttt{FULLERENE\_FACE\_DIAGONAL} & Symmetry-coupled edge, flap, and face modes. \\
\midrule
Face/contact families & Face, hinge, center, and contact-angle combinations. \\
\midrule
Fragment and haptic distances & Fragment and center--atom distance combinations. \\
\midrule
Fragment pose & Relative translation and orientation after rigid-mode removal. \\
\midrule
Noncovalent families & Hydrogen-bond and pseudobond distance, bend, and torsion combinations. \\
\midrule
Distance functions & Inverse, inverse-cubic, and exponential maps. \\
\midrule
Other registered functions & Linear, difference, trigonometric, polar, and elliptic maps. \\
\midrule
\texttt{SYCART} & Symmetry-projected Cartesians with rigid modes removed. \\
\bottomrule
\end{longtable}
\renewcommand{\arraystretch}{1.0}
\setlength{\tabcolsep}{6pt}
\normalsize

Input selects admissible families by domain.  An uncertain region can remain
in SYCART or use simple redundant valence primitives, including all
endocyclic and exocyclic dihedrals.  Mixed charts are revised at certified
checkpoints.

For multiple fragments, SMITH certifies each intrafragment chart and then
completes the relative-fragment tangent space.  Its dimension is $6(F-1)$
for $F$ nonlinear fragments, with the usual reductions for linear and
single-atom fragments.  Pseudobonds remain interfragment measurements and
cannot create cross-fragment valence coordinates.  Sparse selection checks
both fragment and full-chart ranks; exceptional cases and audits appear in
the SI section \textit{Fragment-partitioned SONIC charts}.

Sparse primitive and combination operators act on complete symmetry families;
domain-local preselection and incremental matching certify rank without a
production singular-value decomposition.  Compiled plans preserve unchanged
regions, giving linear storage and construction for bounded domains.  Fullerene
face and edge measurements follow the same contract; formulae and dense-audit
parity are in the SI.

\subsection{Adaptive metrics}
\label{sec:adaptive-metrics}

The working coordinates are selected independently in local domains and evolve
with the accepted geometry.  ORACLE assigns every primitive
family an ordered integer class.  The present hierarchy admits distances
first, followed by valence and linear bends, local out-of-plane, SOOP, and
improper primitives, proper torsions, atom- or bond-local flap and hinge
primitives, and finally pair-center interactions and relative-fragment
rotations. Ring puckering, butterfly, condensed-ring, and fullerene
coordinates are not primitive classes: SMITH derives them by combining these
local rows.  The ranks are deliberately nonconsecutive, so new chemical
families can be inserted without changing existing contracts.  SYCART is
assigned a separate terminal rank and is not interpreted as a chemical
primitive.  Input can override the admissible families and their ordering for
a named domain.

Three partitions are immutable throughout selection.  Coordinates belonging
to different irreducible representations are never mixed.  Intrafragment and
interfragment coordinates are also never mixed, even at the most delocalized
level.  Finally, ORACLE local-domain identifiers restrict delocalization to the
spatial region whose conditioning is being repaired.  Thus a poorly
conditioned ring cannot cause unrelated valence coordinates or a
relative-fragment pose to be delocalized.  This strict direct-sum structure is
also what permits certified regions to remain in SONIC while a different
region uses SYCART.

Within one allowed block, primitive rows are first normalized geometrically,
\begin{equation}
 \widetilde{\mathbf B}=\mathbf D\mathbf B,\qquad
 D_{ii}=\|\mathbf B_i\|_2^{-1},\qquad
 \mathbf G=\widetilde{\mathbf B}\widetilde{\mathbf B}^{\mathrm T}.
 \label{eq:adaptive-geometric-metric}
\end{equation}
For a chart of target rank $r$, its normalized condition number is defined as
\begin{equation}
 \kappa(\widetilde{\mathbf B})=
 \frac{\sigma_{\max}(\widetilde{\mathbf B})}
      {\sigma_{\min}^{(r)}(\widetilde{\mathbf B})}
 =\left[
 \frac{\lambda_{\max}(\mathbf G)}
      {\lambda_{\min}^{(r)}(\mathbf G)}
 \right]^{1/2},
 \label{eq:sonic-condition-number}
\end{equation}
where $\sigma_{\max}$ and $\lambda_{\max}$ are the largest singular value
and metric eigenvalue, respectively, while
$\sigma_{\min}^{(r)}$ and $\lambda_{\min}^{(r)}$ are the smallest retained
nonzero values in the required rank-$r$ space.  A rank-deficient chart has
$\kappa=\infty$.  Row normalization makes Eq.~\ref{eq:sonic-condition-number}
independent of the arbitrary units and scales of the primitive families.
This definition removes the arbitrary scale of heterogeneous primitive
functions but introduces no mass weighting.  The default \texttt{GEOMETRY}
policy constructs local modes from Eq.~\ref{eq:adaptive-geometric-metric}.
An input-selectable \texttt{GEOMETRY\_SH} policy additionally uses the positive
primitive-diagonal SH curvature matrix,
\begin{equation}
 \mathbf G_{\rm SH}=\mathbf K_{\rm SH}^{1/2}\mathbf G
                    \mathbf K_{\rm SH}^{1/2},
 \label{eq:adaptive-sh-metric}
\end{equation}
so that the basis reflects both geometrical independence and estimated
energetic stiffness.  Here $\mathbf K_{\rm SH}=\operatorname{diag}(k_\alpha)$
is defined once by ORACLE over all selected local redundant primitives; it is
not refitted or reconstructed for a particular SONIC family.  SH is analytic
and requires no gradient evaluations.
The purely geometrical alternative remains available both as the conservative
default and as a control for assessing the value of curvature information.

SMITH first removes redundant rows by sparse structural matching, keeping
primitive classes separate.  It then evaluates
Eq.~\ref{eq:sonic-condition-number} automatically.  If the most local chart
exceeds the prescribed threshold, SMITH changes the chart internally: it
admits class mixing in increasing class order only inside the failing ORACLE
domain, constructs bounded partially delocalized blocks, and accepts the
first exact-rank chart that satisfies the condition gate.  No external flag
or optimizer request is needed.  Intrafragment and interfragment spaces
remain separate throughout this repair.  The limiting chemical chart is a
fully delocalized nonredundant internal basis within the affected domain; an
uncertified topology, or failure of every admissible chemical chart, selects
SYCART.  The hierarchy is
\begin{equation}
 \begin{aligned}
 \text{local redundant primitives}
   &\longrightarrow \text{local/class-preserving SONICs}\\
   &\longrightarrow \text{progressively delocalized SONICs}\\
   &\longrightarrow \text{fully delocalized internals}
    \longrightarrow \text{SYCART},
 \end{aligned}
 \label{eq:adaptive-hierarchy}
\end{equation}
where the primitive set and SH remain fixed during delocalization.

Matching certifies structural rank, and sparse extremal singular values monitor
conditioning.  Automatic repair is restricted to bounded local blocks and
never factorizes a global molecular metric, preserving linear storage and work.

Chart updates occur at accepted geometry checkpoints after a displacement or
topology trigger; unchanged domains reuse their compiled operators.  A changed
topology calls ORACLE again and replaces the primitive set.  With unchanged
topology, SMITH alone may revise its nonredundant selection.  LINK then
transforms the last gradient and Hessian model into the new chart, restarts
geometry direct inversion in the iterative subspace (GDIIS), and clears
incompatible secant history.  The boundary energy and gradient are reused.
The SI specifies the selection, refresh, and restart records.

\subsection{LINK: matrix-free optimization and adaptive charts}

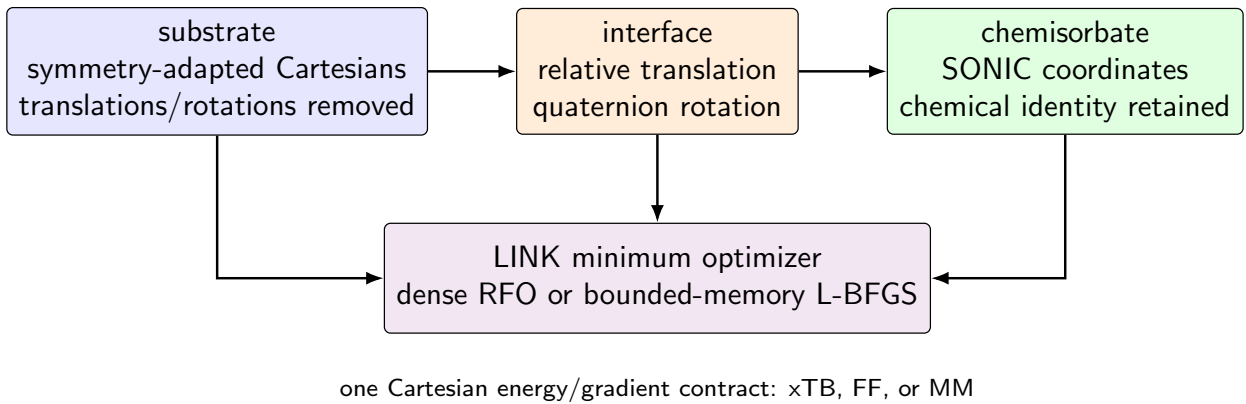
\begin{figure}[t]
\centering
\resizebox{\linewidth}{!}{\input{figures/concept.tex}}
\caption{Composite substrate--interface--adsorbate chart with a common Cartesian contract.}
\label{fig:concept}
\end{figure}

LINK verifies the ORACLE and SMITH identities, presents a Cartesian interface
to every calculator, and applies the selected representation.  Table~\ref{tab:link-representations}
summarizes the available minimum-search choices; LINK consumes coordinate
contracts rather than inferring chemical families.
The composite substrate--adsorbate layout is illustrated in
Figure~\ref{fig:concept}.

\begin{table}[H]
\caption{LINK representations and their intended minimum-search roles.}
\label{tab:link-representations}
\centering
\footnotesize
\begin{tabularx}{\linewidth}{@{}lYY@{}}
\toprule
Representation & Input/chemical regime & Numerical realization \\
\midrule
SYCART & Undefined or deliberately topology-free domain & Totally symmetric
Cartesian subspace, with translations and rotations removed \\
\midrule
Simple redundant primitives & Problematic domain with usable local
connectivity & Standard valence pool, including all endo/exocyclic torsions,
without special or ring-derived coordinates \\
\midrule
SONIC & Certified topology and well-conditioned chart & Nonredundant local,
optionally symmetry-adapted internal coordinates \\
\midrule
Mixed/adaptive & Different domains or stages require different descriptions &
SYCART or simple primitives in problematic regions, SONIC elsewhere, followed
by certified handoff \\
\midrule
Partial & Frozen environment or selected active region & The same operators
restricted by an explicit active-variable mask \\
\bottomrule
\end{tabularx}
\end{table}

For large minimum searches, maps, metrics, symmetry projection, curvature
application, and nonlinear realization are sparse or matrix-free actions.
Compiled plans and accepted state are reused between checkpoints. The common
optimization model is limited-memory Broyden--Fletcher--Goldfarb--Shanno
(L-BFGS) \cite{Nocedal1980}. Its preconditioner is the ORACLE SH diagonal,
which may incorporate calculator-specific local curvatures. This operator
route requires no global dense Hessian, metric, or Cartesian reconstruction.
Exact all-pair xTB parity is an audit mode; production uses bounded local
candidate lists.

The multilevel rule is $L_i$/SYCART or $L_i$/SONIC followed by
$L_{i+1}$/SONIC.  LINK changes chart only after topology, rank, conditioning,
and predicted reduction pass; frozen atoms and active regions use the same
operators.  Trust, refresh, and restart controls are specified in the SI.

\subsection{Composite energy and gradient assembly}

Resident calculators can be combined without changing the optimizer-facing
interface.  For a subtractive two-level construction with a full system $R$
and a model region $M\subset R$, the assembled energy is
\begin{equation}
 E_{\mathrm{comp}}(R,M)=E_{\mathrm{low}}(R)
 +E_{\mathrm{high}}(M)-E_{\mathrm{low}}(M).
\end{equation}
Here $E_{\mathrm{high}}(M)$ may itself be supplied by a composite or
multilevel method; LINK only requires its assembled Cartesian energy and
gradient at the model geometry.
The Cartesian gradient is assembled with the identical signs after scattering
the two model-region contributions onto the atoms of $R$,
\begin{equation}
 \bm g_{\mathrm{comp}}=
 \bm g_{\mathrm{low}}^{R}
 +\mathcal S_M^{\mathrm T}
 \left(\bm g_{\mathrm{high}}^{M}-\bm g_{\mathrm{low}}^{M}\right),
\end{equation}
where \(\mathcal S_M\) selects the model-region Cartesian components.  LINK
therefore receives one energy and one Cartesian gradient regardless of whether
the source is a single method, QM1/QM2, QM/MM, or an ONIOM-type subtractive
composition.  It subsequently applies the SYCART or SONIC transpose map once;
coordinate transformations are never performed separately by the component
calculators.  Shared geometry and atom-map fingerprints prevent incompatible
component orderings, and fixed or link-atom conventions are part of the
composite-calculator contract.  The present extension concerns energy and
gradient composition; high-level SONIC construction and optimization use the
same certified route.  Unit conversion, atom mapping,
boundary metadata, and failure checks are specified in the Supporting
Information section \textit{Composite calculator assembly}.

\begin{figure}[t]
\centering
\includegraphics[width=0.78\linewidth]{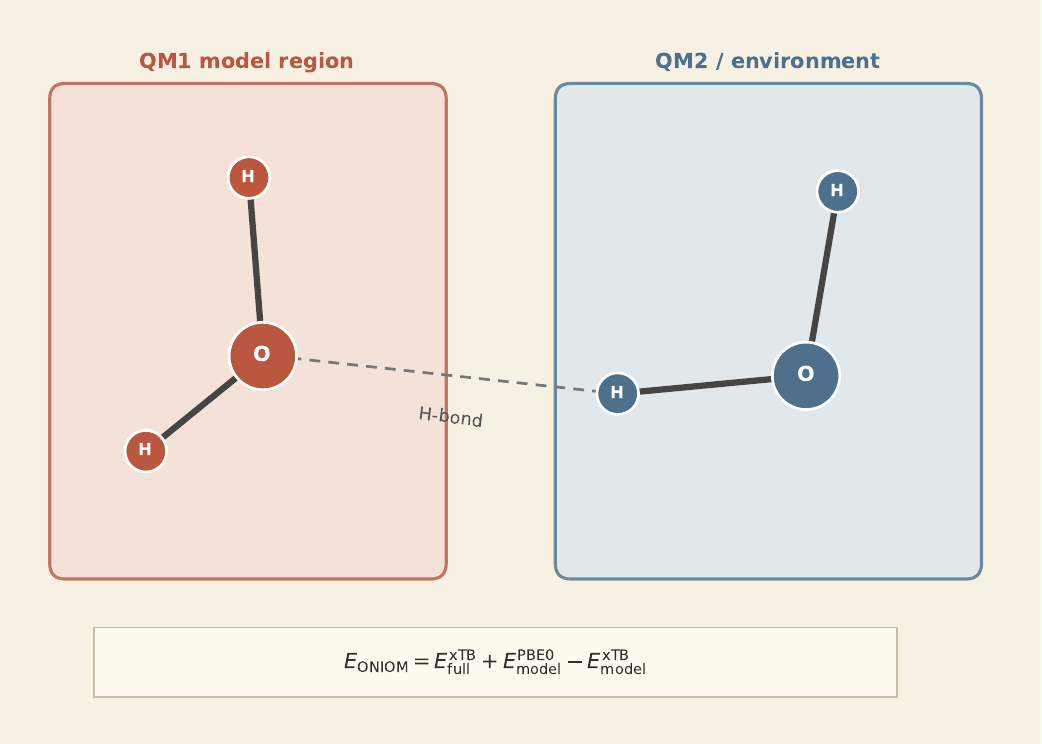}
\caption{QM1/QM2 water-dimer demonstration. The model water uses
PBE0/def2-SVP and GFN2-xTB; the environment uses resident GFN2-xTB/TBLite.}
\label{fig:oniom-water-dimer}
\end{figure}

As a concrete QM1/QM2 example, we use a water dimer with one water molecule
as the model region. LINK requests the full-dimer energy and gradient from
resident GFN2-xTB/TBLite and evaluates the model water twice at the same
geometry, with GFN2-xTB/TBLite and with ORCA at PBE0/def2-SVP. It then forms
the subtractive energy and gradient above and continues the optimization
through the same Cartesian interface (Figure~\ref{fig:oniom-water-dimer}).
This noncovalent example needs no cap; the same runtime uses the hydrogen-link
construction and affine gradient projection for a covalent model boundary.

\subsection{TBLite-LS certified local-density backend}

The small-molecule controls use optimized dense \texttt{tblite} 0.7.0
GFN2-xTB \cite{Bannwarth2019}; the large-system path uses the resident
implementation developed here. Both preserve the same calculator interface,
and the dense implementation remains the numerical reference. The distinction
is deliberate: xTB/ANCopt is retained as a physics and gradient-parity
control, whereas TBLite-LS is the production calculator for the scaling claim.

The electronic structure is organized in bounded atom-centered AO domains,
coupled by a common chemical potential.  Sparse density, response, dispersion,
and gradient contractions remove global overlap inverses, eigenvector matrices,
and dense atom-pair intermediates from the large-system route.  Bounded AO
domains, rather than in-process linkage alone, produce the measured near-linear
recurrent update; kernel details and parity tests are in the SI.

Domain enlargement is fail-closed: electron count, density residuals,
weighted-density consistency, and gradient parity must pass before admission;
the complete kernel certification is in the SI.

The public calculator uses the same certified domains and neighbor lists.
Linear-scaling status requires measured bounded-domain behavior of energy and
gradient phases, not merely sparse storage; unsupported periodic or dispersion
variants fail their capability gate rather than falling through silently.

GFN-FF addresses a different bottleneck. Its local functional form is already
compatible with sparse evaluation, but an external force-field executable
would reintroduce process startup, serialization, and interprocess data
movement at every inexpensive optimization step. The resident GFN-FF path
compiles topology-dependent records once and evaluates bonded, nonbonded,
charge, dispersion, and directional terms in the same process as LINK.

The low-level protocol is topology aware. With a validated topology, GFN-FF
provides the inexpensive Cartesian preoptimisation and TBLite/SONIC follows
before the final DFT/SONIC refinement. If the initial Cartesian structure has
an uncertain or reactive topology, TBLite/SYCART is used first and the route
switches to TBLite/SONIC only after the topology and SONIC quality gates pass.
Thus the same implementation supports both QM preoptimisation and linear-
scaling QM1/QM2 or QM/MM workflows. In both cases, resident execution is part
of the method: avoiding a new process and data transfer at every cheap
energy--gradient step is necessary for the coordinate overhead measured by
SYCART to remain meaningful.

\subsection{Quadraticized SONIC steps}

The quadraticizer described in the accompanying Communication
\cite{Barone2026Quadraticized} is implemented in LINK as an explicit
minimum-search option for direct, separable SONIC charts. It changes neither
the electronic potential nor the SONIC chart; it changes only the local step
variables and their pullback metric. For a physical SONIC displacement $s_i$
and a positive local scale $\alpha_i$, the Morse, Gaussian, and periodic maps
are defined below; implementation details are given in the SI.
\begin{equation}
 u_i^{\rm P}=\frac{2}{\omega_i}\sin\left(\frac{\omega_i s_i}{2}\right),qquad
 \omega_i=\frac{2\pi}{P_i}.
\end{equation}
The explicit maps and pullback metric are given in the SI. All maps have unit
Jacobian at the anchor. LINK solves the existing RFO
problem in $\bm u$, applies the analytic inverse on one common finite-image
branch, and evaluates the trust ratio after realizing the physical SONIC
step. The pullback metric keeps the trust region constant in the transformed
variables. The corresponding metric expression is given in the SI.
\begin{equation}
 \widehat{\bm G}(\bm s)=\bm D\bm F(\bm s)^{\mathsf T}
 \bm M\bm D\bm F(\bm s),
\end{equation}
so the trust region is constant in the quadraticized variables. The existing
GDIIS history, Hessian connection term, acceptance tests, and Cartesian
realization are retained. The option is restricted to minimum searches and
direct separable SONIC coordinates; projected SYCART, mixed charts, and
transition-state routes continue to use their qualified step models.

The production implementation uses a geometry-local GDIIS memory for the
quadraticized branch. The user-specified history remains available to standard
SONIC, whereas the quadraticized branch retains at most the three most recent
admissible points. This prevents points represented by different local
quadratic maps from being mixed and is applied uniformly to all systems. A
chart change or a refreshed quadraticized map clears the transported history
and restarts GDIIS from the new anchor. Delocalized charts are not
quadraticized, because their coordinates do not provide a separable local map.

\subsection{Coordinate alternatives and mixed-domain use}

Redundant primitive optimization through a generalized inverse is effective
for cages and bridged polycycles \cite{PulayFogarasi1992,
FogarasiZhouTaylorPulay1992,PengAyalaSchlegelFrisch1996,BakerKessiDelley1996,
BakerPulay1996Inverse,EckertPulayWerner1997,BakerPulay2000}.  Sparse screened
transformations, local delocalized internals, and Hessian eigenvectors offer
other routes to large systems \cite{FarkasSchlegel1998,
FarkasSchlegelFrisch2003,BilleterTurnerThiel2000,BakerKinghornPulay1999,
vonArnimAhlrichs1999,LiangWangHungLiFrisch2010}.  Global eigenvectors,
however, lose locality and need not remain adapted as the geometry changes.

Gaussian also permits generalized internal coordinates (GICs) as functions of
primitive geometries \cite{Marenich2025GIC}; the present registry constructs
such families automatically and SMITH selects an exact-rank chart.  The
comparison therefore includes redundant, delocalized, modal, SYCART, SONIC,
and mixed representations under matched calculators and stopping criteria.
Certified regions use SONIC and uncertain regions SYCART, with
intra- and interfragment blocks kept separate.  For SONIC, accepted iteration
count is primary because each iteration may require an expensive
energy--gradient evaluation; wall time is secondary for low-cost routes.

For minima, a bounded L-BFGS history acts in the selected active chart. SH is
defined exactly once, in ORACLE, on the complete selected vector of local
redundant primitives $\bm p$:
\begin{equation}
 \bm K_p^{\rm SH}=\operatorname{diag}(k_1,\ldots,k_{n_p}).
 \label{eq:oracle-primitive-sh}
\end{equation}
The xTB \texttt{ddvopt} rules provide the standard local entries, including
the pseudobond distances required for completeness; other ORACLE primitives
such as SOOP receive their curvature through the same registry.
\cite{xtb671ModelHessian} Derived ring, cage, and fragment SONICs never receive
independent force constants. If
$\bm T_{p\leftarrow q}=\partial\bm p/\partial\bm q$ is the tangent map from
the selected SMITH chart to primitive displacements, SMITH constructs
\begin{equation}
 \bm H_q^{\rm SH}=\bm T_{p\leftarrow q}^{\mathrm T}
                  \bm K_p^{\rm SH}\bm T_{p\leftarrow q}.
 \label{eq:oracle-smith-sh}
\end{equation}
Thus a diagonal local primitive model generates all required off-diagonal
couplings in a SONIC basis. The same ORACLE diagonal is transformed to a
class-preserving local SONIC chart, a progressively more delocalized chart, or
the terminal SYCART representation; LINK defines no second curvature model.
Forward and transpose sparse actions evaluate this congruence without a dense
Cartesian Hessian or gradient differences. Geometry and atom identities are
checked before an existing ORACLE model is reused. The construction and its
parity audits are detailed in the Supporting Information.

Composite supported systems retain distinct coordinate semantics: symmetric
Cartesian substrate motion, SONIC adsorbate deformation, and relative
translation/quaternion rotation.  Ring and fullerene coordinates, torsional
multiplicity, redundancy normalization, nonlinear realization, and partial
constraints obey the same sparse contract.  Definitions, equations,
curvatures, safeguards, and analytic derivative checks are collected in the
Supporting Information sections \textit{Bounded-memory step model},
\textit{Initial Hessian ownership and SONIC transformation}, \textit{Symmetry-adapted Cartesian
substrate block}, \textit{Composite supported-system chart}, and
\textit{Partial optimization}.

\section{Results and Discussion}

\subsection{Coordinate efficiency and basin identity}

We compared SYCART and SONIC on eight molecular geometries from the LINK xTB
stress archive: cubane, xylose, glucose, fructose, saccharin, ferrocene,
testosterone, and the water dimer. Both LINK routes used the same resident
GFN2-xTB energy and analytic gradient, dense RFO optimizer, analytic
Berny--Swart starting model, trust policy, and convergence thresholds; only
the coordinates differed. Their active ranks agreed, and the spectral-norm
difference between their Cartesian tangent projectors was
$3.1\times10^{-9}$. Each route required a median of 7 iterations, compared
with 8.5 for native xTB/ANCopt. The constant SYCART map reduced median LINK
wall time from 3.71 to 2.20~s. After rigid alignment, the LINK endpoints
differed by at most 0.00553~\AA{} and 3.71~$\mu E_\mathrm h$. Case-level
counts and endpoint checks are in the Supporting Information.

The quadraticized SONIC path is available through the same LINK
runtime as an explicit step-model choice. We compared it with the preceding
dense-RFO SONIC implementation over small, medium, and difficult geometries
using the same resident TBLite/GFN2-xTB gradients and convergence thresholds.
The expanded comparison is summarized in Table~\ref{tab:quadraticized-sonic};
the settings and wall-time diagnostics are provided in the Supporting
Information. The analytic-map, inverse-branch, metric, and
integration tests pass in the resident LINK test suite.

As an independent runtime check, we repeated the comparison on glycidol,
indane, and pagodane with resident TBLite gradients. The standard and
quadraticized SONIC routes converged in 8/8, 9/9, and 5/5 accepted iterations,
respectively. These checks show that the new step map preserves the established
SONIC convergence on additional molecular geometries while retaining the
iteration reductions listed in Table~\ref{tab:quadraticized-sonic}.

To probe the nonquadratic regime, we also displaced the indane structure by a
controlled affine expansion and bounded Cartesian perturbation before rebuilding
the chart. From this far-start geometry, standard SONIC required 32 accepted
iterations, whereas quadraticized SONIC converged in 22; both routes used the
same resident TBLite gradient and convergence thresholds.

\begin{table}[t]
\caption{Accepted iterations for standard and quadraticized SONIC with resident
GFN2-xTB. Wall-time diagnostics are reported in the SI.}
\label{tab:quadraticized-sonic}
\centering
\small
\begin{tabular}{@{}lrr@{}}
\toprule
Molecule & Standard SONIC & Quadraticized SONIC \\
\midrule
Water & 4 & 4 \\
Ammonia & 4 & 4 \\
Benzidine & 14 & 18 \\
Tryptophan & 53 & 29 \\
QIQGOM & 65 & 64 \\
ISOL24\_i4p & 173 & 131 \\
Indane & 8 & 9 \\
Pagodane & 5 & 11 \\
Glycidol & 8 & 8 \\
Glycine & 6 & 6 \\
Formaldehyde & 4 & 4 \\
\bottomrule
\end{tabular}
\end{table}

\begin{figure}[t]
\centering
\includegraphics[width=\linewidth]{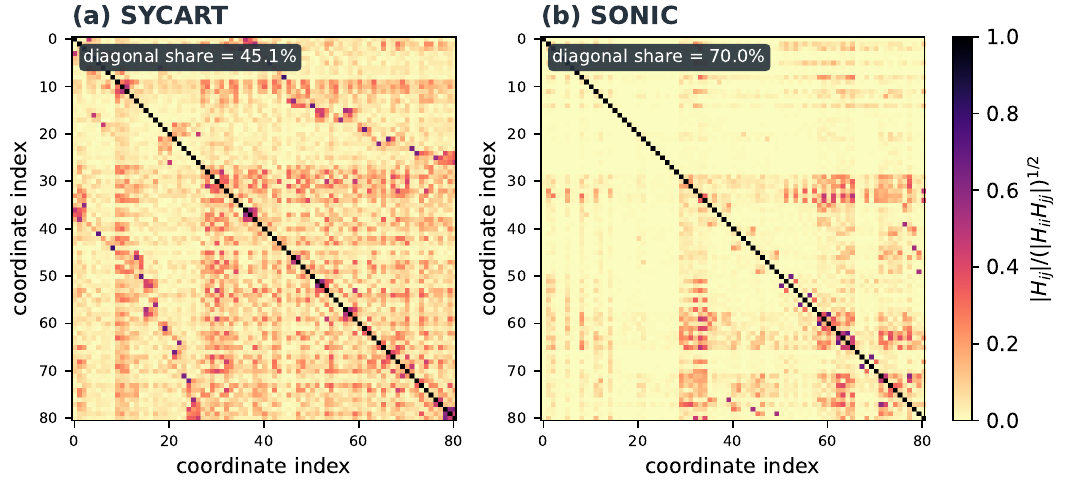}
\caption{Normalized effective xTB/ANC Hessian for menthone in SYCART and SONIC.}
\label{fig:hessian-locality}
\end{figure}

The same effective xTB/ANC Hessian was also transformed at one fixed menthone
geometry into the rank-81 SYCART and SONIC bases (Figure~\ref{fig:hessian-locality}).
After scaling every element as
$H_{ij}/(|H_{ii}H_{jj}|)^{1/2}$, the squared Frobenius contribution of the
diagonal is 45.1\% in SYCART and 70.0\% in SONIC. SONIC is thus more nearly
diagonal in this example, an advantage for vibrational interpretation and
reduced-dimensional anharmonic models. It does not alone guarantee faster
quasi-Newton optimization: away from a minimum, curvilinear coordinates can
concentrate higher-order terms and shorten the locally quadratic region.
SYCART distributes this nonlinearity over Cartesian components. SONIC retains
the locality of redundant primitives without their redundancy, whereas fully
delocalized internals can mix distant families. Provenance and normalization
are given in the Supporting Information.

We therefore assess accepted iterations and cost per iteration separately.
The former matters most for expensive electronic gradients; the latter can
dominate low-cost calculations. Profiling locates the principal SONIC
prefactor in recurrent nonlinear realization. Kernel-level timings and
trajectory-preserving reuse tests are given in the Supporting Information.

\begin{figure}[t]
\centering
\includegraphics[width=0.86\linewidth]{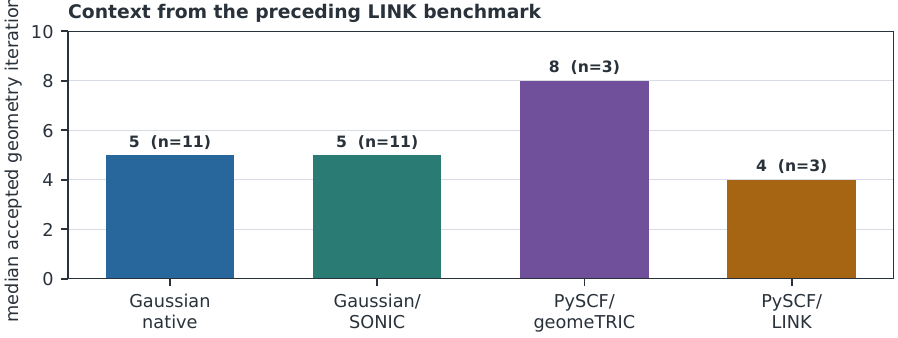}
\caption{Preceding LINK optimizer comparisons on small systems.}
\label{fig:previous-link}
\end{figure}

The preceding LINK study compared established external optimizers on small
systems (Figure~\ref{fig:previous-link}): median accepted iterations were 5
for both Gaussian's native optimizer and Gaussian with SONIC (11 minima), and
8 for PySCF/geomeTRIC versus 4 for PySCF/LINK (three minima). Here native
xTB/ANCopt is the control for the new large-system, low-cost regime, where
coordinate construction and data movement can dominate the calculation.

On three ACONF $n$-hexane geometries, the matched one-thread, resident
GFN2-xTB tests gave 9--12 SONIC, 15--18 SYCART, and 22--26 native
xTB/ANCopt iterations (Table~\ref{tab:current-hexane}). LINK times refer to
the optimizer kernel; one-time preparation is reported separately.

\begin{table}[t]
\caption{One-thread ACONF results: iterations / kernel seconds (milliseconds per accepted step).}
\label{tab:current-hexane}
\centering
\small
\begin{tabular}{@{}lrrr@{}}
\toprule
Conformer & SYCART & SONIC & xTB/ANCopt \\
\midrule
$g^+t^+g^-$ & 18 / 0.051 (2.82) & 12 / 0.307 (25.58) & 26 / 0.099 (3.81) \\
\midrule
$g^+x^-g^-$ & 15 / 0.047 (3.14) & 12 / 0.262 (21.83) & 22 / 0.085 (3.86) \\
\midrule
$g^+x^-t^+$ & 15 / 0.044 (2.92) & 9 / 0.186 (20.67) & 23 / 0.087 (3.78) \\
\bottomrule
\end{tabular}
\end{table}

SYCART takes 2.82--3.14~ms per accepted step, below ANCopt's
3.78--3.86~ms; SONIC takes 20.67--25.58~ms while requiring fewer gradients.
The analytic initial-Hessian preparation costs 0.047--0.049~s once. These
measurements favor SYCART for inexpensive gradients and SONIC when gradient
cost dominates. Mixed routes can exploit both regimes, and a final monitored
SONIC step supplies interpretable coordinates without a SONIC trajectory.
Timing uncertainties are reported as medians and median absolute deviations
from three identical one-thread repetitions; the tables retain the individual
system values so that the spread across chemical cases is not hidden by a
single aggregate statistic. The comparisons are therefore regime diagnostics,
not claims of statistical significance from a small replicate sample.

\subsection{Scaling and resource accounting}

The complete coordinate path was measured with one thread in fresh processes
for bounded-degree covalent chains. From 256 to 8192 atoms (762 to 24,570
active coordinates), elapsed time increased from 0.208 to 7.862~s. Stored
Wilson nonzeros increased from 6852 to 221,124. Log--log fits give exponents
of about 1.05 for total ORACLE--SMITH--LINK coordinate time, 1.00 for Wilson
storage, 1.04 for inverse-operator storage, and 0.39 for peak resident memory.
The largest
right-inverse residual was $7.18\times10^{-16}$. All four predefined scaling
gates are met over the sampled range. These data support an end-to-end
linear-scaling interpretation for the
coordinate infrastructure; calculator evaluation was deliberately excluded
and is not assigned the same asymptotic claim. Machine-readable records and the
complete stage timing are supplied in the SI. A restart is accepted only when
the protocol, chart, active region, electronic/classical state, and cached
accepted geometry reproduce within the declared numeric tolerances.
Regression standard errors and fit windows are supplied with the machine-readable
records; they quantify the uncertainty of the reported exponents without
altering the stated qualification gates.

We additionally profiled stored molecular geometries, including an
uracil--12-water cluster and systems of 3410 and 6996 atoms.  The complete
molecular coordinate path has a fitted exponent of about 1.25, while Wilson
storage and peak memory remain near-linear.  Neighbor-list reuse and sparse
conditioning keep the prefactor bounded; cold/warm and component timings are
reported in the SI.

A complementary control used the same one-thread surface with dense and
matrix-free SYCART and native xTB/ANCopt.  The complete comparison is
tabulated in the SI.  The qualified claim is near-linear recurrent scaling
after initialization: the bounded-domain resident calculator supplies the
local electronic layer, while external xTB remains a nonlinear numerical
control and is excluded from the production scaling claim.

The asymptotic accounting is summarized in Table~\ref{tab:complexity}.
\begin{table}[t]
\caption{Asymptotic scope of the resident production path.}
\label{tab:complexity}
\centering
\footnotesize
\begin{tabularx}{\linewidth}{@{}lXXX@{}}
\toprule Layer & Stored object & Recurrent work & Qualified scope \\
\midrule
ORACLE & Neighbor lists and local primitive records & $O(N)$ & Bounded local domains \\
\midrule
SMITH & Sparse combinations and Wilson CSR rows & $O(N)$ & Certified minimum charts \\
\midrule
LINK & Matrix-free maps, metrics, and L-BFGS history & $O(N)$ & No production dense global object \\
\midrule
TBLite-LS & Bounded AO domains and local contractions & Near $O(N)$ & Resident energy--gradient iteration \\
\midrule
xTB/ANCopt & External reference implementation & Nonlinear & Control only \\
\bottomrule
\end{tabularx}
\end{table}

For every benchmark we report four quantities separately: accepted optimizer
steps, energy--gradient calls, scientific wall time, and resident peak memory.
The latter two are split into calculator, coordinate construction/realization,
optimizer, and process-start components.  This convention prevents a fast
calculator from being penalized for one-time initialization and, conversely,
prevents a favorable total time from concealing a superlinear coordinate
kernel.  Uncertainties are the median and median absolute deviation of three
identical one-thread repetitions; campaign tables retain the individual
records rather than only the aggregate.

The domain policy is summarized in Table~\ref{tab:decision-domains}.
\begin{table}[t]
\caption{Decision domains and permitted coordinate handoffs.}
\label{tab:decision-domains}
\centering
\footnotesize
\begin{tabularx}{\linewidth}{@{}lXXX@{}}
\toprule
Domain & Diagnostic signature & Initial route & Permitted handoff \\
\midrule
Certified topology & Sparse rank/support and condition gates pass; local
gradient is compatible with the chart & Resident FF/TB with SONIC when the
surface is expensive, otherwise SYCART & SONIC after the prospective-step gate \\
\midrule
Uncertain topology & Bond-order or support ambiguity, large local residual, or
poor conditioning & Resident TBLite/SYCART with simple primitives & Re-perceive
only affected neighborhoods, then SONIC on certified blocks \\
\midrule
No chemical chart & Dissociative, Lennard--Jones-like, or otherwise unresolved
connectivity & SYCART or Cartesian/simple-redundant control & No SONIC claim until
all certification tests pass \\
\bottomrule
\end{tabularx}
\end{table}

The common convergence contract uses identical energy, maximum-force,
RMS-force, maximum-displacement, and RMS-displacement thresholds for all
coordinate models.  The chart gate additionally requires the declared sparse
rank, a bounded normalized condition number, and one successful prospective
SONIC model step; numerical values are tabulated in the SI.

The broad one-thread campaign on the reference architecture used the same
geometry-seed convergence profile for all four routes.  SYCART and xTB/ANCopt
converged in all 20 cases; direct SONIC converged in 14 and the adaptive hybrid
in 17.  This robustness set is retained in the Supporting Information, while
its detailed timings are separated from the current compiled control below.
AD6 retains the certified Cartesian fallback.  QIQGOM and GOPPEG now undergo
fresh perception after the Cartesian stage; the resulting hybrid handoffs
converge with a terminal SONIC step.  Per-case iteration, calculator,
coordinate-realization, and process times are tabulated in the Supporting
Information.

\begin{figure}[t]
\centering
\includegraphics[width=\linewidth]{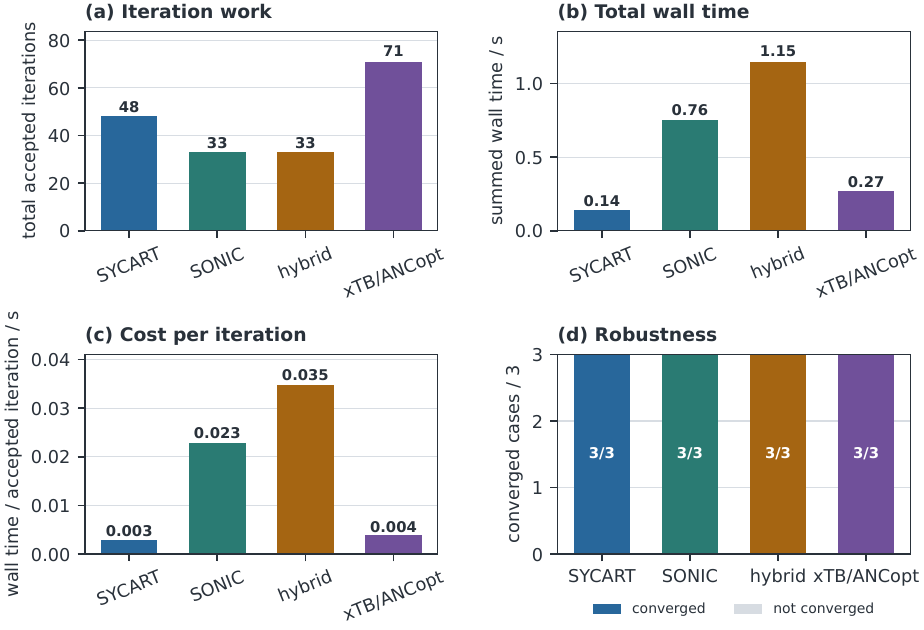}
\caption{Compiled three-conformer control: accepted iterations, wall time,
per-iteration cost, and convergence.}
\label{fig:fourway}
\end{figure}

The latest compiled three-conformer control used for the cost comparison is
shown in Figure~\ref{fig:fourway}.

We then repeated the same four-way protocol on the complete 30-molecule
Baker minimum set. All 90 optimizations converged without a reduced-symmetry
fallback, and the two LINK coordinate models passed a case-by-case Cartesian
tangent-space identity audit (largest projector residual
$1.30\times10^{-6}$). Median iteration counts are 5, 5, and 6 for SYCART,
SONIC, and native ANCopt, respectively. The corresponding totals are 232, 193,
and 260 iterations: SYCART and SONIC tie in 21 systems, SONIC is lower in seven,
and SYCART is lower in two. Despite its larger total, the constant SYCART map
reduces median LINK wall time from 1.26 to 0.88~s. The largest SYCART--SONIC
raw screening-endpoint separation is 0.0163~\AA{} (histidine), whereas the
largest absolute energy difference is 11.43~$\mu E_\mathrm h$
(dimethylpentane). Because RMSD alone cannot establish minimum identity, the
histidine endpoints were cross-restarted with tight thresholds. The resulting
representatives differ by only 0.00204~\AA{} and
0.00559~$\mu E_\mathrm h$; their sampled connecting profile is a single
barrierless well, and independent-SCF finite-difference tblite Hessians have
index zero with lowest frequencies of 33.66 and 33.67~cm$^{-1}$. The original
separation therefore reflects termination at different positions along one
soft basin, rather than convergence to distinct minima. Full case-level data,
the histidine certification, and the acetone source-header normalization are
given in the Supporting Information.

\begin{figure}[t]
\centering
\includegraphics[width=\linewidth]{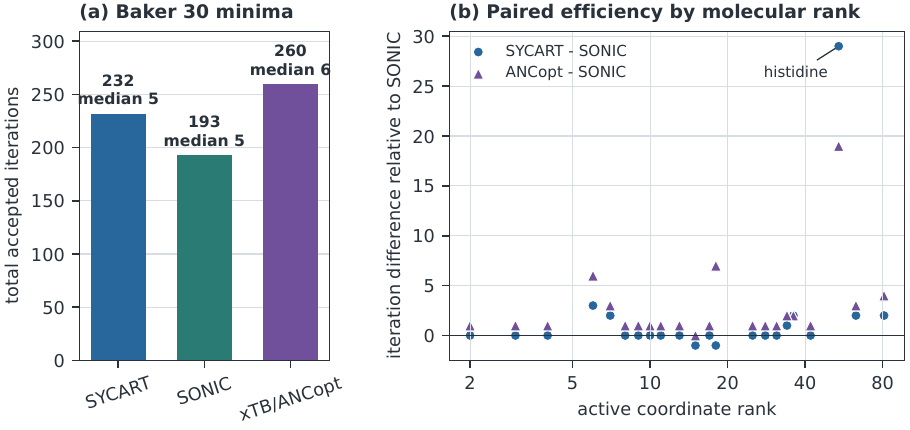}
\caption{Accepted-iteration comparison for the Baker 30-molecule minimum set.}
\label{fig:baker30-iterations}
\end{figure}

Figure~\ref{fig:baker30-iterations} summarizes the 30-molecule iteration comparison.

The coordinate-model prefactor was then tested on a separate 12-molecule set ranging from
benzene and pyrene to testosterone, artemisinin, corrin, and benzylpenicillin
(6--51 atoms).  All four LINK representations and the native ANCopt control
converged for every member of this set.  The complete iteration record is given
in Table~\ref{tab:large-campaign}; times, calculator calls, and resident-memory
measurements are reported in the Supporting Information.  The smaller entries
are curated regression structures or LCB26 reference records. Starting-
geometry provenance is given in the SI.
SONIC required the
fewest iterations for ferrocene and testosterone; for benzylpenicillin, the
definitive frozen-chart validation required 37 iterations, matching ANCopt and
remaining within one iteration of SYCART.  The Hessian-mode and SYCART routes
provided the lowest LINK wall time for most
systems because they avoid nonlinear coordinate realization.  The direct
resident-calculator cost remained small compared with the coordinate and
optimizer kernels, which is the regime in which the compiled Cartesian and
modal routes are most useful.

\begin{table}[t]
\caption{Accepted iterations for the 6--51-atom campaign; delocalized
internals are comparative controls.}
\label{tab:large-campaign}
\centering
\scriptsize
\begin{tabular}{@{}lrrrrrrr@{}}
\toprule
System & $N$ & Redundant & Modes & Delocalized & SYCART & SONIC & ANCopt \\
\midrule
Water dimer & 6 & 6 & 6 & -- & 6 & 6 & 7 \\
Ethanol & 9 & 4 & 5 & -- & 5 & 4 & 6 \\
Benzene & 12 & 2 & 2 & -- & 2 & 2 & 3 \\
Saccharin & 17 & 5 & 5 & -- & 5 & 5 & 6 \\
Ferrocene & 21 & 7 & 7 & -- & 7 & 6 & 9 \\
Pyrene & 26 & 3 & 3 & -- & 3 & 3 & 4 \\
Artemisinin & 42 & 10 & 10 & -- & 10 & 10 & 13 \\
Benzylpenicillin & 41 & 37 & 69 & -- & 36 & 37 & 37 \\
Corrin & 45 & 17 & 18 & -- & 18 & 18 & 20 \\
Testosterone & 49 & 15 & 15 & -- & 15 & 13 & 15 \\
Testosterone (LCB26) & 49 & 15 & 15 & -- & 15 & 13 & 15 \\
Androsterone & 51 & 8 & 8 & -- & 8 & 8 & 10 \\
\midrule
Total (12) & -- & 129 & 163 & -- & 130 & 125 & 145 \\
Mean (12) & -- & 10.75 & 13.58 & -- & 10.83 & 10.42 & 12.08 \\
\bottomrule
\end{tabular}
\end{table}

The benzylpenicillin SONIC entry is the definitive frozen-chart run:
37 accepted iterations and 38 resident TBLite/GFN2-xTB energy--gradient
evaluations.  Full provenance is supplied in the SI.

The numerical comparison in this work uses native xTB/ANCopt as a matched
control, not as an independent electronic-structure method: both routes use
the same GFN2-xTB surface and stopping contract. The objective is to quantify
the resident coordinate path and its scaling change under controlled
conditions. Broader optimizer comparisons with Gaussian and geomeTRIC were
reported in the preceding LINK study\cite{Barone2026LINK}; repeating them
here would not isolate the present implementation change.

The campaign therefore uses iteration count as the primary SONIC measure.
SONIC reduces the number of electronic-structure calls in the largest and most
topologically well-defined examples, whereas Hessian-mode and SYCART routes
keep the coordinate overhead bounded.  Wall-time data are reported only to
diagnose the low-cost regime in which coordinate construction can become a
leading contribution.  This complements the 20-case
reference-architecture campaign rather than replacing it: the former probes
the coordinate and optimizer prefactor as molecular size grows, while the
latter probes adaptive routing across difficult starting topologies.

The larger molecular validation is the 20-case adaptive campaign, which spans
20--81 atoms.  Its ten largest members are reported explicitly in
Table~\ref{tab:very-large-campaign}.  The five C$_{60}$ and two 81-atom organic
structures originate from the C60ISO and ISOL24 subsets of
GMTKN55, respectively \cite{Goerigk2017GMTKN55}; QIQGOM, ZANMOQ, and GOPPEG
originate from the tmQM transition-metal-complex data set
\cite{Balcells2020tmQM}.  The calculations start from the deterministically
distorted Cartesian benchmark geometries archived with the campaign, not from
the corresponding optimized reference structures.  The identical starting
file, charge, multiplicity, calculator, and stopping criteria are used for all
four routes; no route-specific preoptimization is applied.

\begin{table}[t]
\caption{Accepted iterations for the 60--81-atom validation cases.}
\label{tab:very-large-campaign}
\centering
\scriptsize
\begin{tabular}{@{}lrrrrr@{}}
\toprule
System & $N$ & SYCART & SONIC & Hybrid & ANCopt \\
\midrule
C$_{60}$ iso-1  & 60 & 18 & 25   & 19  & 20 \\
C$_{60}$ iso-10 & 60 & 19 & 120* & 22  & 19 \\
C$_{60}$ iso-2  & 60 & 18 & 23   & 21  & 22 \\
C$_{60}$ iso-3  & 60 & 22 & 23   & 21  & 21 \\
C$_{60}$ iso-4  & 60 & 17 & 27   & 14  & 22 \\
QIQGOM           & 80 & 33 & 0*   & 47 & 31 \\
ZANMOQ           & 80 & 28 & 42   & 35  & 30 \\
ISOL24 i4e       & 81 & 31 & 39   & 27  & 44 \\
ISOL24 i4p       & 81 & 56 & 25   & 37  & 70 \\
GOPPEG           & 81 & 32 & 120* & 95 & 38 \\
\midrule
Total (10) & -- & 274 & 444 & 317 & 317 \\
Mean (10; SONIC 9 attempted) & -- & 27.40 & 49.33 & 31.70 & 31.70 \\
\bottomrule
\end{tabular}
\end{table}

An asterisk marks a direct-SONIC route that was not certified or reached the
120-step cap; its cap is included in the SONIC total and attempted-route mean.
SYCART converges for all ten large cases. Fresh perception
after Cartesian preparation gives converged hybrid handoffs for QIQGOM and
GOPPEG, each including a terminal SONIC step; direct SONIC remains uncertified
for QIQGOM and capped for GOPPEG.  The final GOPPEG record contains 95
accepted adaptive steps across eight cycles.  The complete table and route
provenance are given in the SI.

These comparisons define the size-dependent production regimes summarized in
the Methods.  In the C$_{60}$ iso-1 control,
SYCART converged in 18 accepted iterations compared with 20 for xTB/ANCopt.
The sparse SONIC condition number changed from 5.928 at the initial geometry to
5.776 at the SYCART minimum, and the terminal SONIC step reduced the maximum
Cartesian gradient from $1.4695\times10^{-4}$ to
$9.0286\times10^{-5}\ E_{\rm h}\,a_0^{-1}$.  The full sparse certificate and
timing decomposition are reported in the Supporting Information.

The preceding systems are controlled coordinate-model benchmarks designed to
isolate optimizer and representation prefactors.  We also tested the
workflow in the form used for a practical large-system preparation.  A
12-residue polyalanine model was assembled from PCS2 fragments in the
LCB25 archive through the unified LCB26 interface. ORACLE certified the
complete 123-atom topology, validating the adaptive handoff on a fragment-built
geometry; detailed provenance is given in the SI.

\subsection{Intermolecular optimization and microsolvation}
Intermolecular systems use a staged protocol because the solute and solvent
have different useful coordinate descriptions. We first refine the solute in
SONIC coordinates, or in SYCART followed by SONIC when its initial topology is
uncertain. The solute is then held fixed while the solvent is relaxed in
SYCART coordinates with relative fragment translations and quaternions. This
stage is a preparation step and is not subjected to a rigid-fragment
certification gate. After the solvent gradient and prospective SONIC
conditioning criterion are satisfactory, ORACLE perceives the accepted
geometry again and SMITH builds the complete inter- and intrafragment chart.
LINK then performs the final all-SONIC refinement.

For uracil--$(\mathrm{H_2O})_{12}$ (48 atoms), the hybrid
route comprised 22 SONIC steps for the solute, 41 SYCART steps for the
fixed-solute solvent preparation, and 20 final SONIC steps after refreshed
perception (83 accepted steps in total). The final TBLite/GFN2-xTB gradient
infinity norm was $4.19\times10^{-4}\ E_{\rm h}\,a_0^{-1}$. The complete stage
record is given in the SI. The same block
construction applies directly to QM/MM and subtractive ONIOM models: the
high-level region uses SONIC coordinates while the surrounding region can
remain Cartesian until its topology is certified.

To separate the behavior of the solute--solvent interface from the size of the
solvent shell, the reproducibility package also defines a graded uracil--water
panel with one, two, and four waters (15, 18, and 24 atoms). These compact
clusters use the same calculator and convergence criteria as the 12-water
validation. SYCART converged in 21, 56, and 77 steps, respectively, while the
integrated SONIC handoffs required 1, 1, and 8 additional accepted steps. The
corresponding direct SONIC controls required 15, 25, and 77 steps. Each hybrid handoff
includes at least one SONIC step; for the four-water case ORACLE and SMITH were
reperceived after the SYCART geometry change. Starting
structures and timing records are listed in the Supporting Information. The
panel provides a direct intermolecular scaling control rather
than replacing the larger microsolvation example.

Table~\ref{tab:intermolecular} collects the staged counts for the complete
uracil--water series. The hybrid column includes every accepted SYCART step
and the required terminal SONIC stage; the direct-SONIC column is included as
a control.

\begin{table}[t]
\caption{Staged optimization of uracil--water clusters.}
\label{tab:intermolecular}
\centering
\small
\begin{tabular}{@{}lrrrr@{}}
\toprule
System & Atoms & SYCART & Hybrid & Direct SONIC \\
\midrule
Uracil--H$_2$O & 15 & 21 & 22 & 15 \\
Uracil--(H$_2$O)$_2$ & 18 & 56 & 57 & 25 \\
Uracil--(H$_2$O)$_4$ & 24 & 77 & 85 & 77 \\
Uracil--(H$_2$O)$_{12}$ & 48 & 63 & 83 & -- \\
\bottomrule
\end{tabular}
\end{table}

The Supporting Information reports realization profiles, chart diagnostics,
and basin checks for the difficult cases. In particular, its definitive
benzylpenicillin validation gives 37 SONIC iterations and 38 resident
energy--gradient evaluations, matching ANCopt and lying within one iteration
of SYCART. For supported systems, the same mixed chart combines a Cartesian
substrate, SONIC adsorbate, and quaternion relative pose; the active-region
contract retains boundary forces when atoms are fixed.

\subsection{Robustness boundary}

The 20-case automatic campaign defines the present chemical boundary. SYCART
converges in all 20 cases, direct SONIC in 14, and the certified hybrid in 17.
For aldehyde and nectaryl, SONIC requires 11 and
15 iterations versus 32 and 60 for ANCopt; the corresponding LINK times are
0.323 and 0.444~s versus 0.383 and 0.869~s.  Eleven cases fail the production
gate for a direct SONIC start and are handled by the certified Cartesian or
hybrid route.  The chart, prospective-step, realization, and topology-typing
diagnostics therefore provide a defined operating boundary without preventing
convergence of the general SYCART path.

\begin{figure}[t]
\centering
\includegraphics[width=\linewidth]{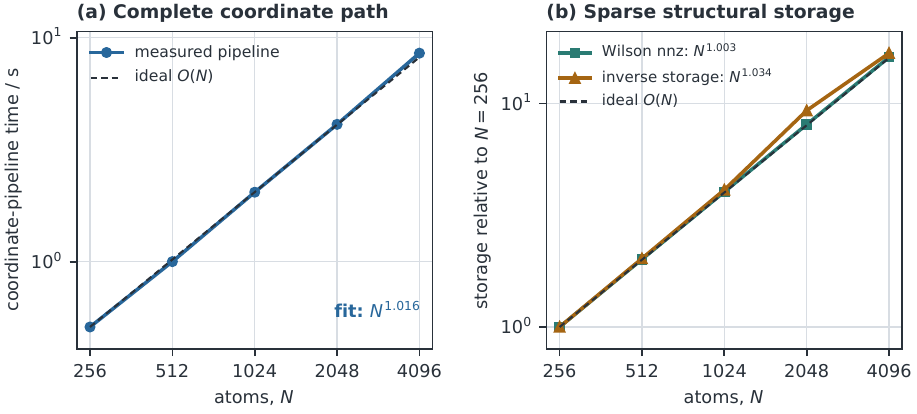}
\caption{One-thread scaling of the calculator-independent ORACLE--SMITH--LINK path.}
\label{fig:oracle-scaling}
\end{figure}

The measured scaling record is shown in Figure~\ref{fig:oracle-scaling}.

Taken together, these results give a chemically transparent rule for choosing a
coordinate system.  For inexpensive force-field or tight-binding gradients,
SYCART is the default because coordinate construction and back-transformation
can dominate; one final monitored SONIC step supplies chemically interpretable
coordinates for the reported structure.  Once a costly quantum-mechanical
surface is reached and topology is certified, SONIC is preferred because the
number of energy--gradient calls is the decisive cost.  Quadraticized SONIC
improves this count for localized anharmonic charts, while mixed SYCART/SONIC
routes cover initially uncertain topologies and other complex cases.  Timing
remains a diagnostic of the low-cost regime rather than the primary SONIC
criterion.  Partial relaxations and chemisorption add a
third benefit: the substrate may remain Cartesian while the adsorbate and its
relative pose use chemically meaningful SONIC coordinates.

\subsection{Practical recommendations}

For routine use we recommend the following order.  Begin with one resident
TBLite energy--gradient evaluation when the input is Cartesian and its
connectivity is uncertain.  Use FF/SYCART only when a chemically complete
topology is supplied (for example by a trusted SMILES or an existing chart).
For inexpensive FF/TB gradients and large systems, retain SYCART through the
optimization because nonlinear coordinate construction and realization can
cost more than the calculator. If chemical interpretation is required, build
SONIC at the final SYCART geometry and take one monitored SONIC step; retain
the handoff only when the Cartesian gradient does not increase. For expensive
QM gradients with certified topology, use SONIC near the minimum because the
number of electronic evaluations is decisive; activate quadraticization for a
localized anharmonic chart. For a mixed system, apply SONIC only to certified
blocks and keep uncertain or reactive blocks in SYCART. The same blockwise
construction applies to transition states: the chemically stable environment
can remain SONIC while the reactive region is treated in SYCART until its
topology becomes certifiable.
For QM/MM and subtractive ONIOM, the same rule is applied by level and region.
The full low-level system is propagated in SYCART, while the certified
high-level region uses a non-overlapping SONIC block.  Link atoms and boundary
directions belong to the low-level SYCART block.  The real-system low-level,
model-system high-level, and model-system low-level gradients are first
assembled in the common Cartesian frame and are then projected once onto the
mixed coordinate map.  At the final geometry ORACLE and SMITH construct a
global SONIC chart and LINK performs the same one-step gradient test, providing
an all-SONIC representation for interpretation without paying its nonlinear
realization cost throughout the low-level optimization.

Parallel kernels are available for independent local domains and calculator
contractions.  To keep the comparison reproducible, all timings reported here
use one thread; multi-thread runs are treated as a throughput study and are
not mixed with the single-thread scaling fits.  The reported exponents
therefore measure algorithmic work rather than a change in thread count or
scheduling policy.

\subsection{Scope and limitations}

The strongest scaling statement in this work concerns the recurrent resident
energy--gradient iteration and the ORACLE--SMITH--LINK coordinate path under
bounded local coordination.  It does not include cold-start SCC
initialization, the nonlinear external xTB implementation, or unqualified
periodic variants.  SONIC robustness is asserted only after the topology,
rank, conditioning, and prospective-step gates pass; failed gates are retained
as explicit boundary cases rather than hidden by fallback data.

\FloatBarrier
\section{Conclusions}

The central result is a self-configuring workflow in which chemical information
and computational cost determine the coordinate route.  One initial energy
and Cartesian gradient, together with bonding and chart diagnostics, selects
direct chemical coordinates or topology-free Cartesian preparation.

For inexpensive gradients, SYCART limits transformation overhead and a final
monitored SONIC step supplies chemical interpretability.  For expensive
gradients with certified topology, SONIC reduces energy--gradient calls;
quadraticization improves localized anharmonic cases, and mixed SYCART/SONIC
charts handle partially certified or reactive systems.  The more diagonal
SONIC Hessian is especially useful for interpretation and reduced-dimensional
anharmonic models.  When topology is uncertain, resident xTB/SYCART prepares a
reliable region before definitive perception and SONIC construction.

The quadraticized SONIC step is the complementary refinement for the regime in
which the topology is certified but the local potential is appreciably
anharmonic. It maps selected nonperiodic coordinates with Morse or Gaussian
functions, retains the physical trust-region test through an analytic pullback
metric, and limits the transported GDIIS history to three local points. The
procedure is activated only for directly separable SONIC charts; delocalized,
SYCART, mixed, and transition-state charts retain their qualified step models.
Across the expanded production campaign it preserved convergence and reduced the
iteration count for difficult minima.  Standard SONIC is used for stable local
charts, quadraticized SONIC for certified anharmonic minima, and SYCART or a
mixed route when Cartesian overhead is the smaller cost.

Extending the workflow to large systems required changes at every layer.
ORACLE now supplies a broader extensible primitive vocabulary and updates
local chemical state selectively.  SMITH constructs, certifies, caches, and
updates nonredundant charts through sparse local operations.  LINK applies the
map, transpose, metric, symmetry projector, and quasi-Newton model without
materializing global dense objects in the production large-minimum route.
Resident GFN-FF removes process and data-transfer overhead that would
otherwise dominate inexpensive Cartesian steps. Resident TBLite-LS also
removes the global dense electronic algebra of conventional xTB, providing
sparse, bounded-domain energy--gradient kernels. The measured coordinate and
resident-electronic path is consistent with near-unit scaling through
8192 atoms over the sampled range, and the independent real-molecule
certificate reaches 6996 atoms with a fitted exponent of 1.250 for the
complete measured coordinate path.  Resident 192--1536-atom
energy--gradient updates scale with exponent 1.04. Cold-start SCC
initialization remains superlinear and is reported separately rather than
hidden in the qualified recurrent-work claim. Native xTB/ANCopt remains the
nonlinear numerical control used to verify energy, gradient, and optimization
behavior; it is not included in the near-linear production claim.

Because every component returns the same Cartesian energy--gradient contract,
the procedure extends naturally from preoptimization to large low-level
regions in QM1/QM2 and QM/MM calculations.  The subtractive composition added
here also supports ONIOM-type energies and gradients, after which the common
LINK transformation is applied only once.  The high-level region is optimized
in SONIC while the low-level real system remains in SYCART; a final global
SONIC construction and one monitored step place the complete result in the
chemical coordinate representation used for interpretation.  The new
contribution is therefore a robust path from large, inexpensive, or
topologically ambiguous starting problems to the established high-level SONIC
description.  This combination of automatic qualification, chemically
expressive coordinates, and linear-scaling infrastructure is intended to let
users focus on the system being studied rather than on the machinery required
to optimize it.

\begin{acknowledgement}
Funding and computational resources are listed in the repository metadata; the
author list for this manuscript comprises Vincenzo Barone.
\end{acknowledgement}

\begin{suppinfo}
Coordinate definitions and rank audits; optimizer and calculator contracts;
benchmark systems and convergence thresholds; complete timing, memory,
iteration, gradient, and restart records; full and partial optimization
comparisons; Cartesian coordinates of all reported structures.
\end{suppinfo}

\section{Data and software availability}

The nano-MATRIX implementation is available in the
\href{https://github.com/yogibubu/nano-MATRIX}{nano-MATRIX repository}
under the BSD 3-Clause License. The ORACLE chemical-state implementation is
identified by \href{https://doi.org/10.1021/acs.jctc.6c01539}{its DOI},
and the SMITH/SONIC coordinate construction by
\href{https://arxiv.org/abs/2607.16550}{arXiv record}. The
\href{https://github.com/yogibubu/LINK}{LINK benchmark archive} contains the
input geometries, protocol definitions, optimization traces, provenance
manifests, analysis scripts, and machine-readable tables used here.  Campaign
directories record the computational route, calculator and machine manifest,
and implementation commit required to reproduce the reported iteration counts
and timings.  Resident calculator tests use the documented Python environment;
records obtained with Gaussian, ORCA, and NWChem identify the corresponding
external installations.

Artificial-intelligence tools were used to organize author-generated numerical
results, transfer them into the Supporting Information, format the manuscript,
and correct its English.  They were not used to generate, modify, or interpret
the data.  All calculations, scientific decisions, and final content were
checked and approved by the author.

\bibliography{references}

\end{document}

The tools used here are ORACLE, SMITH, and LINK.  SONIC
denotes symmetry-oriented nonredundant internal coordinates, SYCART
symmetry-adapted Cartesian coordinates, TBLite the in-process tight-binding
implementation, and ANCopt the native xTB optimizer.

\section{Reproducibility contract}

Each run is identified by the LINK commit, optimizer protocol hash, immutable
coordinate artifact, calculator contract, input geometry hash, active/frozen
selection, and machine record. Timing comparisons are admissible only within a
declared host/runtime combination. Cross-host results may establish
portability, but are not combined into one wall-time speedup.

Required calculator fields are name, version, executable or implementation
hash, energy and gradient units, atom ordering, charge and multiplicity where
applicable, topology/parameter fingerprint for classical methods, supported
elements, cutoff and boundary policy, deterministic settings, and restart
state. A calculator must return the energy and a finite Cartesian gradient for
the supplied geometry without changing atom identity.

\input{coordinate_families.tex}

\section{Revised continuous atomic-synthon contract}
\label{sec:revised-synthons}

The production synthon contract stores four fields: charge $q$, geometric
compactness $C$, incident bond-order product $D$, and angular strain $S$.
These are the only components used in the synthon PCA and effective-number
calibration.

The coordination evidence is evaluated before acceptance of a discrete graph,
\begin{equation}
 \mathrm{CN}_i=\sum_{j\ne i}\frac12\left[1+
 \operatorname{erf}\{\alpha_{\rm CN}(R_{ij}^{0}-R_{ij})\}\right],
\end{equation}
where the coordination-dependent Pyykk\"o radii entering $R_{ij}^{0}$ are
obtained by $C^1$ Hermite interpolation.  A complete CM5 charge vector and
Mayer bond-order matrix are used when present; otherwise charge and bond order
fall back independently to the scaled electronegativity and Pauling estimates,
respectively.

With $R_i^{\rm cov}$ the interpolated covalent radius and $\bar R_i$ the mean
incident distance, the compactness and radial-dispersion descriptors are
defined next:
\begin{equation}
 C_i=\frac{1}{|\mathcal N(i)|}\sum_{j\in\mathcal N(i)}
 \frac{R_i^{\rm cov}+R_j^{\rm cov}}
 {R_{ij}+R_i^{\rm cov}+R_j^{\rm cov}},\qquad
 \Delta_i^R=\frac{1}{|\mathcal N(i)|\bar R_i}
 \sum_{j\in\mathcal N(i)}|R_{ij}-\bar R_i|.
\end{equation}

For a non-isolated atom with positive incident bond orders, the revised
descriptor is evaluated in the log domain,
\begin{equation}
 D_i=\exp\left[\sum_{j\in\mathcal N(i)}
 \log(\mathrm{BO}_{ij})\right].
\end{equation}
This expression is numerically the product defined in the main text; an
isolated atom, or a state containing a non-positive incident bond order, is
assigned $D_i=0$.
The bond orders and therefore $D_i$ vary continuously while the certified
neighbor set is unchanged.  Crossing a topology-acceptance boundary produces
a new versioned chemical state and invalidates dependent atom classes and
primitive-curvature records. The xTB-compatible entries of the current
ORACLE primitive-diagonal SH do not use this descriptor. The radial dispersion
shown above is retained as an explicit
geometric diagnostic and does not enter $D_i$.

The electronic-domain count is
$N_{\rm ED}=\mathrm{CN}+N_{\rm LP}+N_{\rm OS}$.  Reference mean angles are
tabulated for domain counts 2, 3, 4, 5, 6, 8, and 12 as
$180.0^\circ$, $120.0^\circ$, $109.47122^\circ$, $120.0^\circ$,
$90.0^\circ$, $90.0^\circ$, and $63.43^\circ$, respectively.  Hermite
interpolation supplies every intermediate count; values below 2 or above 12
use the nearest endpoint.  Consequently the main-text strain expression is
the same for ordinary, hypervalent,
transition-metal, and high-coordinate environments.  Atoms with fewer than
two accepted neighbors have zero angular strain.  Its normalization uses the
actual number $|\mathcal P_i|=|\mathcal N(i)|[|\mathcal N(i)|-1]/2$ of accepted
neighbor pairs; the continuous $N_{\rm ED}$ enters only the reference angle.

The effective-atomic-number map is interpreted as two contributions.  The
electrostatic contribution is $\Delta Z^{\rm el}=-q$.  The shape contribution
is obtained from a common first principal component of the standardized
$(C,D,S)$ descriptors.  The fit uses 328 LCB25 PCS2 molecules with CM5
charges and Mayer bond orders, covering 1507 carbon, 184 nitrogen and 180
oxygen centers.  This is one data-driven calibration, not a unique choice:
the reference library and the bounded map can be refit when the chemical
domain is extended.
\begin{equation}
 \begin{aligned}
 \mathbf{x}^{\rm CDS}_i={}&((C_i,D_i,S_i)-\boldsymbol{\mu}_{\rm CDS})
 \oslash\boldsymbol{\sigma}_{\rm CDS},
 \\
 \xi_i^{\rm shape}={}&\mathbf{v}_{\rm CDS}^{\mathsf T}\mathbf{x}^{\rm CDS}_i,
 \\
 Z_i^{\rm eff}={}&Z_i-q_i+\Delta Z_i^{\rm shape}.
 \end{aligned}
\end{equation}
The pooled shape-PCA parameters are
\begin{equation*}
\begin{aligned}
\boldsymbol\mu_{\rm CDS}&=(0.483,1.719,0.180),\\
\boldsymbol\sigma_{\rm CDS}&=(0.016,0.675,0.159),\\
\mathbf v_{\rm CDS}&=(0.613,0.607,-0.506).
\end{aligned}
\end{equation*}
Its first component
explains 73.5\% of the shape variance and the first two explain 93.1\%.
Only $q$, $C$, $D$, and $S$ enter this metric.  A scalar effective number can be degenerate when
charge and shape compensate; because $q_i$ is stored independently, the shape
contribution is recovered exactly as
$\Delta Z_i^{\rm shape}=Z_i^{\rm eff}-Z_i+q_i$.
Hydrogen is handled separately as $Z_H^{\rm eff}=1-q_H$; the LCB25
distributions are reported in the main text. The shape-PCA is deliberately
calibrated only for C, N, and O because these are the elements represented with
sufficient population and consistent electronic annotations in the curated
LCB25 archive. This is a coverage limit of the reference data rather than a
limitation of the descriptor definition. For sulfur, transition metals, and
other elements outside the calibrated domain, $C$, $D$, and $S$ remain available
for structural typing, while the present effective-number map uses only
$Z^{\rm eff}=Z-q$ and does not extrapolate a shape score. The tmQM cases in the
main text therefore test topology and coordinate handling for transition-metal
systems, not an unsupported PCA calibration.
Atom similarity
uses the effective-atomic-number difference together with the synthon-vector
distance.  Regression cases include acetylene-like ($D=3$), central cumulene
($D=4$), and aromatic carbon ($D=2.25$) ideal bond-order patterns, together
with explicit preservation of the radial descriptor.

\section{Sparse ORACLE--SMITH construction and restart}
\label{sec:sparse-oracle-smith}

At each accepted topology checkpoint ORACLE serializes atom order, geometry
fingerprint, continuous and discrete connectivity, symmetry operations,
fragments, local references, and redundant primitive definitions.  Candidate
pairs are stored in a persistent Verlet list and rebuilt after half of the skin
has been consumed.  Atomic displacements accumulate relative to the last
actually perceived frame.  Atoms exceeding the update threshold are expanded
by a bounded number of graph shells; an empty domain skips perception, whereas
a nonempty domain renews the affected pair list and then applies the global
semantic certificate.  Connected components, cycle candidates, and reference
searches use iterative graph traversal.  For systems of at least 256 atoms,
linear-bend references are restricted to two graph shells.

Primitive generation obeys a closed support grammar. For an atom-centered
primitive $p_\alpha$,
$\operatorname{supp}(p_\alpha)\subseteq\{i\}\cup\mathcal N(i)$. For a
bond-centered primitive,
$\operatorname{supp}(p_\alpha)\subseteq\{i,j\}\cup\mathcal N(i)\cup
\mathcal N(j)$ for the central bond $i$--$j$. ORACLE never combines internal
distortions located in distinct parts of a fragment. The only special support
records are pair interactions between two fragment centers, with either center
optionally represented by one atom, and rigid fragment rotations. Larger ring,
face, cage, and other collective coordinates are therefore SMITH combinations,
not ORACLE primitives.

Geometry provenance is independent of chemical perception.  Ordered element
labels and rounded Cartesian arrays are hashed, followed by proper Kabsch
alignment.  Reflections and atom permutations are rejected.  The small-system
audit uses pair-distance discrepancies; above 256 atoms the rigorous bound of
twice the maximum aligned atomic displacement avoids two dense distance
matrices.  Electronic continuous bond orders are acquired at the initial
checkpoint and frozen for subsequent SH queries.  When no electronic order is
available and a force-field geometry cannot support a reliable continuous
estimate, integer orders from the accepted connectivity are used explicitly.

SMITH fingerprints the sparse Wilson operator from its dimensions and
canonical IEEE-754 CSR entries.  Its row--Cartesian support graph is separated
into support-disjoint leaf components and locality-crossing separator rows.
Structural rank, unsupported rows, connected components, and incremental
bipartite matching are evaluated in one compiled kernel.  A failed chart
receives a Hall-deficient witness from an alternating forest with polynomial
pruning; accepted pools do not construct a failure witness.  Only rows whose
support intersects a topology invalidation domain are reevaluated.

The compiled primitive plan, GIC combination matrix, matching state, and CSR
Wilson arrays are persistent artifacts.  Cache entries are written atomically,
carry ABI and scientific-state fingerprints plus per-array checksums, and are
restored as read-only memory maps.  Deterministic byte-budget pruning uses
least-recently-used order.  Corrupt, incomplete, or scientifically
incompatible entries are rejected rather than partially reused.  The dense
SVD path is retained only as a small-system parity oracle.

\section{Adaptive-metric selection and chart epochs}
\label{sec:si-adaptive-metrics}

ORACLE serializes, for every primitive, its family, ordered class rank,
fragment-domain kind, local-domain identifier, and symmetry ownership.  Class
ranks are extensible metadata and are independent of force constants.  SMITH
partitions the rows by the tuple (intrafragment/interfragment ownership, local
domain, irrep); no subsequent operation may join distinct tuples.  A local
topology certificate may be supplied for every domain.  Rows from an
uncertified domain are omitted from chemical-coordinate construction and the
missing local rank is assigned to the matrix-free SYCART block.

For each certified partition, SMITH normalizes the sparse primitive Wilson rows and
constructs either the geometrical metric
$\mathbf{G}=\widetilde{\mathbf{B}}\widetilde{\mathbf{B}}^{\mathrm T}$ or its
SH-weighted alternative based on ORACLE's single primitive diagonal,
$\mathbf{G}_{\mathrm{SH}}=\mathbf{K}_{\mathrm{SH}}^{1/2}\mathbf{G}
\mathbf{K}_{\mathrm{SH}}^{1/2}$, as defined in the main text. No ring,
fullerene, or other derived SONIC family defines a second curvature. At level zero, eigensolutions
are formed separately for each class.  Subsequent levels cumulatively merge
classes in increasing ORACLE rank.  Structural row independence is selected
by incremental bipartite matching; numerical acceptance requires the target
rank and the requested normalized condition threshold.  Selection stops at
the first passing level separately for every partition.  Dense algebra is
bounded by the configured maximum local block size.  Larger production
problems use sparse products and extremal-spectrum iteration; exceeding the
local dense budget is a failed domain certificate rather than permission to
construct a global dense matrix.

The resulting certificate records metric policy, selected level and maximum
class rank, structural and numerical rank, geometrical and selection condition
numbers, number of local blocks, domains in which classes were mixed, and
domains delegated to SYCART.  The two public policies are
\texttt{GEOMETRY} and \texttt{GEOMETRY\_SH}; the latter requires one positive
SH curvature for every participating primitive.  No atomic mass enters either
policy.

ORACLE schedules reassessment only after accepted displacements consume the
local persistence margin or after a chemical-state trigger.  A topology
change replaces the primitive payload and invokes ORACLE followed by SMITH.  A
topology-preserving conditioning change invokes SMITH alone on the frozen
primitive payload, so that the nonredundant SONIC selection can change without
repeating chemical perception.  If a new basis is accepted, LINK increments
the chart epoch, reuses the already available Cartesian energy and gradient,
and transports the gradient and Hessian seed by congruence through the old and
new Cartesian tangent maps.  GDIIS history and quasi-Newton/L-BFGS secant
pairs are then cleared, while the transported Hessian becomes the new seed.
Atomic order, Cartesian frame, topology
fingerprint, target rank, direct-sum ownership, and finite-condition checks
must all pass before the transition is committed.  Failure leaves the current
accepted geometry unchanged and routes only the affected domain to its next
allowed level or to SYCART.

\section{Refinement orchestration boundary}

The three tools exchange versioned structures but retain disjoint ownership.
ORACLE owns perceived chemical state, molecular structural elements, and the
redundant primitive registry. SMITH owns the selected nonredundant SONIC chart
and its degree of delocalization. LINK owns the state of a local optimization
and coordinates the external programs or resident calculators that provide
energies, gradients, and auxiliary molecular properties. LINK receives a
versioned structure and returns a refined geometry, score, convergence
certificate, and chemical-state identity. This boundary keeps local refinement
independent of the program that supplies candidate structures.

\section{Composite calculator assembly}
\label{sec:composite-calculator}

For full-system Cartesian vector $\bm x_R$ and model-region selector
$\mathcal S_M$, the resident subtractive calculator evaluates
\begin{align}
 E_{\mathrm{comp}} &=
 E_{\mathrm{low}}(R)+E_{\mathrm{high}}(M)-E_{\mathrm{low}}(M),\\
 \bm g_{\mathrm{comp}} &=
 \bm g_{\mathrm{low}}^R+
 \mathcal S_M^{\mathrm T}
 \left(\bm g_{\mathrm{high}}^M-\bm g_{\mathrm{low}}^M\right).
\end{align}
All component gradients are converted to the common Cartesian units before
assembly.  Atom-map, geometry, charge, multiplicity, boundary, and calculator
fingerprints are checked at every composite checkpoint.  Model-region
contributions are scattered only after the atom map passes.  Fixed atoms,
link-atom definitions, and force redistribution are serialized parts of this
contract.  LINK receives one energy and one Cartesian gradient and therefore
applies the selected SYCART or SONIC transpose only once.  The same interface
covers QM1/QM2, QM/MM, and ONIOM-type subtractive constructions.

\subsection{Water-dimer QM1/QM2 demonstration}
\label{sec:water-dimer-qm1qm2}

The reproducible ONIOM-type example is a six-atom water dimer. The first
water is the model region and the second is the low-level environment. At each
LINK point the three Cartesian jobs are evaluated at identical coordinates:
the full dimer with resident GFN2-xTB/TBLite, the model water with the same
resident calculator, and the model water with ORCA using
\texttt{PBE0 def2-SVP EnGrad}. LINK assembles
$E_{\rm low}(R)+E_{\rm high}(M)-E_{\rm low}(M)$ and the corresponding
gradient, then continues the optimization without a coordinate or geometry
reset. The example is deliberately noncovalent, so no cap is needed.

For a covalent model boundary, the same calculator session constructs one
hydrogen link atom for each cut bond. Its position is the fixed affine
interpolation between the model and environment atoms; the transpose of this
map scatters both the model correction gradient and Hessian back to the real
system. Boundary atom maps, cap scales, charge, multiplicity, and the
coordinates used by all three jobs are serialized in the composite checkpoint.

\section{Bounded-memory step model}

For accepted points $k$, L-BFGS retains at most $m$ pairs $(\bm s_k,\bm y_k)$.
The inverse-Hessian action is formed by the two-loop recursion. The initial
scalar is
\begin{equation}
 \gamma_k=\frac{\bm s_k^{\mathrm T}\bm y_k}
 {\bm y_k^{\mathrm T}\bm y_k},
\end{equation}
for the newest admissible pair. Nonfinite pairs and pairs failing the
scale-aware positive-curvature threshold are discarded. The state is cleared
when the coordinate-chart epoch or calculator accuracy tier changes. A strict
$\bm g^{\mathrm T}\bm p<0$ test guards the final proposal.

The production record must contain history length, accepted/skipped updates,
directional derivative, fallback status, realized Cartesian displacement,
predicted and actual energy decrease, trust update, and time spent in direction
construction, nonlinear realization, calculator evaluation, gradient
projection, and checkpoint serialization.

Function-level profiling is used because resident \texttt{tblite}, analytic
force fields, and MM calculators make coordinate work a measurable part of the
runtime. The benchmark therefore uses function-
level profiling to identify repeated primitive values, Wilson rows, coordinate
maps, and history covectors. Optimizations are accepted only when they preserve
the frozen-chart numerical result; time saved by changing convergence criteria
or omitting nonlinear realization is not counted as an overhead reduction.

The final implementation contains five additional controls. First, a cached
GICForge value plan stores the sparse SONIC--primitive coefficient matrix and
the reference primitive vector; each requested active subset is evaluated by
one compiled primitive batch rather than one call per SONIC. Principal and
continuation torsions are unwrapped vectorially before the sparse matrix
product. Second, the accepted-point sparse Wilson inverse survives the
phase-sensitive cache reset, and the already known accepted SONIC values are
passed into the corrector. Third, the same operator can be reused for ten
microiterations, subject to an earlier refresh at relative secant defect 0.15
or after a stalled solve. Fourth, the nonlinear Newton cap is
$0.25\sqrt{N}$ in global Cartesian norm so that it represents a constant
per-atom RMS displacement rather than decreasing spuriously with system size.
Fifth, limited-memory SONIC optimization uses the controlled GDIIS construction
of Farkas and Schlegel~\cite{FarkasSchlegel2002GDIIS} on a bounded history. One
current RFO-shifted effective Hessian supplies the inverse action for the
reference displacement and every historical residual. Points are added from
newest to oldest, and LINK retains the last prefix passing the four published
tests: the history-size-dependent angular cutoff, the factor-10 displacement
bound, the bound of 15 on the positive or negative coefficient sum, and the
$10^8$ limit on the scaled inverse-overlap solution. All residuals are scaled
by the smallest residual norm before solving the overlap equation. Only a
candidate more than $90^\circ$ from the reference discards the offending point
and all older points. No additional force-monotonicity rule or empirical trust
contraction is applied. Promotion of the calculator accuracy tier or a chart
change clears the history.

For the quadraticized SONIC branch the production cap is three retained GDIIS
points, independently of the larger user-visible history setting. This is a
representation-level rule: it applies to every direct separable quadraticized
SONIC calculation, while delocalized charts, SYCART, mixed charts, and
transition-state calculations retain their qualified settings. The comparison
records used in the main text contain the previous dense-RFO route and the
quadraticized route with their complete iteration and timing manifests.

On the one-thread ORACLE profiles these changes reduce the Python call count
from 6.2 to 3.7 million, sparse Wilson builds from 27 to 17, and active-value
calls from 91 to 66 for the profiled trajectory. Three independent timing
replicates preserve both iteration counts and final energies. Menthone remains
at 10 accepted steps/11 evaluations and changes from
0.723/0.572/0.201~s to median 0.484/0.327/0.113~s for optimizer,
non-calculator overhead, and realization. ACONF $n$-hexane remains at 11/12
and changes from 0.468/0.405/0.148~s to 0.333/0.268/0.088~s. The corresponding
overhead reductions are 42.7\% and 33.8\%; realization decreases by 43.8\% and
40.1\%. These figures exclude fresh Python process startup and retain the
resident calculator as a separately measured component.

The automatic implementation has four additional identity-preserving
controls. For an explicit Cartesian input, the initial resident
TBLite/GFN2-xTB energy--gradient evaluation is performed in Cartesian space.
Its Cartesian force, first-point continuous bond-order evidence, and
prospective chart certificate enter a three-branch routing decision. A certified,
low-gradient structure enters TBLite/SONIC directly. A high-gradient structure
with reliable topology enters GFN-FF/SYCART and subsequently TBLite/SONIC. If
topology is doubtful, TBLite/SYCART is used until the topology and chart pass
their block-boundary gates, followed by TBLite/SONIC. This is the instance
$L_i$/SYCART or $L_i$/SONIC $\rightarrow L_{i+1}$/SONIC of the general
multilevel rule. GFN-FF/SONIC is forbidden in the automatic policy because
SONIC overhead exceeds the force-field point cost. A SMILES input supplies
explicit connectivity, whereas the quality of its generated Cartesian
embedding remains a separate gate. ORACLE and SMITH are rebuilt before SONIC
begins; the SYCART stage need not reach its own convergence threshold when the
projected Cartesian force has already entered the SONIC handoff window.
Intermediate chart gates evaluate the frozen
SONIC definition at the accepted checkpoint geometry and do not rerun
ORACLE--SMITH perception at every probe; the definitive chart is still rebuilt
once at handoff. If a rejected trial leaves the accepted Cartesian point,
gradient, chart epoch, electronic tier, and L-BFGS history unchanged, the exact
generalized-RFO direction is cached and only the smaller trust restriction is
reapplied; any identity change invalidates the cache. Finally, LINK reads the
SHA-256 digest already verified while loading the frozen ORACLE payload instead
of converting the complete contract to JSON a second time. The remaining
required serialization walks immutable frozen records without deep-copying
scalar leaves; on the 81-atom i4p contract this operation decreases from 91.7
to 49.4~ms in a 20-repeat one-thread microbenchmark.

The matched F22 automatic route uses 10 SYCART plus 8 SONIC steps with
bitwise-identical energy paths and final Cartesian components; total hybrid wall
time is 4.460~s. Artemisinin and i4p use 20+23 and 15+31 steps, respectively.
SONIC generalized-metric applications and nonlinear back-transformation remain
the dominant overhead relative to the projected Cartesian action used by SYCART.

As a coordinate-only control, the 62-atom eicosane structure has a
rank-180/180 sparse SONIC chart and initial chart condition number 8.54.
Table~\ref{tab:eicosane-coordinates} reports matched one-thread optimizations
from the same generated Cartesian embedding. It does not test the production
SMILES entry protocol, which first applies SYCART/GFN-FF and rebuilds the
ORACLE--SMITH state.

\begin{table}[t]
\centering
\caption{Eicosane coordinate-only control under the common geometry-seed
stopping contract.}
\label{tab:eicosane-coordinates}
\begin{tabular}{lrr}
\hline
Route & Iterations & Wall time/s \\
\hline
Dense SYCART & 32 & 1.85 \\
TBLite-LS SYCART & 31 & 4.34 \\
TBLite-LS SONIC & 8 & 2.11 \\
Native xTB/ANCopt & 46 & 1.56 \\
\hline
\end{tabular}
\end{table}

The complementary production-path run applies 44 GFN-FF/SYCART iterations
(45 force-field gradients) in one persistent compiled-library session and
rebuilds ORACLE--SMITH at the accepted minimum. The GFN-FF topology contributes
61 bond, 120 angle, and 19 central-bond-aggregated proper-torsion analytic
curvatures to the initial matrix-free Hessian action. Multiple proper torsions
on one central bond are summed once; no gradient difference or dense Cartesian
congruence is used. The ensuing matrix-free TBLite-LS/SONIC segment uses the
sparse SH compliance action and converges in 4 iterations and 5 gradients.
Direct TBLite-LS/SONIC requires 7 iterations and 8 gradients. The endpoints
differ by $1.985\times10^{-8}\ E_{\mathrm h}$ and
$9.27\times10^{-4}$~\AA{} after rigid alignment. These counts must be cost
weighted: the preparation gradients are force-field evaluations, whereas the
three gradients removed from the second segment are electronic-structure
evaluations. GFN-FF/SYCART takes 2.840~s and TBLite-LS/SONIC 1.240~s on one
ORACLE thread. The complete hybrid takes 4.080~s, 2.266~s more than the direct
control for this cheap upper level. The archived execution audit records zero process
launches, no dense Hessian, L-BFGS in both coordinate spaces, and
\texttt{solver\_mode=linear} for every upper-level point. The lower session is
the resident \texttt{libxtb}\cite{Bannwarth2019} parity reference; it is not used as evidence for
the separate native ARCHITECT linear-scaling certificate.

The coordinate comparison has two independent outcomes and runs both routes
for every system. Total accepted iterations and energy--gradient evaluations
measure the electronic-work reduction obtained with SONIC. Resident wall time
per accepted iteration measures optimizer overhead and is partitioned into
calculator, coordinate realization, and remaining LINK work. The iteration count is
decisive for expensive QM gradients; time per iteration is decisive for inexpensive
semiempirical, FF, and MM calculators. For the inexpensive reference,
SYCART must be no slower per iteration than native xTB/ANCopt. Process startup
and ORACLE--SMITH construction are retained as separate timing fields.

No full eigendecomposition of the SYCART Hessian is used: within the frozen
totally symmetric subspace it is merely a change of basis, and null external
modes have already been removed. The large-system SYCART path instead combines
matrix--vector products with a bounded Davidson subspace as the initial
preconditioner for L-BFGS. It consumes the same ORACLE primitive diagonal as
every SONIC chart. The registry uses the xTB-compatible local entries by
default and may accept molecule-specific analytic curvatures from a resident
calculator without creating a second downstream model. The sparse congruence
and bounded Davidson deflation are rebuilt
at the accepted geometry after every ten microiterations. Davidson is seeded
from the preceding bounded Ritz space and compatible L-BFGS secants are
retained. SONIC remains a complete
optimization route, not only a postprocessing label.

\section{Initial Hessian ownership and SONIC transformation}
\label{sec:sh-ownership}

\paragraph{ORACLE: one model in the redundant primitive space.}
Let $\bm p$ collect the local redundant primitives selected from the ORACLE
registry and let $\bm B_p=\partial\bm p/\partial\bm x$. ORACLE is the sole
owner of SH and constructs it once as
\begin{equation}
 \bm K_p^{\rm SH}=\operatorname{diag}(k_1,\ldots,k_{n_p}).
 \label{eq:si-primitive-sh}
\end{equation}
Every diagonal record carries the primitive family, support, units, curvature,
parameter provenance, and chemical-state fingerprint. Standard distance,
angle, linear-bend, torsional, and pseudobond entries use the xTB 6.7.1
\texttt{ddvopt} rules.\cite{xtb671ModelHessian} Local extensions such as
out-of-plane heights, SOOPs, triangular flaps, and bond-centered hinges obey
the same atom/bond support grammar. Special entries are limited to pair-center
interactions and rigid fragment rotations. All enter the same registry. ORACLE does not construct a
second model for any downstream coordinate representation, and neither SMITH
nor LINK contains an independent copy of these force-constant rules.

The distinction between primitive and derived coordinates is strict. A SOOP
or a signed triangular flap is a local ORACLE primitive. Ring puckering, butterfly,
condensed-ring, fullerene, and other collective coordinates are SMITH
combinations and therefore have no independent diagonal entry in
Eq.~\ref{eq:si-primitive-sh}. Exact aliases share one primitive record, so
duplicating a candidate cannot multiply its curvature.

\paragraph{SMITH: congruence to the selected nonredundant chart.}
For a selected chart $\bm q$, SMITH certifies the tangent map
\begin{equation}
 \bm T_{p\leftarrow q}=\frac{\partial\bm p}{\partial\bm q}
                       =\bm B_p\frac{\partial\bm x}{\partial\bm q}
\end{equation}
and constructs
\begin{equation}
 \bm H_q^{\rm SH}=\bm T_{p\leftarrow q}^{\mathrm T}
                  \bm K_p^{\rm SH}\bm T_{p\leftarrow q}.
 \label{eq:si-sh-congruence}
\end{equation}
The SONICs themselves are linear combinations of the frozen primitive
functions; Eq.~\ref{eq:si-sh-congruence} is their corresponding tangent-space
congruence. It generates the physically useful off-diagonal couplings between
SONICs without adding springs for derived families. Its matrix-free form is
the composition $\bm T_{p\leftarrow q}$, the diagonal primitive action, and
$\bm T_{p\leftarrow q}^{\mathrm T}$; explicit matrices are confined to bounded
audits.

Exactly the same construction applies along the full adaptive hierarchy. The
first chart is a direct sum of local, class-preserving SONIC blocks. Classes
are then mixed only where required, producing progressively less local
nonredundant SONICs and, in the limiting internal case, traditional fully
delocalized coordinates. SYCART is the terminal family: its tangent map sends
translation/rotation-free Cartesian displacements to the same ORACLE
primitive vector. Intrafragment and interfragment primitive spaces remain a
strict direct sum at every level.

\paragraph{LINK: consumption and precedence.}
LINK consumes the SH action supplied for the active chart and never assigns
force constants. Explicitly supplied Hessians and compatible transported
operators retain precedence. A chart change causes SMITH to recompute only the
congruence in Eq.~\ref{eq:si-sh-congruence}; it does not cause ORACLE to define
a new model. Atom identities, frozen geometry, topology state, primitive
registry version, and units are checked before an SH record is reused.

\paragraph{Parity and scaling scope.}
We compared standard primitives with the installed libxtb Hessians.  Each
Hessian was transformed onto the same frozen chart for 47 molecules. The
median relative Frobenius error is $7.06\times10^{-15}$ and the maximum is
$1.244\times10^{-9}$ (cubane); AD6 gives $3.11\times10^{-13}$.
Each case also checks explicit versus matrix-free congruence. Additional local
primitive families are qualified by analytic-derivative, diagonal-action,
alias-invariance, and transformed-action tests. These are Hessian-only tests,
with no energy or gradient evaluations. Exact all-pair pseudobond parity is a
quadratic audit mode; the production large-system route uses bounded local
candidate lists.

\subsection{Primitive-curvature registry}
The registry below is the single typed interface from which ORACLE constructs
Eq.~\ref{eq:si-primitive-sh}. Values are attached only to elementary ORACLE
measurements. Names of derived SONIC families may be retained as input aliases,
but resolve to their primitive sources and never introduce an additional
spring.

Each atlas call carries the one-based atom identities, atomic numbers, frozen
effective atomic numbers, atomic charges, and the bond orders within the local
support. These fields are mandatory even when the version-1 parameter does not
yet depend on all of them. Stretch equilibrium values use

\begin{equation}
 r_0=(r_{\rm cov,A}+r_{\rm cov,B})b^{-0.16},
\end{equation}

where $b$ is the frozen bond order. Other equilibrium values are currently
family constants. For a proper torsion $i$--$j$--$k$--$l$, the version-1
curvature is
\begin{equation}
 k_{ijkl}=n_j n_k\left[0.0023+0.07\,S_{jk}(b_{jk})\right],
\end{equation}
where $n_j$ and $n_k$ count substituents on the two sides of the central bond
and $S_{jk}$ is the Baker covalent-overlap term evaluated from its frozen
continuous ORACLE bond order. No torsional periodicity is required to define
this local curvature. When an electronic population analysis is unavailable
and a force-field geometry is far from equilibrium, ORACLE freezes integer
orders 1, 2, or 3 from the declared connectivity instead. Local height, SOOP,
and face-flap primitives use \AA{} and
$E_{\rm h}\,{\text{\AA}}^{-2}$ rather than radian units. Table~\ref{tab:sh-atlas}
records the deliberately provisional values used for the first direct tests.
A subsequent refinement may fit typical values to representative QM Hessians
without changing the SH contract.

The catalog may retain derived-family names for diagnostics and compatibility,
but the builder resolves them to elementary ORACLE measurements. Derived SONIC
labels do not introduce independent springs on top of those primitive
contributions.

\scriptsize
\setlength{\tabcolsep}{3pt}
\begin{longtable}{@{}p{5.2cm}p{2.6cm}rp{3.3cm}@{}}
\caption{ORACLE primitive-diagonal SH registry and units.}\label{tab:sh-atlas}\\
\toprule
ORACLE primitive/operator & $k$ or model & $q_0$ & $q_0$ unit/model \\
\midrule
\endfirsthead
\toprule
ORACLE primitive/operator & $k$ or model & $q_0$ & $q_0$ unit/model \\
\midrule
\endhead
\texttt{R} (covalent) & 1.000 & --- & covalent radii/bond order \\
\midrule
\texttt{R} (local X--H) & 1.200 & --- & covalent radii/bond order \\
\midrule
\texttt{A} & 0.250 & 1.91063 & rad \\
\midrule
\texttt{L} & 0.100 & 0 & rad displacement \\
\midrule
\texttt{D} & $n_jn_k k_B(b_{jk})$ & 0 & rad \\
\midrule
\texttt{U}/\texttt{H} & 0.120 & 0 & rad or \AA{} height \\
\midrule
\texttt{IMPD} & 0.120 & 0 & rad \\
\midrule
SOOP & 0.050 & 0 & \AA{} height \\
\midrule
Local angular flap & 0.050 & 0 & rad \\
\midrule
\texttt{FACE\_FLAP} & 0.050 & 0 & \AA \\
\midrule
\texttt{FACE\_HINGE} & 0.050 & 0 & rad \\
\midrule
\texttt{CENTER\_FACE\_DIST} & 0.100 & 2.5 & \AA \\
\midrule
\texttt{FACE\_DIRECTION\_ANGLE} & 0.080 & 0 & rad \\
\midrule
\texttt{CONTACT\_CENTER\_ANGLE} & 0.080 & 1.91063 & rad \\
\midrule
\texttt{FC\_DIST}/\texttt{FCA\_DIST} & 0.100 & 3.0/2.5 & \AA \\
\midrule
\texttt{FTRANS} & 0.080 & 0 & \AA \\
\midrule
\texttt{FROT} & 0.025 & 0 & quaternion exponential map/rad \\
\midrule
\texttt{CENTER\_ATOM\_DIST} & 0.100 & 2.5 & \AA \\
\midrule
\texttt{R} (pseudobond) & 0.080 & 2.5 & \AA \\
\midrule
\texttt{A} (pseudobond) & 0.080 & 1.91063 & rad \\
\midrule
\texttt{D} (pseudobond) & 0.015 & 0 & rad \\
\midrule
\texttt{FLIN\_TRANS} & 0.080 & 0 & \AA \\
\midrule
\texttt{FAXIS} & 0.025 & 0 & stereographic map/rad \\
\midrule
$R^{-1}$ pair function & 0.100 & 0.33333 & \AA$^{-1}$ \\
\bottomrule
\end{longtable}
\normalsize

The typed $(k,q_0)$ records provide the interface for the common starting
model. SH is diagonal in the elementary ORACLE atlas and coupled in every
selected SONIC chart through the admissible congruence;
additional entries or
chemically typed resolvers can be appended without modifying the coordinate or
calculator interfaces. Qualification tests check the local diagonal,
duplicate-weight invariance, mixed-family congruence, certified inverse actions,
substituent-count and
central-bond-order torsional typing, local-height
units, and propagation of charge, $Z_{\rm eff}$, atomic number, and bond order.

\subsection{Globally regular $Z$--$X(Y)_3$ angular chart}

The five-dimensional angular block uses two $E$ components from the three
$Z$--$X$--$Y$ angles, two $E$ components from the three $Y$--$X$--$Y$ angles,
and one $A_1$ component.  The apparently natural symmetric row
\begin{equation}
 q_{A_1}^{YXY}=3^{-1/2}
 \left[\theta(Y_1XY_2)+\theta(Y_1XY_3)+\theta(Y_2XY_3)\right].
\end{equation}
This expression is unsuitable as a global coordinate: at a planar $XY_3$ arrangement the
bracket is exactly $2\pi$, so its first derivative is zero.  The atlas instead
defines
\begin{equation}
 q_{A_1}^{ZXY}=3^{-1/2}\sum_{i=1}^{3}\theta(ZXY_i)
\end{equation}
as the $A_1$ umbrella for every true $Z$--$X(Y)_3$ center.  It is regular at
the planar limit and provides displacement from the frozen reference through
its angle increments.  The definition is independent of the instantaneous
geometry: no local singular-value decomposition, threshold, branch, or chart
change is needed.  Its construction and storage are constant per center.

For the distorted artemisinin test, all three methyl centers use the same
$ZXY$ definition.  At the exceptional center the candidate minimum singular
values of the old and new local blocks were 0.002871 and 1.130905.  Replacing
only that row reduced the largest singular value of the internal-to-Cartesian
map from 646.55 to 66.21; applying the uniform definition to all three methyl
centers lowers it further to 63.79.  Removing every coordinate touching the
peroxide atoms instead left this value at 627.41, proving that the initial
failure was not a peroxide-specific singularity.  The uniform chart converged
in 93 steps and 94 TBLITE evaluations to $-63.670272243\ E_{\mathrm h}$.
The weakest remaining mode mixes ring-deformation coordinates of the
peroxide-containing ring with those of the fused neighbouring ring.  Its four
largest coefficients belong to \texttt{RDef0011}, \texttt{RDef0002},
\texttt{RDef0013}, and \texttt{RDef0003}; individually these rows retain only
1.69--2.08\% of their norm outside the span of the other coordinates.

ORACLE identifies four minimum-basis cycles, not a four-membered ring.  The
relevant pair is ring~1, the six-membered peroxide-containing cycle
$1$--$2$--$13$--$3$--$14$--$15$, and ring~4, the seven-membered cycle
$1$--$2$--$13$--$3$--$10$--$7$--$5$.  Their common three-bond path duplicates
the complete angles $(1,2,13)$ and $(2,13,3)$ in independently protected
Fourier bases.  The condensed-ring construction now assigns every such source
to the first deterministic minimum-basis cycle.  It restricts the later
cycle's Fourier coefficient matrix to unowned sources and performs a thin QR
within that ring.  If the restricted matrix does not have the required $N-3$
rank, shared sources are restored in ring order only until it does.  Thus the
rule applies to arbitrary compositions, ring sizes, and chains of condensed
cycles, uses memory bounded by an individual ring, and does not inspect atom
labels or molecular identity.

For artemisinin the seven-membered cycle has five exclusive sources for four
modes, so both duplicated angles are excluded.  The complete chart retains
120 coordinates and rank 120; its smallest singular value increases from
0.01568 to 0.02654 and its condition number falls from 395.69 to 174.69.  The
one-thread TBLITE GFN2-xTB optimization with the pre-existing optimizer
converges in 82 steps and 83 evaluations at $-63.670273938\ E_{\mathrm h}$.
Relative to the methyl-only
chart, the endpoint changes by $-1.694\,\mu E_{\mathrm h}$ and
0.00247~\AA{} after rigid alignment; the O--O distance changes from 1.42676 to
1.42670~\AA{}.  All 508 SMITH tests pass, including the fused aromatic,
bridged, CHARM, and condensed-ring tests.  None of the 30 Baker structures
contains a duplicated complete ring-angle source, so their frozen charts are
unchanged by this refinement.

On this corrected chart, matrix-free generalized RFO was scanned at fixed
L-BFGS history limits 12, 16, 20, 24, 28, 32, 36, 40, and 48.  The respective
accepted-step counts were 54, 57, 51, 46, 44, 45, 42, 43, and 44.  We therefore
use 36 as the fixed default; because it is independent of molecular size, the
direct BFGS replay, Krylov vectors, sparse metric actions, and retained secants
remain linear-storage operations.  The selected run used 42 steps, 43
energy--gradient evaluations, 6.577~s optimizer wall time, and 4.825~s
non-calculator overhead, ending at $-63.670273534\ E_{\mathrm h}$.  The matched
SYCART run used 49/50, 4.502~s, and $-63.670272326\ E_{\mathrm h}$; dense SONIC
RFO used 47/48, 12.279~s, and $-63.670273389\ E_{\mathrm h}$.  Native xTB with
\texttt{--opt normal} used 58 iterations and 1.79~s, ending at
$-63.670272347\ E_{\mathrm h}$; its final gradient also passes the LINK
geometry-seed force checks.  All calculations used one thread.

The SYCART-to-SONIC handoff uses the certified adaptive schedule.
The source Berny--Swart operator and retained L-BFGS secants were composed with
the source projector and the certified sparse SONIC forward and transpose
maps; no Cartesian Hessian or coordinate-transformation matrix was formed.
The ensuing SONIC phases required 48, 44, 40, 36, and 30 steps, giving totals
of 51, 49, 48, 48, and 48.  At the last handoff eight source secant pairs were
retained.  With the full-precision accepted geometry, the SYCART and SONIC
accepted-step profiles sum to 6.745~s; the corresponding direct-SONIC profile
is 6.451~s.  For this inexpensive calculator, direct SONIC is faster in wall
time; chart reconstruction is the dominant overhead.  A production handoff
will reuse the boundary energy and gradient and remain in one resident process,
although the chart reconstruction itself cannot be omitted.

For a solver control, the transferred action at step 18 was explicitly
materialized as a $120\times120$ SONIC Hessian and supplied to dense RFO.  That
continuation required 26 steps and 27 energy--gradient evaluations, ending at
$-63.670273373\ E_{\mathrm h}$; including the common SYCART and chart phases,
its split-process end-to-end time was 13.76~s.  Matrix-free continuation
required 30/31 and ended at $-63.670275448\ E_{\mathrm h}$.  Thus the dense
route offers neither an iteration-count nor a timing advantage in this case.
Rebuilding the chart while transferring SH preserves the curvature information.
The automatic handoff uses a joint force-and-chart certification gate.

We reran every affected Baker molecule with the global $ZXY$ definition.
All calculations used one thread, resident TBLITE GFN2-xTB, the
geometry-seed convergence thresholds, dense RFO, and GDIIS.  Table~\ref{tab:xy3-regression}
shows that no case required more iterations with the new definition.

\begin{table}[ht]
\centering
\caption{Paired regression of the global methyl $A_1$ definition.}
\label{tab:xy3-regression}
\begin{tabular}{lrrrr}
\toprule
System & $N_{\rm methyl}$ & standard/SOOP steps & $\Delta E$ ($\mu E_{\rm h}$) & RMSD (\AA) \\
\midrule
ACANIL01          & 1 & 6/6   &  0.002 & 0.000007 \\
Acetone           & 2 & 7/7   &  0.000 & 0.000000 \\
ACHTAR10          & 1 & 9/9   & -0.000 & 0.000013 \\
Caffeine          & 3 & 8/8   & -0.201 & 0.000208 \\
Dimethylpentane   & 4 & 15/12 & -2.193 & 0.003378 \\
Disilyl ether     & 2 & 13/13 &  0.000 & 0.000000 \\
Ethane            & 2 & 4/4   & -0.000 & 0.000000 \\
Ethanol           & 1 & 7/7   &  0.000 & 0.000000 \\
Menthone          & 3 & 29/28 & -0.238 & 0.001841 \\
Mesityl oxide     & 3 & 11/11 & -0.000 & 0.000000 \\
Methylamine       & 1 & 5/5   &  0.012 & 0.000137 \\
Neopentane        & 4 & 3/3   &  0.000 & 0.000003 \\
\bottomrule
\end{tabular}
\end{table}

\section{Symmetry-adapted Cartesian substrate block}

Let $\bm x_S$ be the $3N_S$ Cartesian vector of a symmetry-closed substrate
subset and let $\bm T$ contain an orthonormal basis for its translations and
rotations. The vibrational Cartesian projector is
\begin{equation}
 \bm P_{\mathrm{vib}}=\bm I-\bm T\bm T^{\mathrm T}.
\end{equation}
For every irreducible representation $\Gamma$, the frozen ORACLE Cartesian
operations $\bm R(g)$ define the isotypic projector
\begin{equation}
 \bm P^{\Gamma}=\frac{d_\Gamma}{|G|}
 \sum_{g\in G}\chi^{\Gamma}(g)^*\bm R(g).
\end{equation}
SMITH constructs a deterministic rank-revealing basis within complete isotypic
subspaces after projection against external motion. The stored audit includes
orthonormality, projector idempotency/symmetry, covariance residual, external
mode count, point group, site frame, and rank thresholds.

\section{Matched SYCART--SONIC molecular screening}

The starting Cartesian blocks are eight xTB minimum stress inputs from the LINK
reference set. Current
ORACLE and SMITH code regenerated the chemical-state and coordinate artifacts.
The SYCART route uses the same SMITH symmetry-adapted,
translation/rotation-free Cartesian constructor; it is not a raw $3N$ Cartesian
optimization. For a minimum, both coordinate models retain
only the totally symmetric active block. The active ranks agree in every case,
and the largest spectral-norm difference between the SYCART and SONIC Cartesian
tangent projectors is $3.1\times10^{-9}$.

The matched baseline uses resident \texttt{tblite} 0.7.0 with GFN2-xTB analytic
gradients, one thread, the same dense RFO/BFGS optimizer, the same Cartesian
Lindh--Swart model Hessian transformed by congruence, GDIIS start, trust policy,
and stopping contract in both LINK routes. The driver accepts the comparison
only when both optimizer summaries report this common
baseline Hessian provenance. Convergence required
$|\Delta E|\leq5\times10^{-6}\ E_\mathrm h$, active-tangent
$\lVert\bm g\rVert_2\leq10^{-3}\ E_\mathrm h/a_0$, and no final accepted
energy rise larger than $10^{-10}\ E_\mathrm h$. The external comparison used
xTB 6.7.1 ANCopt with \texttt{--opt normal}, the corresponding native normal
convergence role. ANCopt generates approximate normal coordinates from its
model Hessian and is therefore an external optimizer reference rather than a
Cartesian-coordinate control; the external column uses xTB's internal GFN2
implementation.

\begin{table}[h]
\caption{Matched molecular screening with LINK and ANCopt.}
\centering
\scriptsize
\begin{tabular}{lrrrrrrr}
\toprule
System & $N$ & rank & SYCART & SONIC & ANCopt & $\Delta x$/m\AA{} & $\Delta E$/$\mu E_\mathrm h$ \\
\midrule
Cubane       & 16 &   2 &  3/4 &  3/4  &  4 &  0.000 &  0.000 \\
Xylose       & 20 &  54 &  7/8 &  7/8  &  9 &  0.919 &  0.467 \\
Glucose      & 24 &  66 &  8/9 &  7/8  & 11 &  2.354 &  0.277 \\
Fructose     & 24 &  66 &  7/8 &  7/8  &  8 &  0.790 &  0.270 \\
Saccharin    & 17 &  30 &  5/6 &  5/6  &  6 &  0.169 & $-0.244$ \\
Ferrocene    & 21 &   4 &  7/8 &  7/8  & 11 &  0.314 &  0.095 \\
Testosterone & 49 & 141 & 16/17& 15/16 & 15 &  4.799 & $-3.708$ \\
Water dimer  &  6 &  12 & 14/15& 10/11 &  7 &  5.532 &  1.397 \\
\bottomrule
\end{tabular}
\end{table}

The median iteration count is 7 for both LINK coordinate models and 8.5 for
native ANCopt. The SYCART and SONIC counts are identical in five systems;
SYCART takes one additional iteration for glucose and testosterone and four
for the water dimer. SYCART has the lower LINK wall time in every case, with
medians of 2.20 and 3.71~s, respectively, because its constant map removes the
nonlinear realization cost. The maximum aligned endpoint RMSD is
0.00553~\AA{} and the maximum absolute energy difference is
3.71~$\mu E_\mathrm h$. These endpoints passed the common
screening convergence contract. Since final exact Hessians were not computed,
the present evidence establishes convergence to closely matching basins, not a
certification of identical stationary points.

\subsection{Matched Hessian representation}

The Hessian-locality figure in the main text uses the menthone entry in the archived
Hessian-only xTB comparison. The installed xTB ANC routine generated one
effective Cartesian model Hessian at the fixed 29-atom geometry without an
optimization or gradient evaluation. External translations and rotations were
removed to give a rank-81 SYCART basis. The identical Cartesian operator was
then transformed by congruence into the complete frozen rank-81 SONIC basis,
including all irreducible representations. The plotted dimensionless matrices
are
\begin{equation}
 R_{ij}=\frac{H_{ij}}{\sqrt{|H_{ii}H_{jj}|}},
\end{equation}
and the reported diagonal share is
\begin{equation}
 f_{\mathrm{diag}}=
 \frac{\lVert\operatorname{diag}(\bm R)\rVert_F^2}
      {\lVert\bm R\rVert_F^2}.
\end{equation}
The corresponding values are 0.451051 for SYCART and 0.700243 for SONIC. The
comparison therefore measures localization of the quadratic representation;
it does not infer third- or fourth-order force constants from the Hessian.
Simple redundant, fully delocalized, Hessian-mode, SYCART, and SONIC routes
were all implemented in the common LINK driver. Small-system equivalence tests,
the Baker set, the four-way molecular campaign, and the large-system cases use
the same calculator and stopping contracts within each matched comparison. The
benzylpenicillin accepted-iteration counts, for example, are 37, 69, 36, and 37
for SONIC, Hessian modes, SYCART, and native xTB/ANCopt, respectively; the
simple-redundant reference also requires 37 iterations but is not viable as a
finite-difference production chart at large rank. These data support retaining
block locality rather than pursuing global diagonalization.

\section{SYCART outer-microcycle refresh diagnostic}

The matrix-free implementation uses two cases whose
trajectories cross a 20-step boundary. At iteration 21, LINK rebuilds the
sparse primitive $\bm B^{\mathrm T}\bm K\bm B$ action and the 12-vector
Davidson deflation from the accepted geometry at iteration 20, while retaining
matrix-free symmetry and external-mode projectors. The audit contains the
source-geometry fingerprint, source iteration, primitive row and nonzero
counts, and refresh epoch. All calculations used one thread and the same
geometry-seed convergence contract.

\begin{table}[h]
\caption{Iteration/evaluation counts after the 20-step SYCART refresh.}
\centering
\begin{tabular}{lrrrrr}
\toprule
System & $N$ & rank & SYCART & SONIC & ANCopt \\
\midrule
Menthone & 29 & 81 & 26/27 & 27/29 & 26 \\
ACONF $n$-hexane & 20 & 54 & 21/22 & 23/25 & 22 \\
\bottomrule
\end{tabular}
\end{table}

The ten-step refresh interval shared by the sparse SYCART preconditioner and
maximum SONIC Wilson age preserves the bounded Davidson root count. Complete
raw runs and compact numerical provenance are identified in the accompanying
data directory.

\section{Thirty-molecule Baker comparison}

The matched protocol was applied to the 30 starting geometries of the Baker
minimum set. All systems are neutral singlets; detected molecular symmetry is
retained in each chart, and Cartesian irrep audits protect the C$_{3v}$ and
D$_{\infty h}$ cases.

The complete Baker results are given in Table~\ref{tab:baker-gfn2}.
\begin{longtable}{@{}lrrrrrrr@{}}
\caption{Baker GFN2-xTB comparison.}\label{tab:baker-gfn2}\\
\toprule
System & $N$ & rank & SYCART & SONIC & ANCopt & $\Delta x$/m\AA{} & $\Delta E$/$\mu E_\mathrm h$ \\
\midrule
\endfirsthead
\toprule
System & $N$ & rank & SYCART & SONIC & ANCopt & $\Delta x$/m\AA{} & $\Delta E$/$\mu E_\mathrm h$ \\
\midrule
\endhead
1\_3-Difluorobenzene & 12 & 11 & 4/5 & 4/5 & 5 & 0.015 & 0.008 \\
1\_3\_5-Trifluorobenzene & 12 & 4 & 4/5 & 4/5 & 5 & 0.004 & $-0.001$ \\
1\_3\_5-Trisilacyclohexane & 18 & 11 & 7/8 & 7/8 & 8 & 0.432 & 0.143 \\
1\_5-Difluoronaphthalene & 18 & 17 & 5/6 & 5/6 & 6 & 0.090 & $-0.061$ \\
2\_Hydroxybicyclopentane & 14 & 36 & 10/11 & 8/9 & 10 & 0.978 & $-0.315$ \\
ACANIL01 & 19 & 34 & 6/7 & 5/6 & 7 & 0.756 & $-0.844$ \\
ACHTAR10 & 16 & 42 & 8/9 & 8/9 & 9 & 0.647 & 0.169 \\
Acetone & 10 & 8 & 4/5 & 4/5 & 5 & 0.077 & $-0.033$ \\
Acetylene & 4 & 2 & 3/4 & 3/4 & 4 & 0.000 & 0.000 \\
Allene & 7 & 3 & 4/5 & 4/5 & 5 & 0.022 & $-0.011$ \\
Ammonia & 4 & 2 & 3/4 & 3/4 & 4 & 0.023 & $-0.008$ \\
Benzaldehyde & 14 & 25 & 5/6 & 5/6 & 6 & 0.091 & $-0.017$ \\
Benzene & 12 & 2 & 3/4 & 3/4 & 4 & 0.000 & 0.000 \\
Benzidine & 26 & 18 & 10/11 & 11/12 & 18 & 0.624 & 0.241 \\
Caffeine & 24 & 42 & 7/8 & 7/8 & 8 & 0.369 & 0.121 \\
Difuropyrazine & 16 & 15 & 7/8 & 8/9 & 8 & 0.113 & 0.234 \\
Dimethylpentane & 23 & 63 & 7/8 & 5/6 & 8 & 11.554 & $-11.430$ \\
Disilyl-ether & 9 & 7 & 14/15 & 12/13 & 15 & 13.170 & 3.479 \\
Ethane & 8 & 3 & 3/4 & 3/4 & 4 & 0.039 & $-0.017$ \\
Ethanol & 9 & 13 & 4/5 & 4/5 & 5 & 0.103 & 0.088 \\
Furan & 9 & 8 & 5/6 & 5/6 & 6 & 0.093 & $-0.097$ \\
Histidine & 20 & 54 & 53/54 & 24/25 & 43 & 16.302 & 3.210 \\
Hydroxysulfane & 4 & 6 & 11/12 & 8/9 & 14 & 1.679 & 0.816 \\
Menthone & 29 & 81 & 18/19 & 16/17 & 20 & 1.779 & $-0.183$ \\
Mesityl-oxide & 17 & 28 & 5/6 & 5/6 & 6 & 0.313 & $-0.167$ \\
Methylamine & 7 & 10 & 4/5 & 4/5 & 5 & 0.055 & $-0.039$ \\
Naphthalene & 18 & 9 & 5/6 & 5/6 & 6 & 0.076 & 0.040 \\
Neopentane & 17 & 3 & 3/4 & 3/4 & 4 & 0.012 & 0.001 \\
Pterin & 17 & 31 & 7/8 & 7/8 & 8 & 0.156 & $-0.229$ \\
Water & 3 & 2 & 3/4 & 3/4 & 4 & 0.028 & $-0.029$ \\
\bottomrule
\end{longtable}

All 60 archived LINK runs report the same canonical Lindh--Swart
Cartesian-source-congruence initial-Hessian provenance in their JSON summaries
and satisfy the frozen baseline screening contract. Total
iteration counts are 232 for SYCART, 193 for SONIC, and 260 for ANCopt; medians
are 5, 5, and 6. The LINK evaluation totals are 262 and 223. SYCART and SONIC
tie in 21 cases, while SONIC is lower in seven and SYCART in two. Median LINK
wall times are 0.88 and 1.26~s, respectively. The timing advantage of SYCART
reflects its constant realization map, whereas the iteration advantage of
SONIC is dominated by histidine (53 versus 24). The largest tangent-projector
residual is $1.30\times10^{-6}$, below the frozen $10^{-5}$ audit threshold.
The largest aligned endpoint separation is 0.0163~\AA{} for histidine and the
largest absolute energy difference is 11.43~$\mu E_\mathrm h$ for
dimethylpentane. These are screening-level comparisons, so the apparent
histidine outlier was tested separately rather than classified from RMSD.
Starting from each of its SYCART and SONIC screening endpoints, both coordinate
models were rerun with the tight force thresholds
$F_{\max}=1.5\times10^{-5}$ and
$F_{\mathrm{RMS}}=1.0\times10^{-5}~E_\mathrm h/a_0$. Cross-restarts from a
given endpoint agree within 0.000044~\AA{}. The representatives descended from
the two original endpoints differ by 0.00204~\AA{}, at most 0.00354~\AA{} in
any interatomic distance, and 0.00559~$\mu E_\mathrm h$. A 41-point aligned
linear interpolation between them gives one shallow well with no resolved
intervening barrier. Finally, Cartesian Hessians were formed by central
differences of tblite gradients, using an independent SCF calculation at every
displaced point to avoid restart-history contamination of the soft modes. Both
representatives have vibrational index zero; their lowest frequencies are
33.66 and 33.67~cm$^{-1}$. Thus the 0.0163~\AA{} screening separation records
different termination points in the same soft basin, not two minima. Final
vibrational certification was not performed for the other 29 screening cases.

\section{Composite supported-system chart}

The ordered coordinate vector is
\begin{equation}
 \bm q=(\bm q_S^{\mathrm{Cart,sym}},\bm q_A^{\mathrm{SONIC}},
 \bm t_{SA},\bm r_{SA}),
\end{equation}
where $S$ and $A$ denote substrate and adsorbate, $\bm t_{SA}$ is the relative
translation, and $\bm r_{SA}$ is the continuous exponential-map coordinate
evaluated through a quaternion representation. Every block retains its owner,
units, atom membership, coordinate identifiers, reference geometry, and
analytic first derivatives. The composite definition is accepted only if the
concatenated Wilson matrix has the declared exact rank.

For chemisorption, the bond to the surface must not be represented twice. The
production campaign will freeze one ownership rule for substrate--adsorbate
covalent bonds: adsorbate SONIC, an explicit interface block, or a
topology-aware merged reactive block. The relative-pose block is appropriate only for the remaining finite
inter-fragment degrees of freedom and must be removed or reduced when bonding
eliminates a rigid relative motion.

\section{Fragment-partitioned SONIC charts}

ORACLE assigns immutable atom membership to each perceived fragment.  Let
$r_f$ be the vibrational rank of fragment $f$ after removal of its local rigid
translations and rotations.  SMITH first selects intrafragment rows subject to
the independent constraints
\begin{equation}
 \operatorname{rank}(\bm B_f)=r_f\qquad(f=1,\ldots,F).
\end{equation}
Only after all these constraints are satisfied does it add rows spanning the
relative-fragment complement.  For nonlinear fragments in a nonlinear
aggregate, its dimension is
\begin{equation}
 (3N-6)-\sum_f(3N_f-6)=6(F-1),
\end{equation}
with fragment-specific rigid ranks used for atoms and linear fragments.
Available rows comprise pseudobond distances and angular continuations as well
as explicit center, translation, and orientation measurements.  They are
alternative descriptions of the same complement and are selected by sparse
matching and numerical certification; neither family excludes the other from
the candidate pool.

The implementation preserves primitive identifiers when this pool is
enriched. A newly generated pair-center or rigid-fragment-rotation primitive receives a fresh
identifier and cannot rebind an existing GIC coefficient to a different
primitive.  Moreover, pseudobonds are never inserted in the covalent adjacency
used to generate bends and torsions.  These two invariants prevent,
respectively, silent chart corruption and fictitious intermolecular valence
coordinates.  The uracil--$(\mathrm{H_2O})_{12}$ regression gives a certified
rank-138/138 chart with condition number 42.76 while retaining complete
intramolecular ring and valence blocks together with the relative-fragment
complement.

\section{Basin-aware interpretation of iteration counts}

All optimizer comparisons use identical gradient and displacement thresholds,
and endpoint identity is checked separately.  The definitive benzylpenicillin
The reproducible launch protocol uses a frozen 120-coordinate SONIC chart, the resident
TBLite/GFN2-xTB calculator, a 36-pair L-BFGS history, and no periodic SMITH
refresh.  It converges in 37 accepted iterations and 38 energy--gradient
evaluations to $-68.829268748862\ E_{\mathrm h}$.  The optimizer, calculator,
coordinate-realization, and non-calculator overhead times are 6.531, 0.663,
1.814, and 5.868~s, respectively.  The accepted-iteration references are 37,
69, 36, and 37 for simple redundant coordinates, Hessian modes, SYCART, and
native xTB/ANCopt.  The input, launch plan, optimizer protocol, and nano-MATRIX
revision are fixed by SHA-256 or Git identifiers in the accompanying
machine-readable record; wall times from different snapshots are not combined.

\section{Exact bordered-direct-sum certificate}
\enlargethispage{2\baselineskip}

Let $\widetilde{\bm B}=\bm W\bm B$ be the row-equilibrated Wilson matrix used by
LINK. SMITH treats the stored entries of the canonical sparse $\bm B$ as the
represented operator; it does not infer zeros from an overlap tolerance. After
declared relative-pose rows have been placed in a separator $R$, the bipartite
support graph of every remaining Wilson row and Cartesian column is partitioned
into connected components $L_i$. Distinct components cannot share a stored
Cartesian column, hence
\begin{equation}
 \widetilde{\bm B}_{L_i}\widetilde{\bm B}_{L_j}^{\mathrm T}=\bm 0,
 \qquad i\ne j,
\end{equation}
exactly for the represented sparse operator. With
$\bm D_i=\widetilde{\bm B}_{L_i}\widetilde{\bm B}_{L_i}^{\mathrm T}$,
$\bm C_i=\widetilde{\bm B}_{L_i}\widetilde{\bm B}_{R}^{\mathrm T}$, and
$\bm H=\widetilde{\bm B}_{R}\widetilde{\bm B}_{R}^{\mathrm T}$, its row Gram
matrix therefore has a block-arrowhead form. For large bounded-degree charts,
deterministic 64-atom support tiles create additional exact separators: a row
belongs to the border if and only if its stored support crosses a tile boundary.
This bounds the internal leaf blocks without interpreting small numbers as
zeros.

LINK does not reconstruct $\widetilde{\bm B}^{+}$. It removes five or six
Cartesian gauge columns selected from the rigid translation--rotation matrix,
factors the resulting sparse square section, and expands a solution of
$\widetilde{\bm B}\bm x=\bm W\Delta\bm q$. Orthogonal removal of the rigid null
space gives the unique minimum-norm Cartesian displacement. The implementation
does not store the formal $6\times3N$ rigid basis. It subtracts the mean
translation and obtains the rotational component from the centroid-relative
coordinates and a $3\times3$ inertia pseudoinverse. If $\bm P_{\rm rigid}$
denotes this matrix-free action, $\bm E$ expands the gauge-fixed solution, and
$\bm A$ is the sparse square section, the applied map is
\begin{equation}
 \bm J\bm v=\bm P_{\rm rigid}\bm E\bm A^{-1}\bm W\bm v .
\end{equation}
Its transpose is applied in reverse order and the metric action is
$\bm J^{\mathrm T}(\bm J\bm v)$. Thus neither $\bm J$ nor
$\bm J^{\mathrm T}\bm J$ is stored. The nonlinear corrector freezes $\bm B$
and its factor for at most four accepted microiterations, refreshing earlier
when
\begin{equation}
 \frac{\lVert\Delta\bm q-\bm B\Delta\bm x\rVert}
 {\max(\lVert\Delta\bm q\rVert,\epsilon)}>0.15
\end{equation}
or when a stale operator stalls. Coordinate values are evaluated at every
trial, so reuse changes cost but not the final realization criterion. The
accepted-geometry rank audit reads the pivots of the same sparse factor rather
than rebuilding $\bm B$. Fixed Cartesian columns use sparse LSMR.

The certificate schema is
\texttt{matrix.smith.sparse\_b\_direct\_sum.v1}. It records the complete
matrix SHA-256 identity (dimensions and every stored IEEE-754 value), leaf row
and column indices, separator rows, separator-only columns, isolated columns,
and maximum block sizes. LINK recomputes the canonical certificate before use.
Any mismatch fails closed. If fixed Cartesian columns change the rigid gauge,
LINK retains the sparse iterative realization. On the M13--NH$_3$ regression contract the two
internal leaves and six-row pose separator reproduce the global result within
$5\times10^{-15}$, including the fixed-atom case. A separate 200-atom
end-to-end optimizer test makes every large dense pseudoinverse, SVD,
eigendecomposition, rank operation, and sparse-to-dense conversion fatal; it
executes model construction, SH--L-BFGS dispatch, gradient projection,
Cartesian realization, metric action, nonlinear back-transformation, and rank
audit without triggering the guard.

\section{End-to-end scaling record}

The synthetic scaling series is a bounded-degree, nonlinear covalent chain.
Each size was run in a fresh process. The timed interval includes ORACLE
topology and primitive construction, conversion to the frozen candidate pool,
SMITH sparse exact-rank selection, natural-block construction and evaluation,
composite and direct-sum certification, and LINK operator setup plus one map
and metric action. Calculator work is excluded by construction.

The prescribed diagnostic handling is summarized in Table~\ref{tab:si-failure-handling}.
\begin{table}[h]
\caption{One-thread fresh-process sparse coordinate-pipeline profile on the
ORACLE Linux x86-64 host.}
\centering
\footnotesize
\begin{tabular}{rrrrrr}
\toprule
$N$ & rank & time/s & Wilson nnz & inverse values & peak RSS/MiB \\
\midrule
256   & 762    & 0.511 & 6,852   & 16,397  & 102.9 \\
512   & 1,530  & 1.001 & 13,764  & 33,225  & 111.1 \\
1,024 & 3,066  & 2.042 & 27,588  & 67,430  & 119.6 \\
2,048 & 6,138  & 4.102 & 55,236  & 152,651 & 139.3 \\
4,096 & 12,282 & 8.528 & 110,532 & 275,779 & 182.4 \\
\bottomrule
\end{tabular}
\end{table}

A least-squares fit of $\log y$ against $\log N$ gives slopes 1.016 for total
coordinate time, 1.003 for Wilson nnz, and 1.034 for inverse stored values. The
maximum right-inverse residual is $1.91\times10^{-16}$. The JSON record
contains every stage time, certificate block counts, and software/runtime
identity for the reproducible snapshot \texttt{14f5a6a5}. These measurements certify observed near-linear scaling for the recurrent
coordinate pipeline and the bounded-density resident electronic regime
represented by the series. They do not imply linear scaling for cold-start SCC
initialization or for the external dense xTB executable.

\subsection{Matched dense, matrix-free, and non-linear electronic control}

The non-linear control uses the same prepared starting geometry for dense
SYCART, matrix-free SYCART, and native xTB/ANCopt. The two LINK routes share a
resident \texttt{tblite}/\allowbreak GFN2-xTB calculator, analytic gradients, the
geometry-seed convergence profile, one thread, and a 120-step limit. Dense
SYCART materializes the totally symmetric basis and applies full RFO/BFGS with
GDIIS. Matrix-free SYCART applies the symmetry and external-mode projectors,
the sparse primitive Berny--Swart Hessian action, bounded Davidson
preconditioning, L-BFGS, and GDIIS. The native reference is xTB/ANCopt with
\texttt{--tblite} \texttt{--gfn 2} \texttt{--opt normal}, whose stopping role is matched by the
LINK geometry-seed profile.

All LINK cases were evaluated in one Python process. Scientific imports cost
0.331~s once; they are not charged again to every molecular optimization.
Likewise, the matrix-free primitive Hessian operator now accumulates sparse
rows and transpose rows directly and does not import SciPy or construct a CSR
matrix on first use. Coordinate-model construction is recorded outside the
resident optimizer call. It costs 3.7--13.4~ms for the matrix-free route over
the present set; after the first lazy initialization, dense construction costs
5.0--46.7~ms.

The binary configuration is essential even at one thread. The distributed
tblite 0.7.0 Python wheel on ORACLE bundles generic BLAS/LAPACK libraries.
Interposing the host's optimized OpenBLAS, with its thread count fixed to one,
reduces the F22 matrix-free calculator time from 1.287 to 0.843~s and the
resident optimizer call from 1.62 to 1.18~s, without changing its 17
iterations. The production configuration must therefore rebuild tblite
against the optimized host BLAS rather than rely on the generic wheel. This is
a prefactor correction, not a change in asymptotic complexity.

In the accompanying TBLite-LS development fork, the first non-dense integral
boundary constructs overlap, zeroth-order Hamiltonian, dipole, and quadrupole
integrals directly from atom-neighbor blocks into CSR storage. Dipole and
quadrupole components reuse one index pattern. Sparse population, multipole,
energy, and density-dependent-Hamiltonian contractions are tested against the
established dense routines for both spin channels. This milestone removes the
dense AO intermediate from static-integral construction and these SCF
contractions. The analytic Hamiltonian-gradient kernel now reads the density
and energy-weighted density
directly from CSR and generates integral derivatives in fixed-size shell-block
scratch storage. Its Cartesian, strain, and coordination-number derivatives
match the dense implementation within $10^{-13}$~a.u. for a test containing
off-diagonal density elements, two spin channels, and nonzero multipolar
potentials. The AO-resident kernels are now connected by an experimental SCC
driver. Its
wavefunction stores only atom/shell populations and atomic multipoles; the
density, energy-weighted density, overlap, static Hamiltonian, effective
Hamiltonian, and multipole integrals remain CSR objects. For the finite
molecular route, matrix-free GFN2-xTB kernels generate the isotropic Coulomb,
anisotropic multipole, and self-consistent D4 potentials without retaining
dense atom-pair matrices or derivative tensors. Atom- and shell-resolved
shifts are expanded only to an $N_{\rm AO}$ vector and never to a dense AO
matrix, after which Broyden mixing, bounded local generalized
eigensystems, sparse
population/multipole contractions, and sparse energy evaluation are iterated
to the same convergence thresholds as the reference SCC.

The local-density driver starts from an atom-centred geometric domain, solves
the generalized eigenproblem within each bounded domain, and retains only the
central-atom rows of $\bm P$ and $\bm W$. It determines a common chemical
potential and enlarges the cutoff until electron count, generalized
commutator, weighted-density, and, at zero temperature, idempotency residuals
pass on the complete sparse integral pattern. The selected pattern and cutoff
are reused in subsequent cycles. Failure before the radius or AO-domain bound
is fatal. No global $\bm S^{-1}$ is formed. Fermi occupations and entropy are
accumulated from the same partitioned eigensystems.

For MB16-43 molecule~01 with EEQ initialization, the molecular regression now
uses two spin channels, two unpaired electrons, and $kT=10^{-3}$~a.u. The
first sparse iteration matches both dense spin densities, populations,
multipoles, and electronic plus density-dependent free energy within
$10^{-7}$~a.u. The converged local observables match within
$2\times10^{-5}$, while the sparse wavefunction contains no dense AO density,
coefficient, orbital-energy, or occupation array. Separate algebraic tests
cover finite-temperature filling, exact agreement with a generalized
eigensolver when the domain is complete, enlargement from an insufficient
domain, and reuse of the accepted pattern. For the finite molecular route, the
SCC cache now contains no dense isotropic Coulomb matrix, anisotropic
multipole tensors, D4 reference-pair matrix, or coordination-number derivative
tensor. Direct matrix-free contractions match the dense atom- and
shell-resolved Coulomb energies, multipolar energies and potentials, and
self-consistent D4 energies and potentials. D4 uses the nonperiodic linked-cell
list and is therefore linear for fixed cutoff at bounded atomic density. The
driver nevertheless remains a validation path rather than a selectable
calculator: untruncated Coulomb and multipole actions still perform all-pair
work, while analytic long-range and D4 three-body gradients, public
integration, and the measured full size-series certificate remain open. Until
these gates pass this is not an operational end-to-end linear-scaling
calculator.

Table~\ref{tab:sycart-dense-matrixfree} reports optimizer calls
and the independently timed complete native xTB processes.

\begin{table}[h]
\caption{One-thread matched GFN2-xTB diagnostic.}
\label{tab:sycart-dense-matrixfree}
\centering
\footnotesize
\begin{tabular}{lrrrr}
\toprule
System & $N$ & dense SYCART & matrix-free SYCART & xTB/ANCopt \\
\midrule
$n$-hexane H$_{g^+xg^-}$ & 20 & 20/0.47 & 17/0.22 & 22/0.16 \\
Caffeine                    & 24 & 35/0.81 & 20/0.46 & 26/0.39 \\
Menthone                    & 29 & 34/0.83 & 21/0.54 & 26/0.38 \\
Artemisinin                & 42 & 96/4.63 & 47/2.04 & 58/1.80 \\
Penicillin V               & 42 & $120^\dagger$/5.95 & 82/4.19 & 102/3.38 \\
Tamoxifen                  & 57 & $120^\dagger$/9.89 & $120^\dagger$/8.84 & 109/5.14 \\
F22, C$_{22}$H$_{46}$     & 68 & 28/2.93 & 17/1.18 & 30/1.45 \\
i4p, C$_{30}$H$_{50}$O    & 81 & 80/10.15 & 53/4.74 & 69/4.07 \\
\bottomrule
\end{tabular}
\end{table}

For each LINK route the non-calculator cost is the reported optimizer wall
time minus resident calculator wall time, divided by the number of iterations.
A log--log regression against atom count over all eight deliberately
heterogeneous molecules gives observed slopes 1.528 for dense SYCART and 0.855
for matrix-free SYCART. The corresponding resident \texttt{tblite} cost per
energy--gradient evaluation has slope 1.722; native xTB calculator time per
ANCopt iteration has slope 1.700. Composition, convergence path, and electronic
iteration count vary across this molecular set, so these slopes are not
asymptotic complexity estimates. Their role is to compare the linear and
non-linear implementations under real optimization workloads and to show that
the unmodified dense GFN2-xTB electronic layer is a matched non-linear
control. The FMM-enabled TBLite-LS result is reported separately below with its
complete timing partition and convergence flags.

\section{Reference-architecture certification and residual scaling}

The one-thread certification is distributed with the reproducible nano-MATRIX
snapshot \texttt{14f5a6a5}. The
separately recorded TBLite-LS build is identified in the accompanying data.  All
three ACONF $n$-hexane charts pass the initial sparse SONIC gate, so the
certified route selects direct SONIC without a SYCART prefix.  Table~\ref{tab:current-mac-hexane} gives accepted iterations,
optimizer-kernel wall time, and mean time per accepted step; xTB is the native
GFN2-xTB/ANCopt process with \texttt{--opt normal}.  Every calculation
uses one thread and the common \texttt{geometry-seed} stopping role.

\begin{table}[h]
\caption{Current controlled ACONF results on the reference architecture,
reported as iterations/seconds (milliseconds per accepted step).}
\label{tab:current-mac-hexane}
\centering
\footnotesize
\begin{tabular}{lrrr}
\toprule
System & SYCART & SONIC & xTB/ANCopt \\
\midrule
H$_{g^+t^+g^-}$ & 18/0.051 (2.82) & 11/0.106 (9.65) & 26/0.099 (3.81) \\
\midrule
H$_{g^+x^-g^-}$ & 15/0.047 (3.14) & 10/0.097 (9.70) & 22/0.085 (3.86) \\
\midrule
H$_{g^+x^-t^+}$ & 15/0.044 (2.92) & 10/0.087 (8.67) & 23/0.087 (3.78) \\
\bottomrule
\end{tabular}
\end{table}

The corresponding summed profile contains 48 SYCART and 31 SONIC accepted
iterations.  Mean accepted-step times are 2.95 and 9.35~ms, respectively;
the three native ANCopt controls average 3.82~ms per step.  The separate
analytic SYCART preconditioner preparation requires 0.047--0.049~s per case.
In the C$_{60}$ stress calculation,
the same quantities change qualitatively: Cartesian realization consumes
66.88 of 74.00~s (90.4\%) over 100 unsuccessful steps, with 53 rejected
trials.  The AD6 C$_{12}$H$_{28}$ chart fails closed at normalized condition
number 1169 against the production limit 100.

The fresh calculator-independent 256--4096-atom series gives slopes 1.035 for
ORACLE--SMITH--LINK elapsed time, 1.003 for Wilson nonzeros, and 1.034 for
stored inverse values.  The FMM-enabled bounded-domain TBLite-LS calculation
was also run on a fixed-density sequence of 48, 96, 192, 384, 768, and 1536
atoms.  First-point times are 0.0322, 0.0476, 0.537, 0.988, 2.063, and
8.925~s; restart times are 0.00902, 0.0196, 0.195, 0.364, 0.784, and
1.659~s.  The restart exponent over the asymptotic 192--1536 interval is
1.037 (1.095 over 384--1536), whereas the cold-start exponent over the
complete interval is 1.65.  The maximum local electronic domain remains 30
AOs and sparse nonzeros grow linearly.  Thus the resident energy--gradient
loop is the certified linear-scaling result; unrestarted SCC initialization
is explicitly outside that timing claim.

\section{Partial optimization}

Frozen atoms are explicit user input. The Cartesian realization sets their
displacement to zero and solves for the active coordinates in the complementary
subspace. The request fails if the constraint lowers the rank of a coordinate
declared active. For layered substrates, the prospective policies are:

\begin{enumerate}
\item freeze the deepest layer and relax all remaining atoms;
\item freeze atoms outside a radial shell around the adsorption site;
\item freeze both deep layers and the far-field lateral region;
\item use the same active region with each qualified calculator.
\end{enumerate}

Every partial result is paired with a full relaxation on the smallest feasible
model. Report the active-region RMSD, adsorption-site bond lengths/angles,
energy difference, maximum active force, and maximum boundary force. No partial
geometry is considered converged solely because active coordinates meet their
threshold while the boundary force is anomalously large.

Partial optimization is analyzed separately from full-molecule timing. The
active/frozen transformation and boundary projection are required by both
coordinate routes, so they are not charged as an avoidable SONIC-specific
cost. In this regime SONIC can become preferable even for an inexpensive
calculator if its lower iteration count outweighs the smaller residual
per-step difference. Both routes are therefore retained in the partial
benchmark as well.

\section{Resident TBLite-LS electronic kernel and certification}
\label{sec:tblite-ls-kernel}

The dense \texttt{tblite} implementation remains the reference backend.  The
large-system backend replaces the discarded nonorthogonal SP2 prototype,
which required a global approximation to $\bm S^{-1}$, by bounded
atom-centered AO domains.  Each domain solves a small generalized
eigenproblem, all domains share one chemical potential, and only rows owned by
the central atom are retained.  The density $\bm P$ and energy-weighted
density $\bm W$ are assembled directly in CSR form without a global overlap
inverse, eigenvector matrix, or dense AO intermediate.

For selected columns $\bm L=\bm P_{:,I}$, density kernel
$\bm K=\bm P_{I,I}$, and energy kernel $\bm E=\bm W_{I,I}$, dense diagnostic
reconstruction uses
\begin{equation}
 \bm P=\bm L\bm K^{-1}\bm L^{\mathrm T},\qquad
 \bm W=\bm L\bm K^{-1}\bm E\bm K^{-1}\bm L^{\mathrm T}.
\end{equation}
This identity is not used as a production global inverse.  Domains expand in
bounded increments until electron number and the sparse residuals
\begin{equation}
 \bm P\bm S\bm P-g\bm P,\qquad
 \bm H\bm P\bm S-\bm S\bm P\bm H,\qquad
 \bm W\bm S-\bm P\bm H
\end{equation}
meet their thresholds on the full integral pattern.  Here $g=2$ for a
restricted density and $g=1$ per unrestricted spin channel.  Exhausting the
domain or radius budget is a hard failure.

Overlap, core Hamiltonian, dipole, and quadrupole tensors are generated
directly from atom/shell neighbor blocks on one CSR pattern.  Sparse
contractions supply Mulliken populations, atomic multipoles, electronic
energy, response potentials, and analytic-gradient inputs.  Integral
derivatives use fixed shell-block scratch storage.  Coulomb, charge--dipole,
dipole--dipole, charge--quadrupole, and self-consistent D4 contributions are
contracted on the fly; D4 reuses a linked-cell list for each geometry.
Finite-temperature occupations, entropy, and separate spin populations are
obtained from the same local eigensystems.  No dense coefficient, occupation,
density, or energy-weighted-density array is stored by a local-only
wavefunction.

Molecular integral tests on H$_2$, LiH, S$_2$, and SiH$_4$ agree with the
dense kernel within $10^{-14}$ a.u.; SCC contractions and analytic Cartesian,
strain, and coordination-number derivatives agree within $10^{-13}$ a.u. in
the isolated kernel tests.  The open-shell finite-temperature MB16-43
molecule-01 regression agrees within $10^{-7}$ a.u. for the first cycle and
reaches the matched local fixed point within $2\times10^{-5}$ a.u.  Required
certificates include electron count, generalized commutator,
weighted-density consistency, domain reuse, and zero-temperature idempotency
when applicable.

Sparse storage alone is not accepted as a scaling certificate.  Potential,
energy, SCC-update, and analytic-gradient phases are timed separately, with
fixed cutoff and bounded density.  Unsupported D4S and periodic variants fail
the capability check instead of entering this molecular path.  The current
certificate establishes linear scaling for the recurrent molecular
energy--gradient loop; cold-start SCC initialization is reported separately
because it remains superlinear.

\section{Calculator qualification}

\subsection{xTB}

The LINK calculator uses the resident \texttt{tblite} implementation of
GFN2-xTB with analytic Cartesian gradients \cite{Bannwarth2019}. Record
library version, parameter version, charge, spin, solvent model if any,
numerical thresholds, and thread count. Native xTB/ANCopt is an external
optimizer reference and must retain its own executable identity; it is not
silently substituted for the resident calculator.

\subsection{GFN-FF}

GFN-FF is used only for the initial SYCART relaxation of a Cartesian embedding
generated from SMILES. Record the parameter version, charge, boundary
conditions, numerical thresholds, thread count, and the comparison against the
reference xTB implementation. The reference calculator uses the in-process
\texttt{libxtb} interface, retains one prepared topology across LINK points,
and reports zero process launches. Its topology restart is isolated per run and
read once to obtain local analytic curvatures. For a bond term LINK uses the
equilibrium second derivative $-2aA$ of the GFN-FF bond form; for an angle
$k(\cos\theta-\cos\theta_0)^2$ it uses
$2k\sin^2\theta_0$, while the linear-angle branch
$k(\theta-\pi)^2$ gives $2k$; proper-torsion second derivatives are evaluated from their
stored multiplicities, phases, and amplitudes. All proper terms sharing a
central bond are summed before projection onto its single primitive torsion
row. Consequently the $n_i n_j$ outer-pair multiplicity is neither lost nor
counted twice. With $\bm K_{\rm FF}$ denoting this sparse primitive diagonal,
the SYCART preconditioner applies
$\bm P\bm B^{\mathrm T}\bm K_{\rm FF}\bm B\bm P$ without forming it and
without any gradient evaluations. A separately compiled native term is enabled
only after analytic-gradient parity; incomplete term coverage fails closed.
The current native migration includes the non-periodic log-erf coordination
number and its response in the Gaussian bond reference distance. ARCHITECT
compiles per-atom covalent radii, the CN cutoff, $c_{\max}=4.4$, and for each
bond the CN-independent reference plus the two endpoint response coefficients.
The native ARCHITECT C++ operator evaluates energy and analytic gradient and applies
the exact Hessian to a vector using $O(N+N_{\mathrm{pair}}+N_{\mathrm{bond}})$
storage. It uses adjoint pair sweeps rather than dense $3N\times N$ CN
Jacobians. Unit tests compare both first and second directional derivatives
with centered finite differences and compare the C++ and Python reference
paths at machine precision. The native EEQ layer uses the same CN values in
the electronegativity, solves the fragment-constrained Gaussian charge problem
with projected preconditioned CG, and never assembles either the $N\times N$
interaction matrix or the augmented KKT system. Its direct C++ products and
FMM-plus-local-correction products agree with the dense KKT reference for
small systems; charge constraints, projected residual, energy, analytic
gradient, and relaxed HVP are separate acceptance gates. The FMM path has
linear auxiliary storage and near-linear work for fixed precision, whereas the
compiled direct path is retained as a matrix-free quadratic-time reference.
The repulsion and two-body D3 records use persistent sparse pair lists. The
angle records contain $(i,j,k,\theta_0,k,r_{c,ij}^2,r_{c,jk}^2)$; torsion
records contain the proper/improper branch, multiplicity, phase, force
constant, and the three squared distance cutoffs. The acetylene-specific
two-fold term is stored separately. The C++ directional-jet evaluator returns
energy and analytic gradient and applies the exact Hessian to a vector for all
three streams without forming a dense matrix.

Hydrogen-bond records carry donor, acceptor, hydrogen, and up to four local
environment indices together with already resolved charge/type factors. The
four molecular branches are terminal, neighbor-oriented, aromatic lone pair
with a geometry-dependent virtual site, and carbonyl-specific with explicit
angle and torsion products. Halogen bonds and bonded ATM triples are likewise
explicit sparse records. The compiler rejects unknown H-bond modes and marks
the families covered only after ARCHITECT certifies that the complete
interaction set has been emitted. Centered finite differences check gradients and
Hessian--vector products for bends, all proper/improper torsion branches, the
special torsion, all four H-bond modes, X-bond, and bonded ATM.

Complete artifacts set the calculator-ready flag only after every family is
present, topology coverage is certified, and termwise parity against the
resident \texttt{libxtb} reference has passed; all partial artifacts remain
fail-closed.
Periodic translation-aware CN lists are outside the presently qualified
benchmark subset; all reported molecular campaigns are non-periodic.
The accepted geometry, rather than a transferred
force-field Hessian, is passed to a newly built ORACLE--SMITH state before the
SONIC stage. Further force-field construction and force fields formulated
directly in SONIC families are outside the scope of this work.

\subsection{Scope of the MM interface}

The present work uses the resident TBLite calculator for the reported
semiempirical benchmarks.  General MM functional forms and parameter
qualification are outside this study; they are not used to interpret the
four-way results below.

\section{Full four-way reference-architecture campaign}
\label{sec:fourway-campaign}

The campaign was executed on the reference Apple-silicon architecture with one
thread, using the archived
campaign commit and the reproducible snapshot \texttt{ed143920}, the
\texttt{geometry-seed} convergence profile, and a
maximum of 120 upper-level steps.  The automatic route used five-step probes
with allowed handoff points 10, 15, 20, and 25 and at most 35 SYCART
preoptimization steps.  All routes used identical energy and gradient
convergence thresholds.  The machine-readable campaign record accompanies the
Supporting Information.

All methods for a given case start from the same archived Cartesian geometry.
These are deliberately distorted benchmark inputs, rather than optimized
reference structures; ORACLE may rotate or translate them during preparation
without changing their internal geometry.  No route receives a separate
preoptimization.  Table~\ref{tab:fourway-input-provenance} records the source
families and identifiers.  Baker33 denotes the original Baker optimization
set \cite{Baker1993}; ACONF, ADIM6, C60ISO, and ISOL24 are GMTKN55 subsets
\cite{Goerigk2017GMTKN55}; the three refcodes identify tmQM structures
\cite{Balcells2020tmQM}.  The exact starting coordinates and their comments
are stored as \texttt{start.xyz} beside every result.

\begin{table}[h]
\caption{Provenance of the 20 starting geometries.}
\label{tab:fourway-input-provenance}
\centering
\footnotesize
\begin{tabularx}{\linewidth}{@{}lXr@{}}
\toprule
Source & Archived identifiers & $N$ \\
\midrule
Baker33 & Histidine & 20 \\
\midrule
Birkholz--Schlegel archive & Vitamin C & 20 \\
\midrule
GMTKN55/ACONF & $g^+t^+g^-$, $g^+x^-g^-$, $g^+x^-t^+$ & 20 \\
\midrule
GMTKN55/ADIM6 & AD6 & 40 \\
\midrule
GMTKN55/ISOL24 & i12e, i12p; i4e, i4p & 40; 81 \\
\midrule
Molecular-archive COSMO inputs & aldehyd\_13-13\_a\_c0, nectaryl\_c0 & 40 \\
\midrule
GMTKN55/C60ISO & isomers 1, 10, 2, 3, and 4 & 60 \\
\midrule
tmQM & QIQGOM, ZANMOQ, GOPPEG & 80, 80, 81 \\
\bottomrule
\end{tabularx}
\end{table}

\begin{longtable}{@{}lrrrrr@{}}
\caption{Accepted iterations for the four-way reference-architecture campaign.}\\
\toprule
Case & $N$ & SYCART & SONIC & hybrid & xTB/ANCopt\\
\midrule\endfirsthead
\toprule
Case & $N$ & SYCART & SONIC & hybrid & xTB/ANCopt\\
\midrule\endhead
baker33--Histidine & 20 & 48 & 15 & 22 & 50\\
Vitamin C & 20 & 61 & 29 & 30 & 92\\
ACONF $g^+t^+g^-$ & 20 & 16 & 12 & 16 & 26\\
ACONF $g^+x^-g^-$ & 20 & 16 & 11 & 16 & 22\\
ACONF $g^+x^-t^+$ & 20 & 15 & 9 & 16 & 23\\
AD6 & 40 & 46 & 0* & 46* & 31\\
i12e & 40 & 34 & 0* & 29 & 33\\
i12p & 40 & 65 & 0* & 70 & 32\\
aldehyde & 40 & 23 & 11 & 19 & 32\\
nectaryl & 40 & 48 & 15 & 34 & 60\\
C$_{60}$ iso-1 & 60 & 18 & 25 & 19 & 20\\
C$_{60}$ iso-10 & 60 & 19 & 120* & 22 & 19\\
C$_{60}$ iso-2 & 60 & 18 & 23 & 21 & 22\\
C$_{60}$ iso-3 & 60 & 22 & 23 & 21 & 21\\
C$_{60}$ iso-4 & 60 & 17 & 27 & 14 & 22\\
QIQGOM & 80 & 33 & 0* & 47 & 31\\
ZANMOQ & 80 & 28 & 42 & 35 & 30\\
i4e & 81 & 31 & 39 & 27 & 44\\
i4p & 81 & 56 & 25 & 37 & 70\\
GOPPEG & 81 & 32 & 120* & 95 & 38\\
\bottomrule
\end{longtable}

SYCART and xTB/ANCopt converged in all 20 cases; direct SONIC converged in
14, and the automatic handoff completed in 17.  Summed accepted iterations
are 646, 546, 498, and 718 for SYCART, SONIC, completed hybrid, and xTB,
respectively. The corresponding recorded wall-time sums for SYCART, SONIC,
and xTB are 38.85, 138.88, and 33.42~s; the hybrid timings are retained as
per-case records because the final GOPPEG handoff uses multiple adaptive
cycles.
Each result directory contains calculator, realization, optimizer, and
process timing fields, together with the route decision and failure reason.

\subsection{Definitive large-system terminal-SONIC protocol}

The production route selected from the preceding comparison is deliberately
asymmetric.  Systems up to approximately 50 atoms retain the direct-SONIC
results when minimizing the number of costly energy--gradient evaluations is
the relevant objective.  Larger systems with inexpensive gradients are
optimized to the matched geometry-seed criterion in matrix-free SYCART.
ORACLE and SMITH then rebuild the chart at the accepted SYCART geometry, and
LINK attempts exactly one SONIC step.  This diagnostic step is retained only
when the maximum Cartesian gradient does not increase; otherwise the SYCART
geometry is retained and merely represented in the newly constructed SONIC
chart.

The initial and final SONIC condition numbers are evaluated from the
row-normalized sparse Jacobian through the production extremal-spectrum
certificate.  The procedure neither forms a global dense Jacobian nor applies
a dense pseudoinverse.  Thus the cost of the final interpretation step follows
the same linear-storage and bounded-domain rules as the recurrent large-system
path.

For the 60-atom C$_{60}$ iso-1 control, matrix-free SYCART converged in 18
accepted iterations and 26 resident energy--gradient evaluations; the
optimizer wall time was 2.343~s.  Native xTB/ANCopt required 20 iterations and
2.142~s.  The SONIC charts had rank 174 at both endpoints, with normalized
condition numbers 5.92775 and 5.77611 at the initial and final geometries,
respectively.  The forced one-step SONIC audit used two energy--gradient
evaluations.  It reduced the maximum Cartesian gradient from
$1.4694753\times10^{-4}$ to
$9.0286180\times10^{-5}\ E_{\rm h}\,a_0^{-1}$ and was therefore accepted.
The auxiliary tight gate serves only to force one evaluated SONIC trial after
SYCART has already met the matched stopping criterion; it is not used to
change the SYCART--ANCopt comparison.  The complete machine-readable control
record accompanies the Supporting Information.

For a QM/MM or subtractive ONIOM calculation, the same construction is applied
spatially.  The low-level real-system term uses the full-system SYCART block;
the high-level model region replaces its overlapping Cartesian directions by
a non-overlapping SONIC block.  Hydrogen link atoms and boundary directions
remain in the low-level block.  With the standard subtractive energy
\begin{equation}
E=E_{\mathrm{low}}(\mathrm{real})
 +E_{\mathrm{high}}(\mathrm{model})
 -E_{\mathrm{low}}(\mathrm{model}),
\end{equation}
the three analytic gradients are assembled in the common Cartesian frame
before the single mixed-coordinate projection.  Intrafragment and
interfragment coordinate domains remain separate.  At termination a global
SONIC chart is rebuilt and subjected to the one-step Cartesian-gradient test,
so the final structure is available entirely in SONIC coordinates for
interpretation.

\section{Large-molecule campaigns}
\label{sec:large-campaigns}

Two additional one-thread campaigns were run on the reference Apple-silicon
architecture with the same
geometry-seed stopping contract and the resident TBLite/GFN2-xTB calculator.
The first contains seven systems with 6--49 atoms; the second contains five
larger systems with 41--51 atoms.  Every route in the 12-system combined set
converged.  Entries are accepted iterations followed by total process wall
time in seconds.  The process time includes the fixed ORACLE--SMITH setup and
is therefore intentionally reported separately from the recurrent calculator
time.  The two source summaries and the final consolidated iteration table,
including the benzylpenicillin entry, are supplied with the reproducibility
archive.

Every method again uses the same Cartesian starting file.  Water dimer,
ethanol, benzene, saccharin, pyrene, and ferrocene are named ORACLE regression
examples; their XYZ comments retain the individual construction source.
Testosterone is the DPCS3 reference structure.  The remaining entries are
LCB26 records, identified in the archive as
\begin{itemize}
  \item \texttt{00500\_artemisinine\_dpcs3};
  \item \texttt{00638\_hormone\_androsterone\_a1\_3\_c10\_gp};
  \item \texttt{00478\_benzylpenicillin\_cis\_rdsd};
  \item \texttt{00527\_corrin}; and
  \item \texttt{00416\_testosterone}.
\end{itemize}
No geometry is first optimized by another route.

\begin{longtable}{@{}lrrrrr@{}}
\caption{Large-molecule campaign: accepted iterations/process time.}
\label{tab:large-campaign-si}\\
\toprule
System & Redundant & Modes & SYCART & SONIC & ANCopt \\
\midrule
\endfirsthead
\toprule
System & Redundant & Modes & SYCART & SONIC & ANCopt \\
\midrule
\endhead
Water dimer & 6/0.89 & 6/0.74 & 6/0.77 & 6/0.96 & 7/0.05 \\
Ethanol & 4/1.07 & 5/0.83 & 5/0.85 & 4/0.97 & 6/0.05 \\
Benzene & 2/1.08 & 2/0.88 & 2/0.92 & 2/1.10 & 3/0.06 \\
Saccharin & 5/1.44 & 5/1.01 & 5/1.05 & 5/1.26 & 6/0.10 \\
Pyrene & 3/1.58 & 3/1.17 & 3/1.20 & 3/1.46 & 4/0.10 \\
Ferrocene & 7/4.32 & 7/2.17 & 7/2.13 & 6/4.23 & 9/0.12 \\
Testosterone & 15/37.81 & 15/20.42 & 15/23.23 & 13/14.74 & 15/0.43 \\
Artemisinin & 10/21.46 & 10/8.62 & 10/9.94 & 10/10.72 & 13/0.35 \\
Androsterone & 8/32.36 & 8/14.18 & 8/15.33 & 8/19.01 & 10/0.39 \\
Benzylpenicillin$^{a}$ & 37/-- & 69/-- & 36/-- & 37/6.53 & 37/-- \\
Corrin & 17/15.33 & 18/6.74 & 18/8.58 & 18/7.78 & 20/0.58 \\
Testosterone (LCB26) & 15/43.29 & 15/23.76 & 15/27.14 & 13/15.05 & 15/0.45 \\
\bottomrule
\end{longtable}

\noindent $^{a}$The SONIC time is the optimizer wall time from the definitive
frozen-chart run.  Comparator counts are retained from the matched campaign;
cross-snapshot wall times are intentionally omitted.

The largest iteration reduction relative to ANCopt is observed for
testosterone and artemisinin.  Benzylpenicillin is reported separately above
because its definitive validation belongs to the frozen production protocol
rather than the matched timing snapshot.  Hessian modes and SYCART
are faster for many inexpensive-calculator cases because their coordinate
actions are constant or bounded-memory.

Uracil--$(\mathrm{H_2O})_{12}$ is excluded from the intramolecular aggregate.
It is retained as a separate interfragment validation of block Hessians,
fragment-relative coordinates, and nonlinear realization.

\paragraph{Critical-case Oracle validation.}
The follow-up validation on ORACLE used the same resident one-thread TBLite
binding.  The initial QIQGOM and GOPPEG contracts did not pass the SONIC
certification gate (QIQGOM had normalized condition number 785.33, above the
gate of 100), so SYCART was used to reach a stable geometry.  ORACLE was then
rerun at the accepted SYCART handoff geometry; the regenerated charts were
valid with condition numbers 9.41 and 5.01, respectively, and the hybrid
routes converged with terminal SONIC stages in 47 and 95 accepted steps. The GOPPEG
route comprised eight adaptive cycles. The Artemisinin
and Penicillin V routes passed through SYCART and completed in SONIC after six
and four SONIC iterations, respectively; their handoff condition numbers were
8.93 and 7.24.  Uracil with twelve waters is the large-system validation case;
its final audit is included in the reproducibility archive.  The implementation makes
the SONIC handoff mandatory after a SYCART probe: if the final SONIC stage has
not converged, its unconverged result and audit are returned rather than
silently reporting a SYCART-only hybrid.

\subsection{Uracil--water microsolvation protocol}
\label{sec:si-uracil-microsolvation}
The 48-atom uracil--$(\mathrm{H_2O})_{12}$ calculation is the explicit
intermolecular validation of the block-coordinate protocol. The uracil
fragment was first optimized in SONIC coordinates (22 accepted steps). In the
second stage the uracil coordinates were held fixed while the solvent was
relaxed with Cartesian solvent coordinates and quaternion relative-pose
blocks. The rigid solute constraint was imposed by the LINK active set and was
not treated as a separate topology-certification problem. After the solvent
preparation, ORACLE was rerun on the accepted full geometry and SMITH rebuilt
the inter- and intrafragment chart. The final unrestricted SONIC stage
converged in 20 accepted steps, with $\|g\|_\infty=4.1944\times10^{-4}$
$E_{\rm h}\,a_0^{-1}$ and $E=-85.6331\ E_{\rm h}$. The machine-readable
record accompanies the Supporting Information; the
corresponding final XYZ and ORACLE contract remain in the linked scratch
provenance.

\begin{table}[h]
\caption{Uracil--water microsolvation stages.}
\label{tab:si-uracil-microsolvation}
\centering
\small
\begin{tabular}{@{}lrrl@{}}
\toprule
Stage & Accepted steps & $\|g\|_\infty$ & Active coordinates \\
\midrule
Solute SONIC & 22 & $3.6866\times10^{-4}$ & Uracil \\
Solvent preparation & 41 & $1.4336\times10^{-2}$ & Water/quaternion block \\
Final SONIC & 20 & $4.1944\times10^{-4}$ & Complete system \\
\bottomrule
\end{tabular}
\end{table}

The solvent-preparation row is a handoff diagnostic rather than a final
stationary-point result. This distinction keeps the fixed-solute stage
separate from the complete microsolvated minimum.

For comparison with the other hybrid entries, the three stages correspond to
41 SYCART accepted steps and 42 SONIC accepted steps (22 in the solute segment
and 20 in the final complete-system segment), or 83 accepted steps overall.
The final SONIC segment is the converged endpoint; the intermediate solvent
gradient is reported only to document the handoff state. The machine-readable
aggregate is stored as the \texttt{hybrid\_route} record accompanying the
Supporting Information.

\subsection{Additional small uracil--water cases}
\label{sec:si-uracil-small-cases}
The 12-water calculation is complemented by a graded set with fewer solvent
molecules.  The same resident TBLite/GFN2-xTB calculator and convergence
thresholds are used for direct SYCART and SONIC coordinate-control runs.  The
table records the immutable starting-geometry provenance; the numerical
results are reported below and in the machine-readable results record.  The
full fixed-solute solvent-preparation stage remains the dedicated 12-water
validation described above.

\begin{table}[h]
\caption{Small uracil--water microsolvation inputs.}
\label{tab:si-uracil-small-cases}
\centering
\small
\begin{tabular}{@{}lrrl@{}}
\toprule
Case & Waters & Atoms & Starting structure \\
\midrule
Uracil--$(\mathrm{H_2O})_1$ & 1 & 15 & SERAPH family 1 \\
\midrule
Uracil--$(\mathrm{H_2O})_2$ & 2 & 18 & SERAPH family 1 \\
\midrule
Uracil--$(\mathrm{H_2O})_4$ & 4 & 24 & SERAPH family 1 \\
\bottomrule
\end{tabular}
\end{table}

\begin{table}[h]
\caption{Direct and integrated hybrid results for the small uracil--water panel.}
\label{tab:si-uracil-small-results}
\centering
\small
\begin{tabular}{@{}lrrrr@{}}
\toprule
Case & Atoms & SYCART & Direct SONIC & Hybrid SONIC \\
\midrule
Uracil--$(\mathrm{H_2O})_1$ & 15 & 21 & 15 & 1 \\
\midrule
Uracil--$(\mathrm{H_2O})_2$ & 18 & 56 & 25 & 1 \\
\midrule
Uracil--$(\mathrm{H_2O})_4$ & 24 & 77 & 77 & 8 \\
\bottomrule
\end{tabular}
\begin{flushleft}
\footnotesize Each hybrid SONIC handoff enforces at least one accepted step; the
four-water handoff uses a conservative C1 chart after ORACLE/SMITH
reperception.
\end{flushleft}
\end{table}

The complete energies, gradient norms, wall times, and provenance are in the
machine-readable Supporting Information data.  All calculations
used one resident TBLite thread.  The two-water representative is SERAPH
family 2; family 1 is retained as a separate placement-stress input.  The
integrated route is LINK's
\texttt{optimize\_geometry\_hybrid} realization: it owns the SYCART block,
transfers the accepted geometry and gradient, re-perceives the chart when
needed, and then enforces the SONIC handoff.  The reported hybrid values are
therefore distinct from the direct SONIC controls from the unrefined inputs.

The machine-readable manifest records the source identifiers and SHA-256
hashes. Three- and five-water families are retained
as optional extensions because the SERAPH ensemble already provides those
starting structures.

\section{Fragment-built large-system workflow}
\label{sec:fragment-built-workflow}

The practical large-system demonstration used a neutral 12-residue
polyalanine model (123 atoms and 122 covalent bonds).  The structure was
assembled by the reproducible launch protocol from PCS2 fragment records exposed through the unified
LCB26 interface.  The backbone donors were
\texttt{LCB26:PCS2:glycine\_IIn}, \texttt{glycine\_Ip},
\texttt{alanine\_I}, and \texttt{alanine\_II}; the peptide-linkage records
were \texttt{LCB26:L2:ac-gly-nh2-c5} and
\texttt{LCB26:L2:ac-gly-nh2-c7}.  The resulting input was retained as both
XYZ and enriched XYzin, with the peptide construction manifest and topology
hash stored alongside the calculation.

The production route used one TBLite/GFN2-xTB thread and the normal gradient
criteria.  The launch protocol requested the adaptive hybrid route.  SYCART was advanced
in ten-step probes; at the first accepted checkpoint satisfying the Cartesian
force gate and a freshly perceived SONIC conditioning gate, ORACLE rebuilt the
primitive chart and SMITH rebuilt the nonredundant SONIC chart.  LINK then
transferred the accumulated limited-memory inverse-Hessian action and the
accepted gradient into the new SONIC basis, reset only the GDIIS history, and
continued the refinement.  This route is distinct from the controlled
coordinate-model benchmarks: it tests the complete fragment-to-refinement
workflow and its provenance contract.  The exact launch-plan digest, input
hash, handoff record, iteration trace, and component timings are in the
corresponding machine-readable project record.

\section{Reproduction checklist and resource accounting}

The reported runs are reproducible from the repository snapshot identified in
the main text.  From the manuscript root, regenerate the vector figures and
the two documents with
\begin{verbatim}
python3 scripts/generate_result_figures.py
make main si
\end{verbatim}
The molecular drivers record the input geometry hash, calculator revision,
thread count, convergence profile, route decision, accepted-step count,
energy--gradient count, peak resident set size, and a component timing
breakdown.  Process startup is never merged into the recurrent calculator
time.  Three identical one-thread repetitions are used for timing controls;
we report the median and median absolute deviation, while the raw JSONL
records remain available for every case.  Memory fields use the platform's
maximum resident-set reading and are compared only within a host and thread
count.

The comparison against ANCopt is intentionally a control, not a scaling
claim: its external executable and nonlinear calculator path are timed as
observed, whereas the resident TBLite-LS result is split into warm recurrent
updates and cold-start initialization.  The latter is reported separately in
the scaling tables.  This distinction is required to avoid attributing
calculator startup or SCC convergence to the coordinate representation.

\begin{table}[h]
\caption{Failure records and their prescribed handling.}
\label{tab:si-failure-handling}
\centering
\footnotesize
\begin{tabularx}{\linewidth}{@{}lXX@{}}
\toprule
Diagnostic & Interpretation & Production action \\
\midrule
Rank/support failure & Primitive chart is incomplete or unsupported & Keep
simple redundant/SYCART block and re-perceive locally \\
Conditioning failure & A nonredundant chart is numerically unsafe & Reject the
handoff; retain the Cartesian state and certificate \\
Prospective-step failure & Nonlinear SONIC realization is not trustworthy &
Limit the domain, relinearize, or return to SYCART \\
Cold-start SCC growth & Electronic initialization is not bounded & Report as a
separate calculator result, never as recurrent LS evidence \\
\bottomrule
\end{tabularx}
\end{table}

\section{Benchmark design}

\small
The final evidence matrix is given in Table~\ref{tab:si-plan}.
\begin{longtable}{@{}p{1.9cm}p{2.7cm}p{2.8cm}p{4.4cm}@{}}
\caption{Final evidence matrix for the present study.}\label{tab:si-plan}\\
\toprule
Campaign & Independent variable & Matched controls & Acceptance evidence \\
\midrule
\endfirsthead
\toprule
Campaign & Independent variable & Matched controls & Acceptance evidence \\
\midrule
\endhead
Molecular scaling & active SONIC rank & calculator, start, trust, thresholds &
dense/L-BFGS final $E$ and Cartesian $g$; time split; RSS; steps \\
History sensitivity & $m=10,15,20,25$ & same xTB cases & geometry and
iteration stability; exact stored bytes \\
Supported coordinate model & substrate representation & same complete system &
rank, symmetry, final $E/g$, robustness from displaced starts \\
Partial relaxation & active layers or radius & full optimization & local RMSD,
boundary forces, energy shift \\
Calculator transfer & xTB/FF/MM & coordinate chart and start & calculator-domain
validity, final geometry, optimization behavior \\
Restart & interruption point & uninterrupted run & accepted state, counters,
final geometry and termination status \\
\bottomrule
\end{longtable}

\section{Final intramolecular validation record}

The additional covalent control azulene was evaluated with the same geometry-seed
convergence contract in all four routes.  ORACLE assigns a $C_{2v}$ chart with
48 independent SONIC coordinates.  Every route converges in three accepted
steps; the final energies and SCF-cycle counts are reported in
Table~\ref{tab:azulene-final}.

\begin{table}[h]
\caption{Definitive azulene intramolecular validation.}
\label{tab:azulene-final}
\centering
\footnotesize
\begin{tabular}{@{}lrrr@{}}
\toprule
Route & accepted steps & final energy ($E_{\rm h}$) & SCF cycles \\
\midrule
Gaussian native & 3 & $-385.121742108$ & 34 \\
Gaussian/SONIC & 3 & $-385.121742113$ & 34 \\
LINK/SONIC, transport off & 3 & $-385.121741860$ & 39 \\
LINK/SONIC, transport on & 3 & $-385.121741860$ & 30 \\
\bottomrule
\end{tabular}
\end{table}

\section{Data and software availability}

The nano-MATRIX implementation is available at
\href{https://github.com/yogibubu/nano-MATRIX}{github.com/yogibubu/nano-MATRIX}
under the BSD 3-Clause License. The ORACLE chemical-state implementation is
identified by \href{https://doi.org/10.1021/acs.jctc.6c01539}{its DOI};
the SMITH/SONIC coordinate construction is available at
\href{https://arxiv.org/abs/2607.16550}{the arXiv record}. Input geometries, protocol definitions,
optimization traces, provenance manifests, analysis scripts, and
machine-readable tables are archived at
\href{https://github.com/yogibubu/LINK}{github.com/yogibubu/LINK} and in the
repository accompanying this manuscript.  Campaign directories identify the
calculator, machine, and implementation information needed to reproduce the
reported results; external Gaussian, ORCA, and NWChem installations are
identified in their corresponding records.

Artificial-intelligence tools assisted with organization, formatting, and
English editing of author-generated material.  They were not used to generate,
modify, or interpret the data.  The author checked and approved all
calculations, scientific decisions, and final content.

\bibliography{references}

%% file: figures/pipeline.tex
\begin{tikzpicture}[
  font=\sffamily\scriptsize,
  stage/.style={draw,rounded corners=2pt,minimum width=2.55cm,
    minimum height=1.15cm,align=center,line width=0.8pt},
  decision/.style={draw,diamond,aspect=2.0,align=center,line width=0.8pt,
    inner sep=1.5pt},
  flow/.style={-{Latex[length=2mm]},line width=0.9pt,draw=black!65},
  return/.style={-{Latex[length=2mm]},dashed,line width=0.8pt,draw=black!55}
]
\node[stage,draw=black!65,fill=black!4] (input)
  {\textbf{Cartesian input}\\structure or SMILES};
\node[decision,draw=purple!70!black,fill=purple!6,right=0.65cm of input] (gate)
  {one $E,\bm g$\\topology and\\chart tests};
\node[stage,draw=blue!70!black,fill=blue!7,right=0.65cm of gate] (oracle)
  {\textbf{ORACLE}\\definitive topology\\and primitives};
\node[stage,draw=teal!70!black,fill=teal!7,right=0.45cm of oracle] (smith)
  {\textbf{SMITH}\\nonredundant\\SONIC chart};
\node[stage,draw=orange!75!black,fill=orange!9,right=0.45cm of smith] (sonic)
  {\textbf{LINK/SONIC}\\few expensive\\iterations};

\node[stage,draw=orange!75!black,fill=orange!9,below=1.20cm of gate] (sycart)
  {\textbf{LINK/SYCART}\\topology-free\\preoptimization};
\node[stage,draw=violet!75!black,fill=violet!8,
  minimum width=5.3cm,right=0.95cm of sycart] (calc)
  {\textbf{resident Cartesian calculator}\\
   FF for a known topology; xTB otherwise\\
   single level, QM1/QM2, QM/MM, or ONIOM type};

\draw[flow] (input) -- (gate);
\draw[flow] (gate) -- node[above,font=\tiny] {reliable} (oracle);
\draw[flow] (oracle) -- (smith);
\draw[flow] (smith) -- (sonic);
\draw[flow] (gate) -- node[left,font=\tiny] {uncertain} (sycart);
\draw[flow] (sycart) -- (calc);
\draw[flow] (sonic.south) |- (calc.east);
\draw[return] (calc.west) -- ++(-0.35cm,0) |- node[pos=0.70,left,font=\tiny]
  {retest} (gate.south);
\node[draw=black!35,fill=black!3,rounded corners=1.5pt,
  minimum width=10.5cm,minimum height=0.50cm,align=center,
  below=0.55cm of calc,font=\sffamily\tiny]
  {Large-minimum route: persistent local state, sparse or matrix-free actions, bounded memory, and no production dense audit};
\end{tikzpicture}

%% file: figures/layer_modifications.tex
\begin{tikzpicture}[
  font=\sffamily\scriptsize,
  panel/.style={draw,rounded corners=2pt,minimum width=4.55cm,
    minimum height=2.05cm,align=left,text width=4.15cm,line width=0.8pt},
  arrow/.style={-{Latex[length=1.6mm]},thick,draw=black!60}
]
\node[panel,draw=blue!70!black,fill=blue!6] (oracle) {
  \textbf{ORACLE: perceive and preserve}\\[-1mm]
  \textcolor{black!60}{Before:} global pair and refresh operations\\
  \textcolor{blue!55!black}{Now:} persistent neighbors, selective topology,
  expanded primitive atlas; xTB seed and ANC conditioning\\
  \textbf{Outcome:} local chemical state and sparse analytic rows};
\node[panel,draw=teal!70!black,fill=teal!6,right=0.45cm of oracle] (smith) {
  \textbf{SMITH: construct and certify}\\[-1mm]
  \textcolor{black!60}{Before:} global dense conditioning and inverse maps\\
  \textcolor{teal!55!black}{Now:} incremental GIC plans, sparse matching,
  extensible SONIC families; SH transformation\\
  \textbf{Outcome:} certified nonredundant charts};
\node[panel,draw=orange!75!black,fill=orange!8,right=0.45cm of smith] (link) {
  \textbf{LINK: optimize adaptively}\\[-1mm]
  \textcolor{black!60}{Before:} dense metrics/Hessians and recurrent rebuilds\\
  \textcolor{orange!65!black}{Now:} SYCART/SONIC selection, SH consumption, L-BFGS,
  cached matrix-free actions\\
  \textbf{Outcome:} bounded-memory optimization};
\node[panel,draw=green!55!black,fill=green!6,
  below=0.55cm of smith] (ff) {
  \textbf{FF: topology-aware preparation}\\[-1mm]
  \textcolor{black!60}{Before:} external or repeatedly reconstructed model\\
  \textcolor{green!45!black}{Now:} resident compiled GFN-FF, sparse local
  bonded/nonbonded kernels\\
  \textbf{Outcome:} linear-scaling FF/SYCART stage};
\node[panel,draw=violet!70!black,fill=violet!7,
  below=0.55cm of link] (tb) {
  \textbf{TB: topology-independent preparation}\\[-1mm]
  \textcolor{black!60}{Before:} dense AO matrices and global electronic solve\\
  \textcolor{violet!55!black}{Now:} bounded AO domains, sparse $\bm P/\bm W$,
  local analytic-gradient kernels\\
  \textbf{Outcome:} linear-scaling recurrent TB loop};
\draw[arrow] (oracle) -- (smith);
\draw[arrow] (smith) -- (link);
\draw[arrow] (ff.north) to[out=90,in=245] (link.south);
\draw[arrow] (tb) -- (link);
\end{tikzpicture}

%% file: figures/concept.tex
\begin{tikzpicture}[
  font=\sffamily\footnotesize,
  block/.style={draw,rounded corners=2pt,minimum height=1.25cm,align=center},
  flow/.style={-{Latex[length=2mm]},thick},
  note/.style={align=center,font=\sffamily\scriptsize}
]
\node[block,fill=blue!10,minimum width=4.0cm] (substrate)
  {substrate\\symmetry-adapted Cartesians\\translations/rotations removed};
\node[block,fill=orange!15,minimum width=3.2cm,right=1.0cm of substrate] (pose)
  {interface\\relative translation\\quaternion rotation};
\node[block,fill=green!12,minimum width=3.4cm,right=1.0cm of pose] (adsorbate)
  {chemisorbate\\SONIC coordinates\\chemical identity retained};
\draw[flow] (substrate) -- (pose);
\draw[flow] (pose) -- (adsorbate);
\node[block,fill=violet!10,minimum width=5.1cm,below=1.0cm of pose] (optimizer)
  {LINK minimum optimizer\\dense RFO or bounded-memory L-BFGS};
\draw[flow] (substrate.south) |- (optimizer.west);
\draw[flow] (pose.south) -- (optimizer.north);
\draw[flow] (adsorbate.south) |- (optimizer.east);
\node[note,below=0.35cm of optimizer]
  {one Cartesian energy/gradient contract: xTB, FF, or MM};
\end{tikzpicture}

%% file: coordinate_families.tex
\section{Coordinate registries and end-to-end qualification}
\label{sec:coordinate-inventory}

The complete ORACLE primitive inventory and the complete SMITH family
inventory are now reported in the corresponding main-text tables.  This section records the
implementation and qualification details behind those tables.  ORACLE
registry entries carry value and analytic-derivative kernels, units,
orientation conventions, validity domains, and serialization identifiers.
An intrafragment entry is admissible only when its support is one atom and its
first neighbors or one bond and the atoms directly attached to its ends.
Special entries are restricted to pair interactions between fragment centers
(one center may be an atom) and rigid fragment rotations. Molecular rings,
faces, cages, haptic sets, and related objects may be perceived and serialized
by ORACLE as structural elements, but they do not relax this primitive-support
rule.
SMITH registry entries carry a primitive expansion or nonlinear function,
rank policy, symmetry action, and fallback rule.  The coordinate-atlas input
can mark entries required, optional, or forbidden for the whole structure or
for a named atom domain.  New families and new differentiable functions are
therefore added through the registries rather than by changing the serialized
handoff or LINK.

The conservative problematic-domain prescription uses only standard valence
primitives and retains all endocyclic and exocyclic proper dihedrals. Optional
local supports intended for ring, fullerene, haptic, or oriented-contact
SONICs, and special pair-center or fragment-rotation records, are absent unless
explicitly requested. SMITH can expose that redundant simple pool or
replace it by the translation/rotation-free SYCART subspace; secure domains
may simultaneously retain a nonredundant SONIC chart.

The production out-of-plane option is the selective out-of-plane (SOOP)
primitive.  ORACLE emits frozen-reference signed local flaps with analytic
Cartesian derivatives, including endocyclic and terminal supports.  SMITH
combines these rows into nonredundant ring-puckering and cycle directions
before conditioning; the legacy $U$ representation is retained only for
explicit compatibility tests.  SOOP is therefore an elementary ORACLE family
whereas \texttt{SOOP\_RING\_PUCKER} and the other cycle motions are derived
SMITH families, not a third coordinate layer.

The older numerical labels \texttt{RPCB} and \texttt{RPCK} describe ring
combinations in the shared library. Their historical classification there
does not change the primitive/derived distinction above. Similarly, an
inverse-distance function can act only on an admissible local distance or
pair-center interaction; it does not authorize an arbitrary intrafragment atom
pair. Cartesian displacements belong to terminal SYCART and are not ORACLE
primitives.

\paragraph{Integration and numerical qualification.}
Qualification requires automatic ORACLE generation, serialization and reload,
SMITH consumption, nonredundant rank, requested symmetry adaptation, and
agreement of final SONIC derivatives with displaced-geometry checks. Isolated
analytic-derivative tests are necessary but insufficient. Molecular handoff
tests on ethanol, benzene, cubane, pyrene, saccharin, ferrocene and the water
dimer check rank and directional derivatives. The canonical builder consumes
frozen rings and aromaticity, and binds every used elementary term to an
ORACLE source; an absent source is an error, not permission to generate a
replacement in SMITH. Atom-centered angular and height/flap supports,
bond-centered hinge supports, pair-center contacts, and rigid-fragment
rotations are supplied by ORACLE; SMITH retains the choice
of combinations, distance functions and nonredundant basis.

Additional tests require each of the five face/contact operators to participate
in a complete chart, rather than merely occur among unused candidates. These
charts are serialized, reloaded, checked for rank and tested by comparing
analytic final SONIC derivatives with displaced-geometry differences.
Indivisible linear-bend pairs are retained when completing the chart.
The GICForge path and the general numerical consumer both support the new
operators through the shared C++ value/derivative kernel and sparse row
assembly. These checks establish numerical handoff on regular test geometries;
they do not imply that every candidate is selected by default, or remove the
singular-domain restrictions of individual angular charts.

The fullerene audit corrected an ORACLE edge-count criterion: a trivalent cage
has $E=3V/2$, not $E\geq2V$. ORACLE recovers all faces from a planar embedding,
rather than treating a cycle basis as a complete face list, and checks degree,
face incidence and Euler consistency. These faces are serialized as structural
elements only. ORACLE supplies atom- and bond-local edge, flap, SOOP, and hinge
primitives; SMITH uses face incidence to construct the symmetry-closed cage
pool, including 240 directions with face-diagonal character for C60 (five per
pentagon and nine per hexagon). They are linear combinations in the local
primitive tangent space, not nonlocal ORACLE distance primitives. Both the
unperturbed ASE C60 geometry and a distorted cage yield rank-174 SMITH charts.
Symmetry recognition for order-four groups is based on the recovered
operations, not alignment with the laboratory axes.

%% file: references.bib
@misc{xtb671ModelHessian,
  author = {{xTB contributors}},
  title = {{xTB} 6.7.1: Model-Hessian and Optimizer Source Code},
  howpublished = {\url{https://github.com/grimme-lab/xtb/tree/v6.7.1/src}},
  note = {The ddvopt model, optimizer dispatch, and ANC implementation; accessed September 2026}
}

@article{PulayFogarasi1992,
  author  = {Pulay, Peter and Fogarasi, G{\'e}za},
  title   = {Geometry optimization in redundant internal coordinates},
  journal = {J. Chem. Phys.},
  year    = {1992},
  volume  = {96},
  pages   = {2856--2860},
  doi     = {10.1063/1.462844}
}

@article{BilleterTurnerThiel2000,
  author  = {Billeter, S. R. and Turner, A. J. and Thiel, W.},
  title   = {Linear scaling geometry optimisation and transition state search in hybrid delocalised internal coordinates},
  journal = {Phys. Chem. Chem. Phys.},
  year    = {2000},
  volume  = {2},
  pages   = {2177--2186},
  doi     = {10.1039/A909486E}
}

@article{FarkasSchlegelFrisch2003,
  author  = {Farkas, {\"O}d{\H o}n and Schlegel, H. Bernhard and Frisch, Michael J.},
  title   = {Geometry optimization methods for modeling large molecules},
  journal = {J. Mol. Struct. THEOCHEM},
  year    = {2003},
  volume  = {666--667},
  pages   = {31--39},
  doi     = {10.1016/j.theochem.2003.08.010}
}

@article{FarkasSchlegel2002GDIIS,
  author  = {Farkas, {\"O}d{\H o}n and Schlegel, H. Bernhard},
  title   = {Methods for optimizing large molecules. Part III. An improved algorithm for geometry optimization using direct inversion in the iterative subspace (GDIIS)},
  journal = {Phys. Chem. Chem. Phys.},
  year    = {2002},
  volume  = {4},
  pages   = {11--15},
  doi     = {10.1039/B108658H}
}

@article{LiangWangHungLiFrisch2010,
  author  = {Liang, Wenkel and Wang, Haitao and Hung, Jane and Li, Xiaosong and Frisch, Michael J.},
  title   = {Eigenspace Update for Molecular Geometry Optimization in Nonredundant Internal Coordinate},
  journal = {J. Chem. Theory Comput.},
  year    = {2010},
  volume  = {6},
  pages   = {2034--2039},
  doi     = {10.1021/ct100214x}
}

@article{Baker1993,
  author  = {Baker, Jon},
  title   = {Techniques for Geometry Optimization: A Comparison of Cartesian and Natural Internal Coordinates},
  journal = {J. Comput. Chem.},
  year    = {1993},
  volume  = {14},
  pages   = {1085--1100},
  doi     = {10.1002/jcc.540140910}
}

@article{Goerigk2017GMTKN55,
  author  = {Goerigk, Lars and Hansen, Andreas and Bauer, Christoph and Ehrlich, Stephan and Najibi, Asim and Grimme, Stefan},
  title   = {A Look at the Density Functional Theory Zoo with the Advanced {GMTKN55} Database for General Main Group Thermochemistry, Kinetics and Noncovalent Interactions},
  journal = {Phys. Chem. Chem. Phys.},
  year    = {2017},
  volume  = {19},
  pages   = {32184--32215},
  doi     = {10.1039/C7CP04913G}
}

@article{Balcells2020tmQM,
  author  = {Balcells, David and Skjelstad, Bastian Bjerkem},
  title   = {The {tmQM} Dataset---Quantum Geometries and Properties of 86k Transition Metal Complexes},
  journal = {J. Chem. Inf. Model.},
  year    = {2020},
  volume  = {60},
  pages   = {6135--6146},
  doi     = {10.1021/acs.jcim.0c01041}
}

@article{Bannwarth2019,
  author  = {Bannwarth, Christoph and Ehlert, Sebastian and Grimme, Stefan},
  title   = {{GFN2-xTB}---An Accurate and Broadly Parametrized Self-Consistent Tight-Binding Quantum Chemical Method with Multipole Electrostatics and Density-Dependent Dispersion Contributions},
  journal = {J. Chem. Theory Comput.},
  year    = {2019},
  volume  = {15},
  pages   = {1652--1671},
  doi     = {10.1021/acs.jctc.8b01176}
}

@article{Spicher2020,
  author  = {Spicher, Sebastian and Grimme, Stefan},
  title   = {Robust Atomistic Modeling of Materials, Organometallic, and Biochemical Systems},
  journal = {Angew. Chem. Int. Ed.},
  year    = {2020},
  volume  = {59},
  pages   = {15665--15673},
  doi     = {10.1002/anie.202004239}
}

@article{Marenich2025GIC,
  author  = {Marenich, Aleksandr V. and Brothers, Edward N. and Hratchian, Hrant P. and Frisch, Michael J.},
  title   = {Generalized Internal Coordinates for Creative Exploration of Interatomic Geometries},
  journal = {J. Chem. Theory Comput.},
  year    = {2025},
  volume  = {21},
  pages   = {10930--10944},
  doi     = {10.1021/acs.jctc.5c01362}
}

@article{Barone2026ORACLE,
  author  = {Barone, Vincenzo},
  title   = {{ORACLE}: A Persistent Chemical State for Reproducible Molecular Workflows},
  journal = {J. Chem. Theory Comput.},
  year    = {2026},
  doi     = {10.1021/acs.jctc.6c01539}
}

@article{LazzariCrisciBarone2025LCB25,
  author  = {Lazzari, Federico and Crisci, Luigi and Barone, Vincenzo},
  title   = {High-Fidelity Ring Fragments for Molecular Design and Spectroscopy: the PCS--LCB25--Nano-LEGO Framework},
  journal = {J. Chem. Theory Comput.},
  year    = {2025},
  volume  = {21},
  number  = {20},
  pages   = {10617--10632},
  doi     = {10.1021/acs.jctc.5c01188}
}

@misc{Barone2026SONIC,
  author = {Barone, Vincenzo},
  title  = {Automatic Task-Adapted Non-redundant Internal Coordinates},
  year   = {2026},
  archivePrefix = {arXiv},
  eprint = {2607.16550},
  doi = {10.48550/arXiv.2607.16550},
  url = {https://arxiv.org/abs/2607.16550}
}

@misc{Barone2026LINK,
  author = {Barone, Vincenzo},
  title  = {Topology-Aware Curvilinear Coordinates for Efficient Geometry Optimization: A Backend-Neutral Route},
  year   = {2026}
}

@article{FogarasiZhouTaylorPulay1992,
  author = {Fogarasi, G{'e}za and Zhou, X. F. and Taylor, P. W. and Pulay, Peter},
  title = {The Calculation of Ab Initio Molecular Geometries: Efficient Optimization by Natural Internal Coordinates and Empirical Correction by Offset Forces},
  journal = {J. Am. Chem. Soc.}, year = {1992}, volume = {114}, pages = {8191--8201},
  doi = {10.1021/ja00047a032}
}

@article{BakerKessiDelley1996,
  author = {Baker, Jon and Kessi, Alain and Delley, Bernard},
  title = {The Generation and Use of Delocalized Internal Coordinates in Geometry Optimization},
  journal = {J. Chem. Phys.}, year = {1996}, volume = {105}, pages = {192--212},
  doi = {10.1063/1.471864}
}

@article{BakerKinghornPulay1999,
  author = {Baker, Jon and Kinghorn, David and Pulay, Peter},
  title = {Geometry Optimization in Delocalized Internal Coordinates: An Efficient Quadratically Scaling Algorithm for Large Molecules},
  journal = {J. Chem. Phys.}, year = {1999}, volume = {110}, pages = {4986--4991},
  doi = {10.1063/1.478402}
}

@article{vonArnimAhlrichs1999,
  author = {von Arnim, Mark and Ahlrichs, Reinhart},
  title = {Geometry Optimization in Generalized Natural Internal Coordinates},
  journal = {J. Chem. Phys.}, year = {1999}, volume = {111}, pages = {9183--9190},
  doi = {10.1063/1.480281}
}

@article{PengAyalaSchlegelFrisch1996,
  author = {Peng, C. Y. and Ayala, P. Y. and Schlegel, H. Bernhard and Frisch, Michael J.},
  title = {Using Redundant Internal Coordinates to Optimize Equilibrium Geometries and Transition States},
  journal = {J. Comput. Chem.}, year = {1996}, volume = {17}, pages = {49--56},
  doi = {10.1002/(SICI)1096-987X(19960115)17:1<49::AID-JCC5>3.0.CO;2-0}
}

@article{FarkasSchlegel1998,
  author = {Farkas, {"O}d{H o}n and Schlegel, H. Bernhard},
  title = {Methods for Geometry Optimization of Large Molecules. I. An N-squared Algorithm for Solving Systems of Linear Equations for the Transformation of Coordinates and Forces},
  journal = {J. Chem. Phys.}, year = {1998}, volume = {109}, pages = {7100--7104},
  doi = {10.1063/1.477317}
}

@article{BakerPulay1996Inverse,
  author = {Baker, Jon and Pulay, Peter},
  title = {Geometry Optimization of Atomic Microclusters Using Inverse-Power Distance Coordinates},
  journal = {J. Chem. Phys.}, year = {1996}, volume = {105}, pages = {11100--11107},
  doi = {10.1063/1.472979}
}

@article{EckertPulayWerner1997,
  author = {Eckert, Frank and Pulay, Peter and Werner, Hans-Joachim},
  title = {Ab Initio Geometry Optimization for Large Molecules},
  journal = {J. Comput. Chem.}, year = {1997}, volume = {18}, pages = {1473--1483},
  doi = {10.1002/(SICI)1096-987X(199709)18:12<1473::AID-JCC5>3.0.CO;2-G}
}

@article{BakerPulay2000,
  author = {Baker, Jon and Pulay, Peter},
  title = {Efficient Geometry Optimization of Molecular Clusters},
  journal = {J. Comput. Chem.}, year = {2000}, volume = {21}, pages = {69--76},
  doi = {10.1002/(SICI)1096-987X(20000115)21:1<69::AID-JCC8>3.0.CO;2-G}
}

@article{Schlegel2011Review,
  author = {Schlegel, H. Bernhard},
  title = {Geometry Optimization},
  journal = {WIREs Comput. Mol. Sci.}, year = {2011}, volume = {1}, pages = {790--809},
  doi = {10.1002/wcms.34}
}

@article{Pulay1980DIIS,
  author = {Pulay, Peter},
  title = {Convergence Acceleration of Iterative Sequences. The Case of Scf Iteration},
  journal = {Chem. Phys. Lett.}, year = {1980}, volume = {73}, pages = {393--398},
  doi = {10.1016/0009-2614(80)80396-4}
}

@article{Broyden1970,
  author = {Broyden, C. G.},
  title = {The Convergence of a Class of Double-Rank Minimization Algorithms},
  journal = {J. Inst. Math. Appl.}, year = {1970}, volume = {6}, pages = {76--90},
  doi = {10.1093/imamat/6.1.76}
}

@article{Fletcher1970,
  author = {Fletcher, Roger},
  title = {A New Approach to Variable Metric Algorithms},
  journal = {Comput. J.}, year = {1970}, volume = {13}, pages = {317--322},
  doi = {10.1093/comjnl/13.3.317}
}

@article{Goldfarb1970,
  author = {Goldfarb, Donald},
  title = {A Family of Variable-Metric Methods Derived by Variational Means},
  journal = {Math. Comput.}, year = {1970}, volume = {24}, pages = {23--26},
  doi = {10.1090/S0025-5718-1970-0258249-6}
}

@article{Shanno1970,
  author = {Shanno, David F.},
  title = {Conditioning of Quasi-Newton Methods for Function Minimization},
  journal = {Math. Comput.}, year = {1970}, volume = {24}, pages = {647--656},
  doi = {10.1090/S0025-5718-1970-0274029-2}
}

@article{Nocedal1980,
  author = {Nocedal, Jorge},
  title = {Updating Quasi-Newton Matrices with Limited Storage},
  journal = {Math. Comput.}, year = {1980}, volume = {35}, pages = {773--782},
  doi = {10.1090/S0025-5718-1980-0566437-1}
}

@article{DennisMore1977,
  author = {Dennis, J. E. and Mor{\'e}, Jorge J.},
  title = {Quasi-Newton Methods, Motivation and Theory},
  journal = {SIAM Rev.}, year = {1977}, volume = {19}, pages = {46--89},
  doi = {10.1137/1019005}
}

@article{Goedecker1999,
  author = {Goedecker, Stefan},
  title = {Linear Scaling Electronic Structure Methods},
  journal = {Rev. Mod. Phys.}, year = {1999}, volume = {71}, pages = {1085--1123},
  doi = {10.1103/RevModPhys.71.1085}
}

@article{BowlerMiyazakiGillan2002,
  author = {Bowler, David R. and Miyazaki, Tsuyoshi and Gillan, Mike J.},
  title = {Parallel Sparse Matrix Methods for Electronic Structure Calculations},
  journal = {J. Phys.: Condens. Matter}, year = {2002}, volume = {14}, pages = {2781--2798},
  doi = {10.1088/0953-8984/14/11/307}
}

@article{RubenssonRudberg2011,
  author = {Rubensson, Emanuel H. and Rudberg, Elias},
  title = {Bringing about Matrix Sparsity in Linear-Scaling Electronic Structure Calculations},
  journal = {J. Comput. Chem.}, year = {2011}, volume = {32}, pages = {1411--1423},
  doi = {10.1002/jcc.21723}
}

@article{PrachtGrantGrimme2020,
  author = {Pracht, Philipp and Grant, David F. and Grimme, Stefan},
  title = {Comprehensive Assessment of {GFN} Tight-Binding and Composite Density Functional Theory Methods for Calculating Gas-Phase Infrared Spectra},
  journal = {J. Chem. Theory Comput.}, year = {2020}, volume = {16}, pages = {7044--7060},
  doi = {10.1021/acs.jctc.0c00877}
}

@article{Grimme2010D3,
  author = {Grimme, Stefan and Antony, Jens and Ehrlich, Stephan and Krieg, Stephan},
  title = {A Consistent and Accurate Ab Initio Parametrization of Density Functional Dispersion Correction ({DFT-D}) for the 94 Elements H--Pu},
  journal = {J. Chem. Phys.}, year = {2010}, volume = {132}, pages = {154104},
  doi = {10.1063/1.3382344}
}

@article{Grimme2011D2,
  author = {Grimme, Stefan},
  title = {Semiempirical {GGA}-Type Density Functional Constructed with a Long-Range Dispersion Correction},
  journal = {J. Comput. Chem.}, year = {2006}, volume = {27}, pages = {1787--1799},
  doi = {10.1002/jcc.20495}
}

@article{Vreven2006ONIOM,
  author = {Vreven, T. and Byun, K. S. and Kom{\'a}romi, I. and Dapprich, S. and Farkas, O. and Morokuma, K. and Frisch, M. J.},
  title = {Combining Quantum Mechanics Methods with Molecular Mechanics Methods in ONIOM},
  journal = {J. Chem. Theory Comput.}, year = {2006}, volume = {2}, pages = {815--826},
  doi = {10.1021/ct050289g}
}

@article{WarshelLevitt1976,
  author = {Warshel, Arieh and Levitt, Michael},
  title = {Theoretical Studies of Enzymic Reactions: Dielectric, Electrostatic and Steric Stabilization of the Carbonium Ion in the Reaction of Lysozyme},
  journal = {J. Mol. Biol.}, year = {1976}, volume = {103}, pages = {227--249},
  doi = {10.1016/0022-2836(76)90311-9}
}

@article{FieldBashKaru1989,
  author = {Field, Martin J. and Bash, Paul A. and Karplus, Martin},
  title = {A Combined Quantum Mechanical and Molecular Mechanical Potential for Molecular Dynamics Simulations},
  journal = {J. Comput. Chem.}, year = {1990}, volume = {11}, pages = {700--733},
  doi = {10.1002/jcc.540110605}
}

@article{MauriGalliCar1993,
  author = {Mauri, Francesco and Galli, Giulia and Car, Roberto},
  title = {Orbital Formulation of Electronic-Structure Calculations with Linear System-Size Scaling},
  journal = {Phys. Rev. B}, year = {1993}, volume = {47}, pages = {9973--9976},
  doi = {10.1103/PhysRevB.47.9973}
}

@unpublished{Barone2026Quadraticized,
  author = {Barone, Vincenzo},
  title  = {Quadraticizing Internal Coordinates with Chemically Designed Local Metrics},
  year   = {2026},
  note   = {JCP Communication}
}
